\documentclass[article,11pt]{emulateapj}
\usepackage{lineno}
\usepackage{graphicx}
\usepackage{times}
\usepackage{natbib}
\usepackage{amsmath}
\usepackage{amssymb}
\def\jgr{J. Geophys. Res. }

\def\aj{Astron. J.}
\def\apj{Astrophys. J.}
\def\apjl{Astrophys. J. Lett.}
\def\grl{Geophys. Res. Let.}

\def\planss{Planet. \& Space Sci.}

\def\asht{AsH$_3$ }
\def\ashtx{AsH$_3$}
\def\hto{H$_2$O }

\def\icm{cm$^{-1}$ }

\def\deg{$^\circ$ }
\def\degx{$^\circ$}

\def\mum{$\mu$m }
\def\mumx{$\mu$m}
\def\chf{CH$_4$ }
\def\chfx{CH$_4$}
\def\chtd{CH$_3$D }

\def\chisq{$\chi^2$ }
\def\chisqx{$\chi^2$}
\def\herat{He/H$_2$ }

\def\pht{PH$_3$ }
\def\phtx{PH$_3$}
\def\pthf{P$_2$H$_4$ }
\def\pthfx{P$_2$H$_4$}
\def\nht{NH$_3$ }
\def\nhtx{NH$_3$}
\def\nhfsh{NH$_4$SH }
\def\nhfshx{NH$_4$SH}

\def\nthfx{N$_2$H$_4$}

\begin{document}

\title{Saturn's south polar cloud composition and structure inferred from 2006 Cassini/VIMS
spectra and ISS images.}
\author{L.A. Sromovsky$^1$, K. H. Baines$^1$, and P.M. Fry$^1$}
\affil{$^1$Space Science and Engineering Center, University of Wisconsin-Madison,
1225 West Dayton Street, Madison, WI 53706, USA}

\slugcomment{Journal reference: L.A. Sromovsky, K.H. Baines and P.M. Fry, Icarus, https://doi.org/10.1016/j.icarus.2019.113398}
\begin{abstract}
We used 0.85 -- 5.1 \mum 2006 observations by Cassini's Visual and
Infrared Mapping Spectrometer (VIMS) to constrain the unusual vertical
structure and compositions of cloud layers in Saturn's south polar
region, the site of a powerful vortex circulation, shadow-casting
cloud bands, and evidence for ammonia ice clouds without lightning.
Finding ammonia ice spectral signatures in polar regions is surprising
because over most of Saturn an overlying haze layer of unknown
composition but significant optical depth completely obscures it,
unless penetrated by significant convection, confirmed by lightning,
as in the Great Storm of 2010-2011 (Sromovsky et al. 2013, Icarus 226,
402-418), and in the storms of Storm Alley region (Baines et al. 2009,
Planet. \& Space Sci. 57, 1650-1658).  This is clarified by our
radiative transfer modeling of VIMS spectra of the south polar
background and discrete features, using a 4-layer model that includes
(1) a stratospheric haze, (2) a top tropospheric layer of
non-absorbing (possibly diphosphine) particles near 300 mbar, with a
fraction of an optical depth (much less than found elsewhere on
Saturn), (3) a moderately thicker layer (1 -- 2 optical depths) of
\nht ice particles near 900 mbar, and (4) extending from 5 bars up to
2-4 bars, an assumed optically thick layer where \nhfsh and \hto are
likely condensables.  The ammonia layer is the main modulator of
pseudo-continuum I/F in reflected sunlight. That layer has about one
optical depth in background clouds, but about double that in the
brightest clouds, and about half that in discrete dark clouds.  What
makes the 3-\mum absorption unexpectedly apparent in these polar
clouds is the relatively low optical depth of the top tropospheric
cloud layer, which can be an order of magnitude less than in non-polar
regions on Saturn, perhaps because of polar downwelling and/or lower
photochemical production rates. We found changes in the \pht vertical
profile and \asht mixing ratio that support the existence of
downwelling within 2\deg of the pole.  We also found evidence for
step-wise decreases in optical depth of the stratospheric haze near
87.9\degx S and in the putative diphosphine layer near 88.9\degx S,
and evidence against the idea that deep convective eyewalls are
responsible for the shadows observed near the same latitudes.  In
752-nm Cassini images we identified moderately bright features
extending from shadow-producing boundaries when those boundaries
rotated to the opposite side of Saturn's pole.  Under those observing
conditions an illuminated eyewall should produce a bright feature
extending towards the pole. Instead, the features extend away from the
pole, as expected for what we call antishadows, which are bright
features produced by light illuminating a translucent layer from
below.  This provides strong qualitative evidence that both shadows
and antishadows are produced by small step changes in the optical
depth of the overlying translucent aerosol layers.

\end{abstract}
\maketitle
\shortauthors{Sromovsky et al.}
\shorttitle{Saturn's south polar cloud structure.}
\newpage

\section{Introduction}\label{Sec:intro}

In 2006 the Cassini spacecraft made important observations of Saturn's
south polar region using its Visual and Infrared Mapping Spectrometer
(VIMS) and the Imaging Science Subsystem (ISS).  These observations
provided an unprecedented detailed view of the south polar vortex and
sharply defined two shadow-casting concentric rings of cloud
structures surrounding the pole.  Initial papers
\citep{Dyudina2008Sci,Dyudina2009} described these rings as eyewalls,
evoking the image of deep convective walls like those associated with
earthly hurricanes.  However, until our work, reported in this paper,
there had been no quantitative radiative transfer analysis of these
data to determine whether the interpretation of the ring structure is
consistent with the actual cloud structure constrained by VIMS
spectral imaging.  The south polar region is also interesting because
it is a region of generally lower aerosol scattering, which provides an
opportunity to better constrain the composition and vertical structure
of Saturn's main cloud layers, and to help determine the nature of the many
discrete cloud features in this region that have spectral signatures
of ammonia ice.  VIMS observations are especially useful in this
regard because they provide access to a wide range of methane band
strengths for constraining vertical structure, coverage of near-IR
spectral signatures of candidate cloud components, and coverage of the relatively transparent
spectral region near 5 \mumx, where Saturn's thermal emission is visible and
aerosols and gases down to the 4-5 bar region can be constrained.

\begin{figure*}[!htb]\centering
\hspace{-0.15in}\includegraphics[width=3.15in]{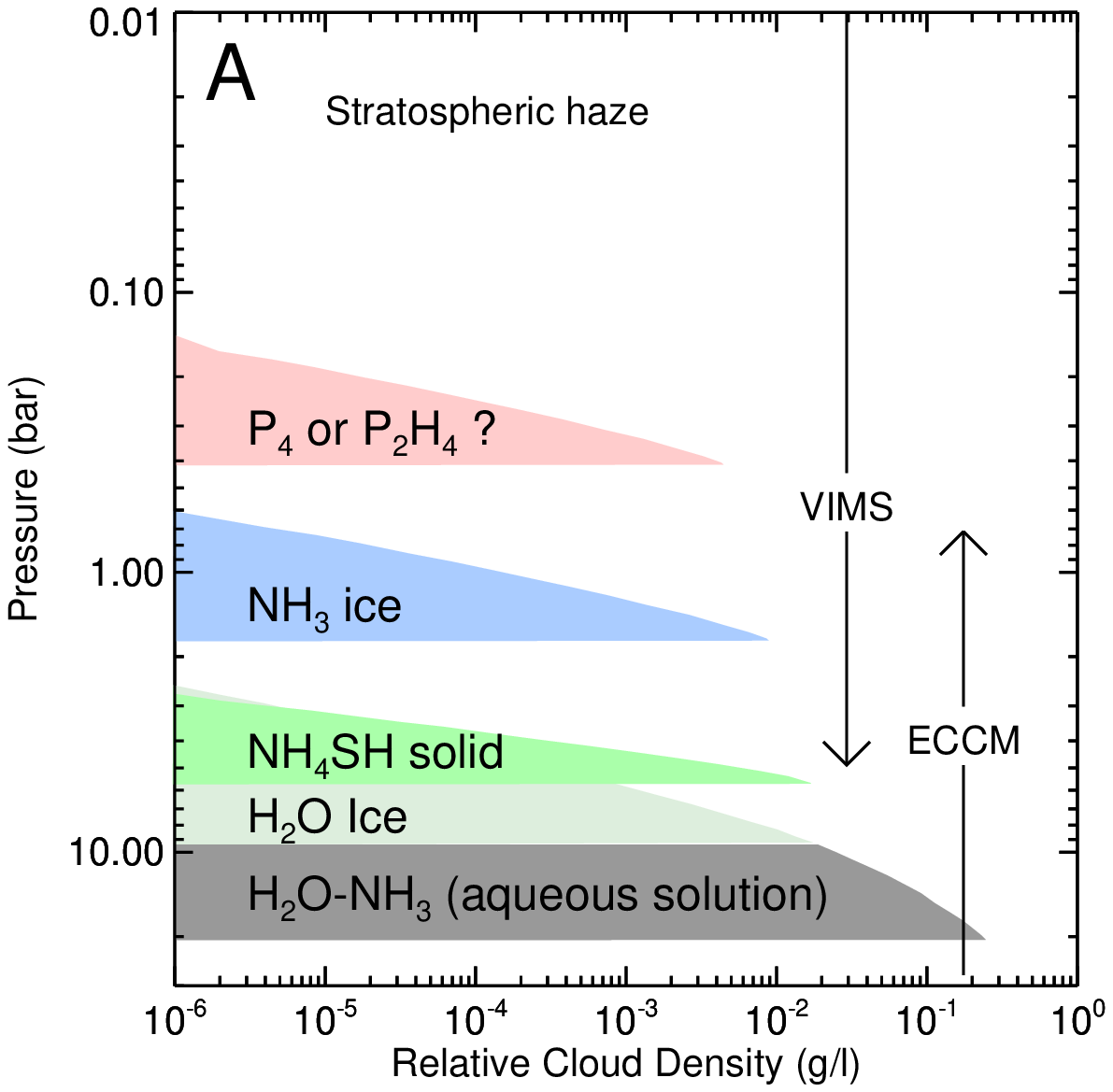}\hspace{-0.15in}
\includegraphics[width=3.15in]{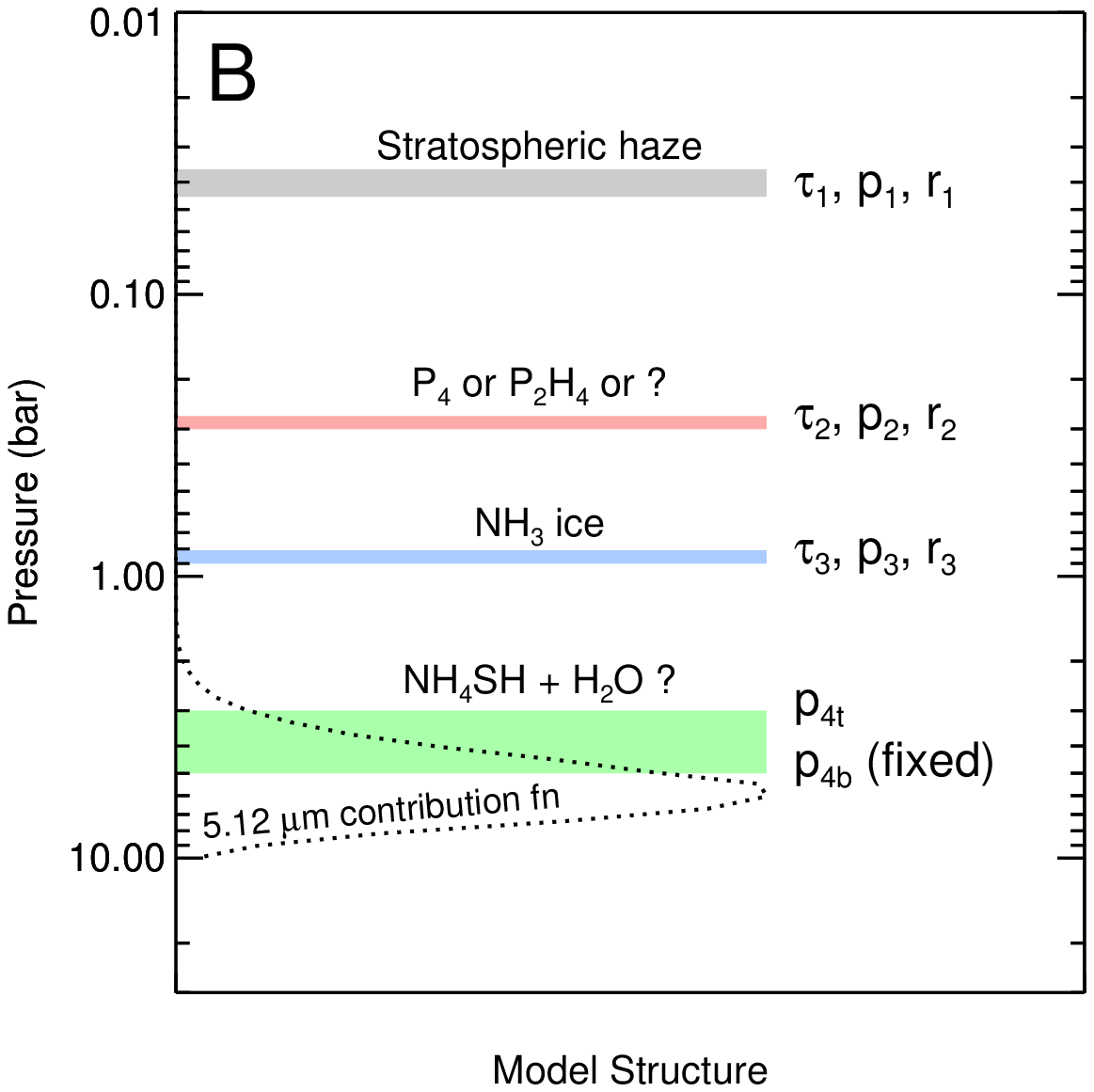}
\caption{{\bf A:} Equilibrium cloud condensation model (ECCM)
 of the composition and
  vertical distribution of clouds on Saturn for P $>$ 500 mbar
  according to \cite{Atreya2005SSR}.  All the ECCM clouds have
  densities expected to greatly exceed actual cloud amounts. For P $<$
  500 mb, a cloud layer of unknown composition (possibly P$_4$ or
  \pthfx) has been inferred from radiation transfer modeling (see
  text).  The density profile of that cloud is chosen for illustrative
  purposes.  According to \cite{Sro2013gws,Sro2018dark} this layer
  does not display any 3-\mum absorption features and thus cannot be
  composed entirely of some combination of the materials expected at
  higher pressures. The downward arrow indicates the pressure depth to
  which VIMS observations might be sensitive to cloud composition. {\bf B:} 
 Our model structure containing four sheet clouds corresponding to the
 expected compositional layers shown in A. Further details are provided
in Section \ref{Sec:param}. Also shown is
  the clear-atmosphere scaled contribution function for emission at 5.12 \mum
  from \cite{Sro2016}.}\label{Fig:eccm}\label{Fig:cloudmodel}
\end{figure*}

Our initial understanding of Saturn's vertical cloud structure is
based on the equilibrium cloud condensation model (ECCM) of
\cite{Weidenschilling1973}, which suggests that the top condensation
cloud on Saturn should be composed of \nht ice and, according to
\cite{Atreya2005SSR}, its cloud base should be near 1.7 bars
(Fig.\ \ref{Fig:eccm}).  However, this cloud layer is almost never
observed from above because most of Saturn is covered by an overlying
cloud layer of unknown composition but significant optical depth
\citep{Sro2013gws}.  From a limb-darkening analysis of Hubble Space
Telescope Wide Field Planetary Camera 2 images over the 1994-2003
period \cite{Perez-Hoyos2005} found that this upper tropospheric haze
layer had a strong latitudinal dependence in optical thickness,
reaching 20 -- 40 at the equator, but as low as 4 at 86\degx S, which
dropped to 2.5 in 2003, all at a wavelength of 814 nm.  For 2002
observations, they found a bottom pressure close to 400 mbar at all
latitudes and a top pressure increasing from about 50 mbar at the south pole to 100
mbar at the equator.  In a 5\deg wide latitude band centered at
36\degx S, referred to as Storm Alley, \cite{Sro2018dark} found that
in the background cloud structure this layer extended from about 200
mbar to 400 mbar with an optical depth of 4-6 at a wavelength of 2
\mumx.  A similar layer was found in the cleared out region that
developed in the wake of the Great Storm of 2010--2011 after the storm
itself dissipated \citep{Sro2016}. There the layer was found in the
140 mbar to 400 mbar region with an optical depth of about 3 shortly
after the storm ended, and growing slowly by about an optical depth
over seven months. According to \cite{Fouchet2009} a leading
candidate for this upper tropospheric haze is diphosphine (\pthfx),
which is so far not a spectrally testable hypothesis because very
little is known about the optical properties of \pthfx.

\begin{figure*}[!htb]\centering
\includegraphics[width=5.15in]{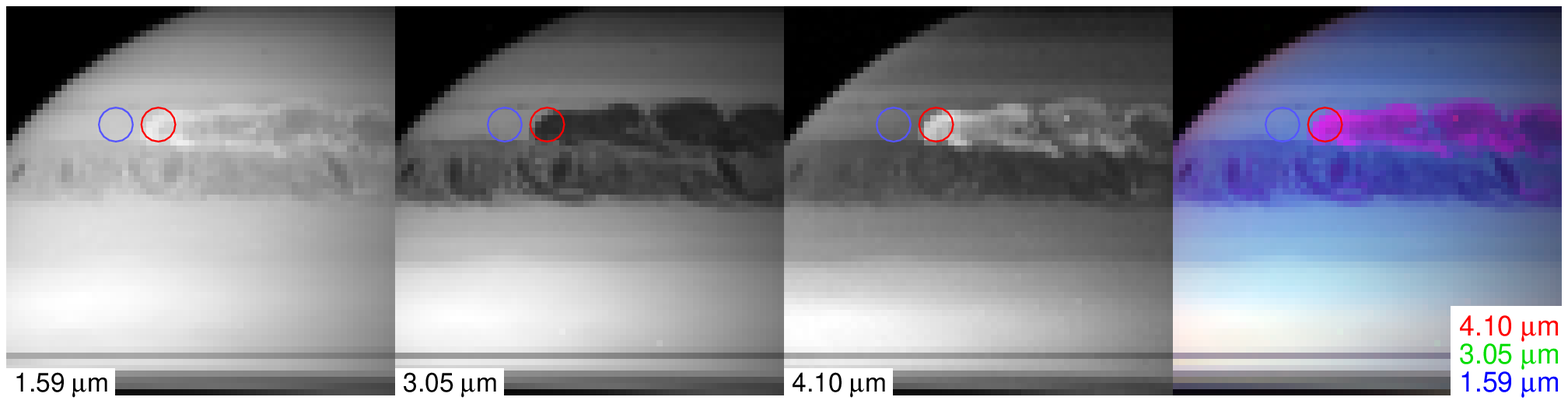}
\includegraphics[width=5.1in]{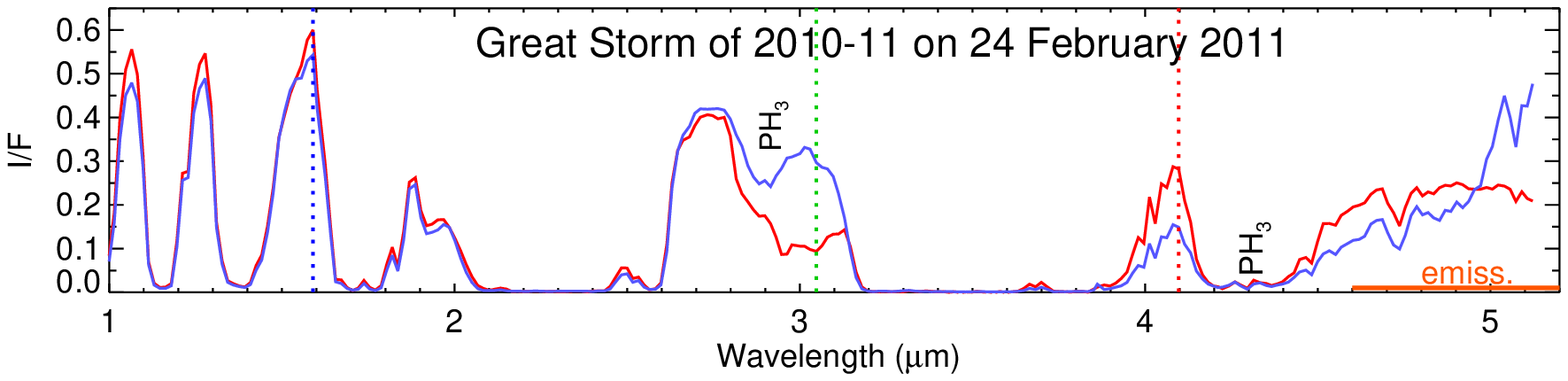}
\includegraphics[width=5.15in]{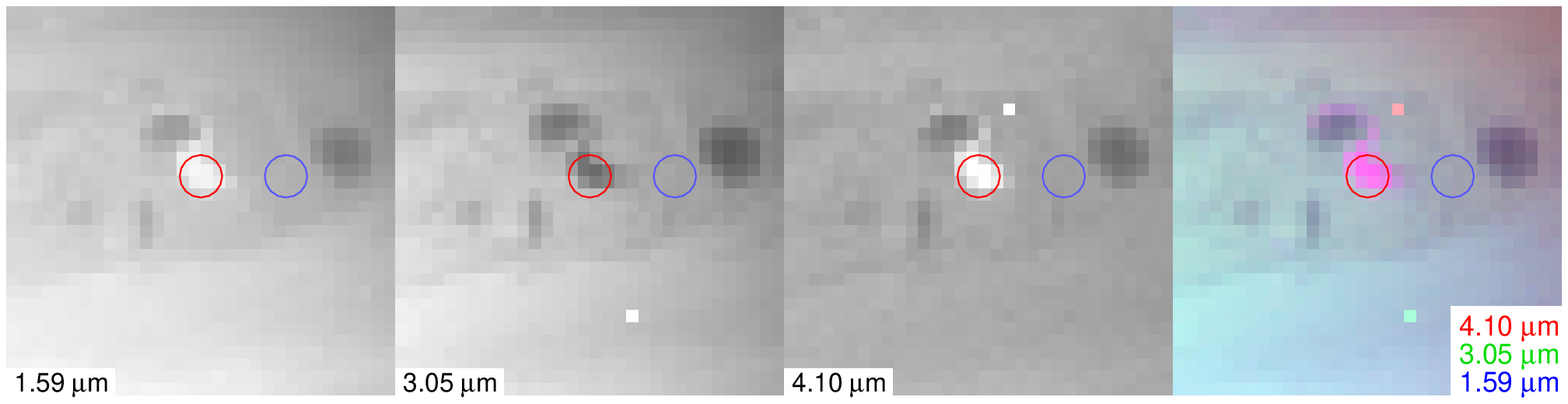}
\includegraphics[width=5.1in]{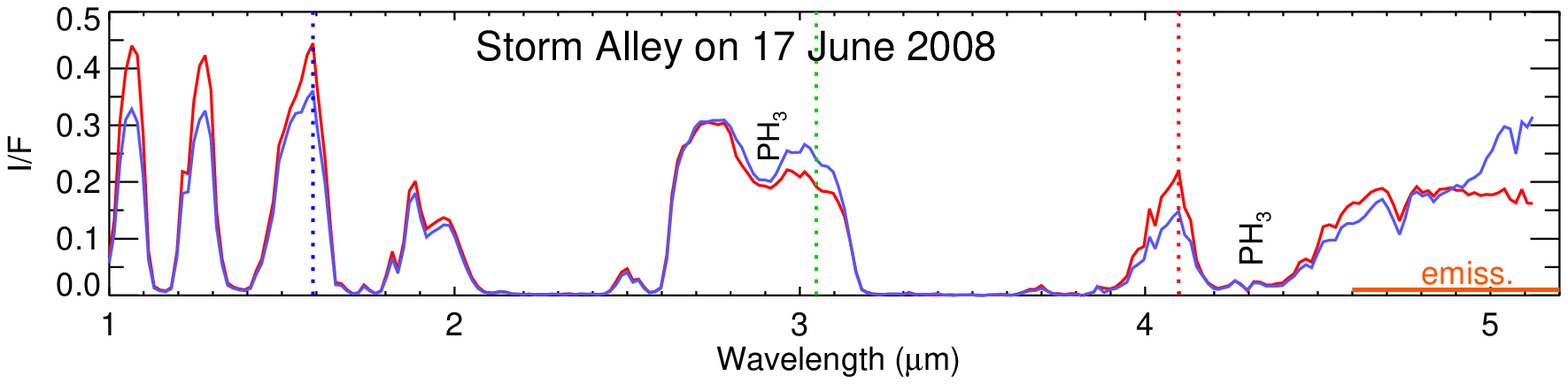}
\caption{VIMS observations of the Great Storm (top) and Storm Alley (bottom) illustrate
rarely seen spectral signatures of ammonia ice. Blue and
red circles identify background and storm clouds respectively.  
Corresponding spectra are plotted using the same color assignments. The
color composite images produce a distinctive magenta
color for features with the \nht absorption signatures, which include
brighter continuum peaks between 1 and 1.6 \mum and at 4.1 \mumx, combined with
absorption by \nht ice near 3 \mumx. }
\label{Fig:GWScase16}
\end{figure*}

Evidence of Saturn's underlying \nht ice
cloud (its strong 3-\mum absorption signature) has only been evident
under unusual circumstances.  It has been seen in
association with lightning storms, including the Great Storm of
2010-2011 \citep{Sro2013gws} near 35\degx N planetocentric latitude,
and in much smaller storms located near 36\degx S in the Storm Alley
region \citep{Baines2009stormclouds}.  In the Great Storm, strong convection
apparently brought \nht ice particles to
 the visible cloud tops. As illustrated in
Fig.\ \ref{Fig:GWScase16}, compared to background clouds, the bright
convective cloud at the head of the Great Storm is much darker at 3
\mum than it is at shorter and longer wavelengths.   Just
where the cloud is brightest at continuum wavelengths, it is also the
darkest at 3 \mumx.  A similar, though less dramatic signature is seen in the bright storm
clouds in Storm Alley, also illustrated in Fig.\ \ref{Fig:GWScase16}.
This is consistent with ammonia ice
reaching into but not fully penetrating the upper cloud
\citep{Sro2018dark}.  In both of these examples, the
existence of strong convection was confirmed by the observation of
associated lightning.  Because no lightning has ever been detected in
Saturn's polar regions, indicating a lack of strong convection, it was
surprising to see 3-\mum absorption features in discrete clouds within
the eye of the north polar vortex \citep{Baines2018GeoRL}, as well as
in the south polar region that is the subject of this paper.  A hint
of what might be responsible is the \cite{Perez-Hoyos2005} finding of
a dramatically lower optical depth of the upper tropospheric cloud in
the south polar region.

Here we describe quantitative radiative transfer models of both bright
and background cloud features.  We use their spectral signatures to
constrain both composition and vertical structure.  We use the
difference between observed spectra and spectra computed from aerosol
models as constraints on those models. The discussion is organized as
follows.  We first describe our data selection approach.  We then describe the Visual and Infrared
Mapping Spectrometer (VIMS) and the observations that we use to gain
new insights into the nature of these cloud features. That is followed
by descriptions of our approach to radiation transfer modeling, our
parameterization of aerosol and gas profiles, and our approach to
fitting the observations.  Next, we describe the results of fitting
background clouds, bright clouds, and dark features.  Finally we
discuss the implications of these results and summarize our
conclusions. 
  
\section{South polar observations}

\subsection{Data selection}

We selected VIMS observations to take advantage of the near-IR methane bands
that can be used to constrain vertical aerosol structures, aerosol composition,
and the abundance of arsine and vertical profile for phosphine.  Although VIMS also
provides visible spectral observations, we decided not to include these in the
analysis because they have significant stray light and striping artifacts as well
as imaging pointing offsets that create difficulties in analysis of discrete features.
The particular VIMS data set we selected provided an optimal compromise
between spatial resolution and viewing geometry.  The VIMS observations
also were made when valuable ISS imaging was available to
help understand the latitudinal variations at an even higher spatial
resolution than available from VIMS.  This is also during the period
when ISS imaging was revealing putative eyewalls and cloud shadows, which
we hoped to investigate for the first time with quantitative radiation
transfer modeling. There are additional observations near this time
that provide similar advantages, and could be productively analyzed.

\subsection{VIMS instrument characteristics and data reduction}

As described by \cite{Brown2004}, the VIMS instrument combines two
mapping spectrometers, one covering the 0.35-1.0 \mum spectral range,
and the second covering an overlapping near-IR range of 0.85-5.12
\mumx. The near-IR spectrometer use 256 contiguous channels sampling
the spectrum at intervals of approximately 0.016 \mumx.  The
instantaneous field of view of each pixel pair as combined in the
observations we used covers 0.5 $\times$ 0.5 milliradians, and a
typical frame has dimensions of 64 pixels by 64 pixels.  The
responsivity of the VIMS near-IR detector array is significantly
reduced at joints between order sorting filters, which occur at
wavelengths of 1.64 \mumx, 2.98 \mumx, and 3.85 \mum
\citep{Miller1996SPIE,Brown2004}, causing potential calibration
uncertainties.  We avoided comparisons between models and observations
at the most strongly affected regions near 1.64 \mum and 3.85 \mumx,
but found that the effects of the 2.98-\mum joint were not a
significant issue.

The VIMS data sets were reduced using the USGS ISIS3
\citep{Anderson2004} vimscal program, which was derived from the
software provided by the VIMS team (and is available on PDS archive
volumes). The radiometric calibration utilized the RC17 calibration
\citep{Clark2012}, and conversion to I/F (reflectivity relative to
Lambertian reflector normally illuminated) used the solar spectrum
packaged with the ISIS3 and PDS-supplied software, which is based on
the \cite{Drummond1973} solar spectrum. The spectra were
converted to the more recent RC19 calibration following
\cite{Sro2018dark}.  Navigation of VIMS cubes
(computation of planet coordinates and illumination parameters for
each pixel of a VIMS cube) utilized kernels supplied by JPL NAIF
system and SPICELIB software \citep{Acton1996}. 

Our detailed analysis is based primarily on selected VIMS near-IR
observations obtained at 19:35 UT and 20:05 UT on 11 October 2006,
which are described more completely in Table\ \ref{Tbl:obslist}, along
with corresponding observing conditions.  For the chosen 2006
observations, a typical solar zenith angle is 75.5\deg with a typical
observer zenith angle of 57\degx.  These observations include
measurements of reflected sunlight as well as thermal emission, which
we found to provide the best combination of constraints on upper and
lower cloud parameters and on \nht and phosphine
(\phtx). Table\ \ref{Tbl:obslist} also includes information about VIMS
observations displayed in Fig.\ \ref{Fig:GWScase16}.

\begin{table*}[!htb]\centering
\caption{Observing conditions for VIMS IR data cubes and ISS images used to constrain
cloud structure models.}
\vspace{0.1in}
\setlength\tabcolsep{3pt}
\begin{tabular}{ l c c c c c r}
                             &                   &     UT Date  &        Start &   Pixel &    Phase & Figure \\
              Observation ID$^1$ &     Cube Version &   {\footnotesize YYYY-MM-DD} &      Time &     size & angle & Ref. \\
\hline
{\small 030SA\_FEATRACK005} & {\small V1539288419\_1} & 2006-10-11 & 19:35:01.6  & 145 km & 37.2\deg & \ref{Fig:innereye}, \ref{Fig:innershort}\\
{\small 030SA\_FEATRACK005} & {\small V1539290255\_1} & 2006-10-11 & 20:05:37.6  & 142 km & 33.7\deg & \ref{Fig:outereye} \\
{\small 072SA\_DYNMOVIE} &  {\small V1592396725\_1}  & 2008-06-17 & 11:47:35.1   & 403 km & 24.9\deg & \ref{Fig:GWScase16}\\
{\small 145SA\_WIND5HR001} & {\small V1677201862\_3}  & 2011-02-24 & 00:36:35.5  & 883 km & 52.0\deg & \ref{Fig:GWScase16}\\
\hline
ISS image ID               &  ISS Filter                  &            &              &             &     \\
\hline
  W1539293298 &    MT3 (890 nm)  &  2006-10-11 &   20:56:08  & 33.6  km & 27.3\deg &\ref{Fig:isstriple} \\
  W1539293315 &    CB2 (752 nm)  &  2006-10-11 & 20:56:37  & 16.8 km & 27.3\deg & \ref{Fig:polecontext}, \ref{Fig:isstriple}\\
  W1539293355  &    MT2 (728 nm) &  2006-10-11 &   20:57:14  & 16.8 km & 27.3\deg &\ref{Fig:isstriple} \\
\hline
\end{tabular}\label{Tbl:obslist}
\parbox{4.5in}{$^1$The full observation ID has a leading VIMS\_ and suffix \_PRIME for rows 3 and 4. \par
}
\end{table*}
\renewcommand{\baselinestretch}{1.3}\normalsize

\subsection{ISS instrument characteristics and data reduction}

We also made use of 2006 observations by the Cassini Imaging Science
Subsystem (ISS).  The ISS \citep{Porco2004SSR} has a
narrow angle camera (NAC) with a field of view 0.35\deg across, and a
the wide angle camera (WAC) with a FOV of 3.5\degx, both using
1024-pixel square CCD arrays with pixel scales of 1.24 and 12.4
arcseconds/pixel respectively (in the unbinned imaging mode).  For our
analysis we used WAC images (identified in Table\ \ref{Tbl:obslist}),
as no relevant NAC images were available.  The image files were
retrieved from the NASA Planetary Data System’s Imaging Node and
processed with the USGS ISIS 3 cisscal application, which is derived
from the IDL cisscal application developed by the Cassini Imaging
Central Laboratory for Operations (CICLOPS). This cisscal application
produces images in I/F units (where a Lambertian reflector,
illuminated and viewed normally at the same distance from sun as the
target, has an I/F of 1). Ephemeris and pointing data allowing
transformations between image and planet coordinates are disseminated
by NASA's Navigation and Ancillary Information Facility
\citep{Acton1996}.
 
\subsection{Cassini observations of south polar clouds in 2006}

\begin{figure*}[!htb]\centering
\includegraphics[width=4.2in]{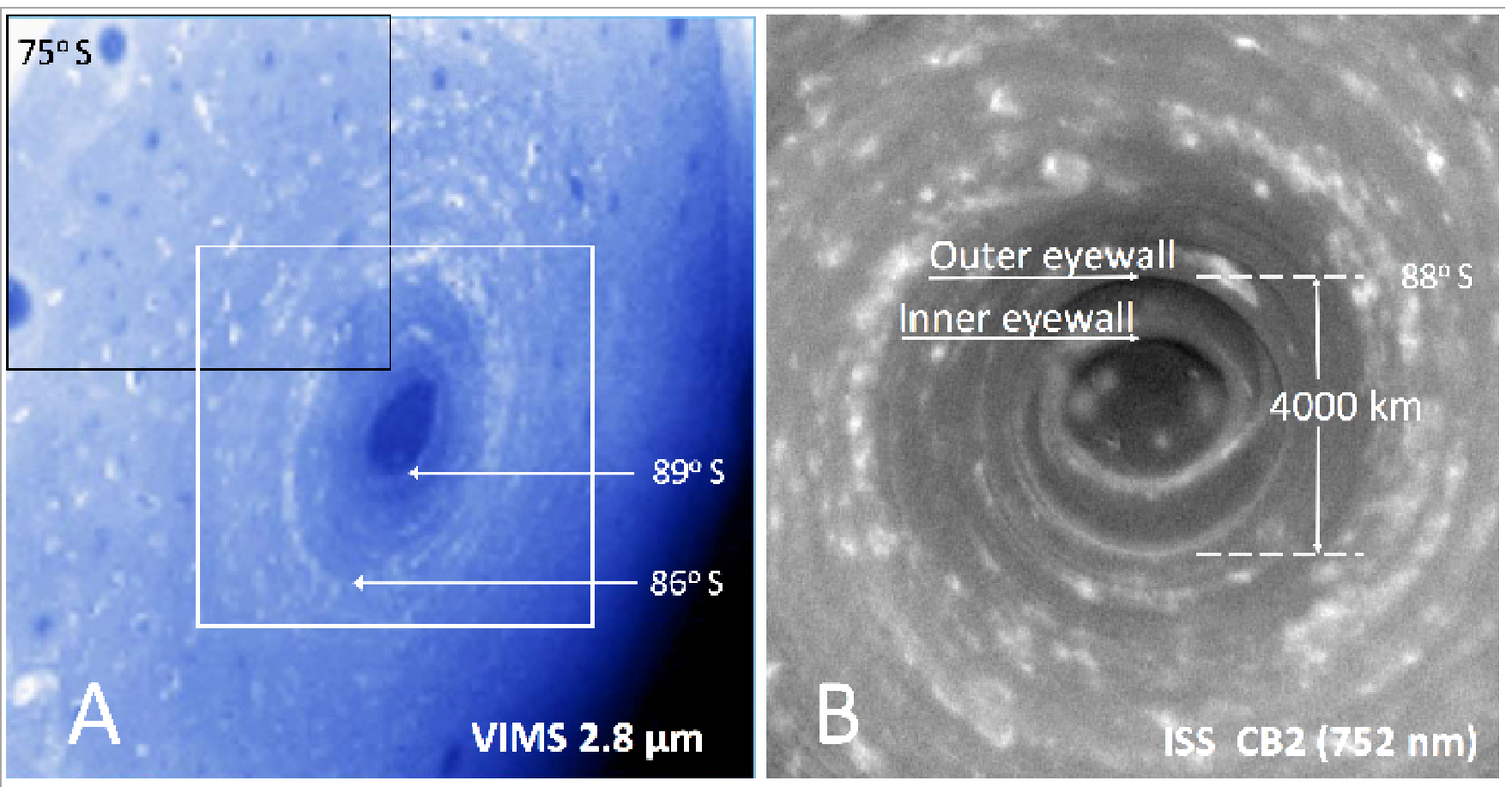}
\includegraphics[width=2.0in]{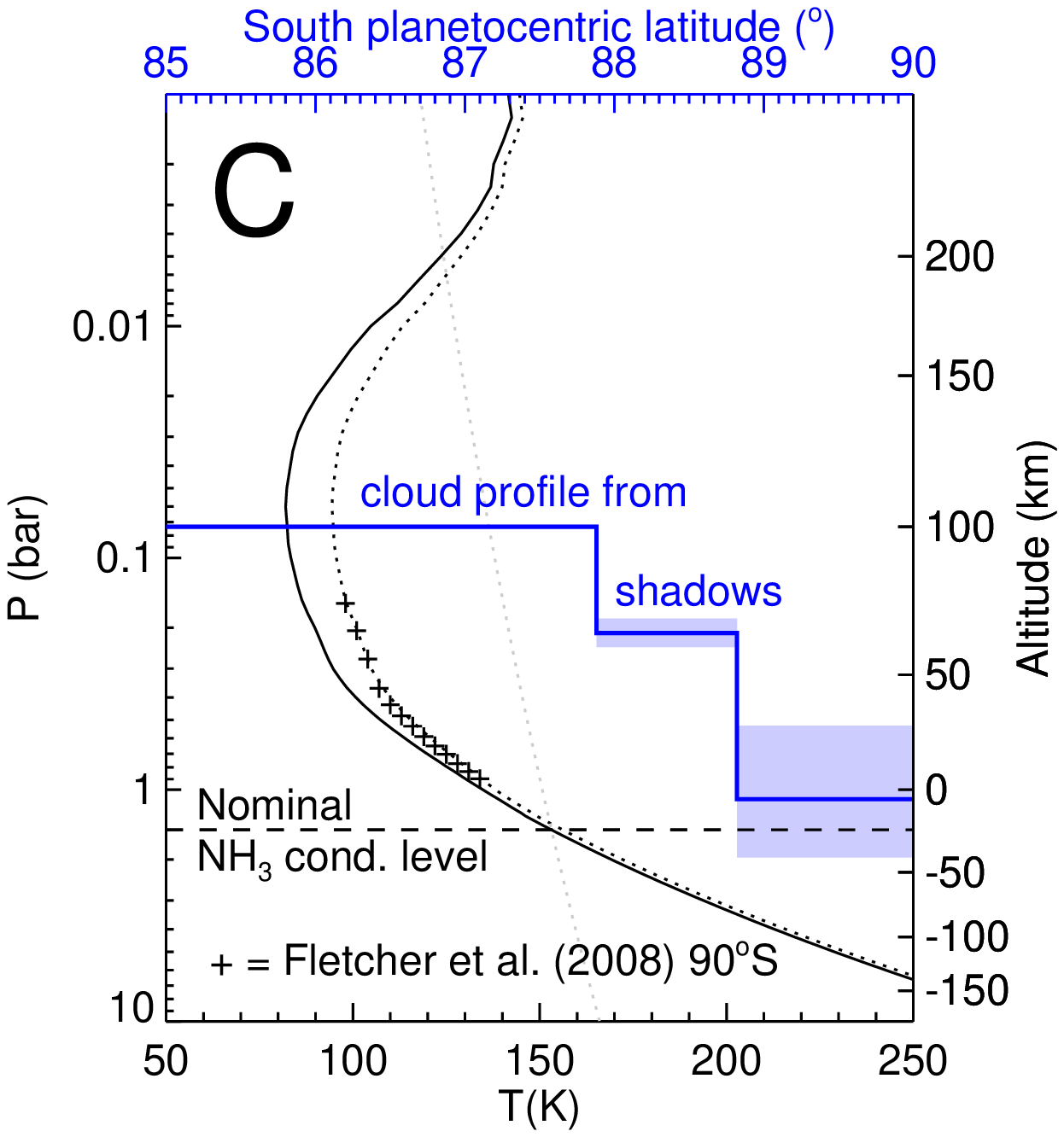}\\
\caption{Views of Saturn's south polar region on 11 October 2006. The VIMS image
  at 2.8 \mum (A) is a 7-image mosaic from
  \cite{Baines2008}.  The black and white squares indicate the
  approximate coverage of individual data cubes used in our analysis. 
The Cassini ISS image (B) is  a polar projection of an
image taken at 20:56:37.0 UT (following
the VIMS observations by about 45 minutes) 
using a CB2 (752 nm) filter. Its identification number
  is W1539293315. The very dark region centered in the south pole in the VIMS image
  corresponds to the region inside what is indicated as the inner ``eyewall''
   in the ISS image (B).  Note the apparent shadows at both the
  inner and outer ``eyewall'' boundaries in the ISS image (sunlight is incident
from the top). The vertical
cloud profile (C) is a possible interpretation of cloud heights inferred from the shadows
lengths measured  by \cite{Dyudina2009}.  The temperature profile shown as a solid curve
is from \cite{Lindal1985} compressed to account for higher gravity at polar latitudes. 
The dotted curve is the same profile with temperatures scaled to match the 90\degx S
profile of \cite{Fletcher2008}
}\label{Fig:polecontext}
\end{figure*}

The morphology of Saturn's south polar region in October of 2006 is
illustrated in Fig.\ \ref{Fig:polecontext}, where a mosaic of VIMS
observations at 2.8 \mum (panel A) is compared with an ISS image taken
with a 752-nm filter (panel B).  The region from 89\degx S to the pole
is quite dark in the VIMS image and bounded by a cloud band edge
labeled as the ``inner eyewall'' in the ISS image.  Also note the
apparent shadow seen below the top edge of the inner eyewall and the
second shadow seen extending from a similar feature labeled as the
``outer eyewall''.  These features were first analyzed by
\cite{Dyudina2008Sci} who inferred, from the shadow geometry, cloud
height differences of 40$\pm$20 km at the outer boundary and 70$\pm$30
km for the inner boundary. They also inferred that the top cloud layer
extended up into the stratosphere because of boundaries seen in MT-2
and MT-3 methane band images.  A possible interpretation of these
measurements in the form of a complete cloud profile is shown in
Fig.\ \ref{Fig:polecontext}C.  If the innermost region of the eye, out
to about 89\degx, is assumed to be roughly at the level of ammonia
condensation, suggesting that the overlying aerosols are either absent
or of very low optical depth, then a second cloud top, which would
produce the inner shadow, would be expected near the 200 mbar level, which
would likely be of the same unknown composition as the ubiquitous
upper tropospheric layer commonly found in this region.  A second
step, up to about 70 mbar, might be at the level of the stratospheric haze.  

The
bright cloud feature just outside the outer wall and intersected by
the left end of the dashed line is the same feature seen above the
darkest region in the VIMS image.  This is in a region of intermediate
darkness extending from 89\degx S to 86\degx S.
Further from the pole is a somewhat brighter region containing
many even brighter small discrete features in both images and a smaller number of 
small dark features, which are only seen clearly in the VIMS image. 
There is another step change in cloud properties near 75\degx S (barely
visible in the upper left corner of the left panel in Fig.\ \ref{Fig:polecontext}), where
cloud reflectivity increases, indicating a likely increase in optical
depth at that point.  

\begin{figure*}[!hbt]\centering
\includegraphics[width=6in]{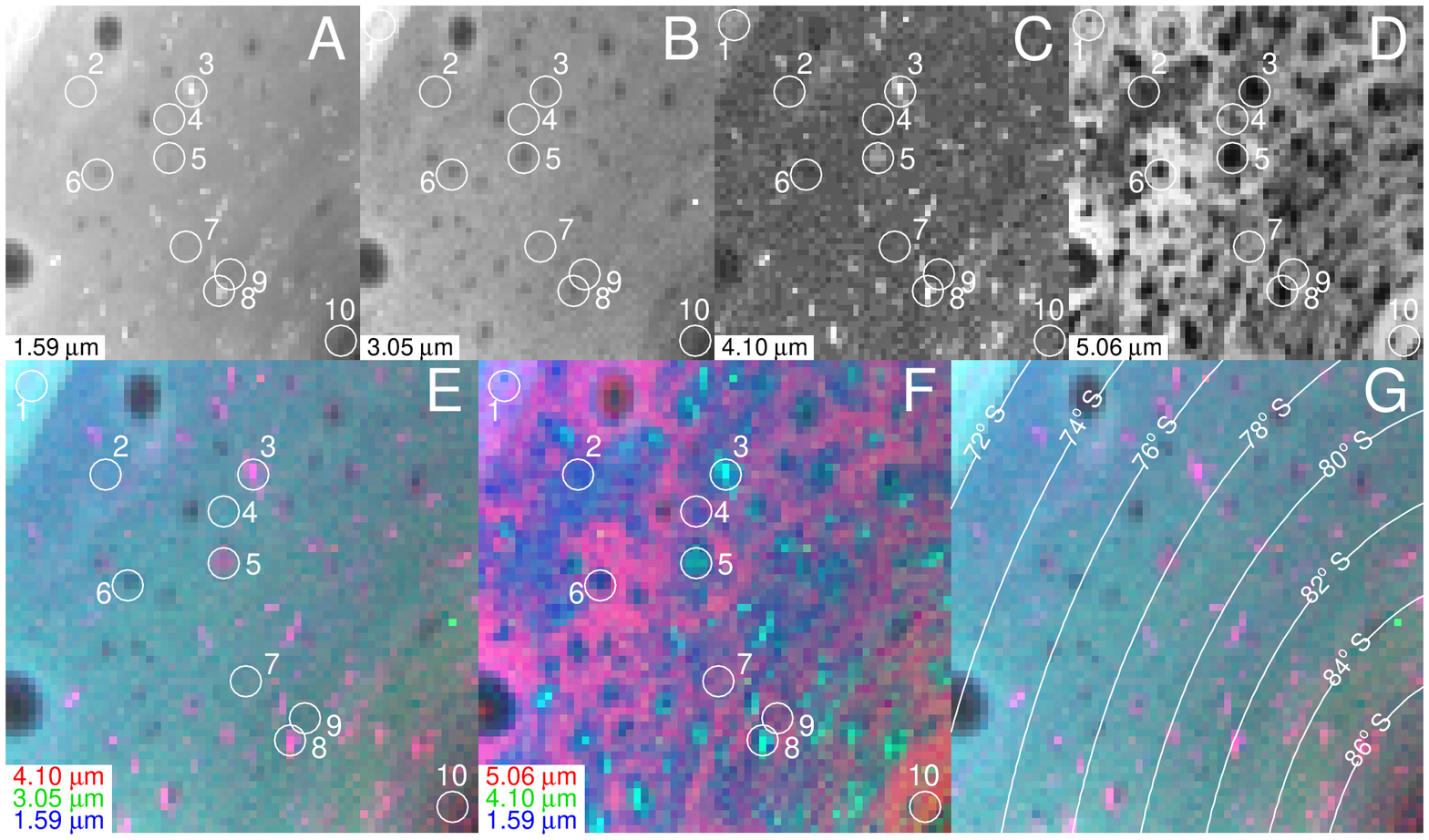}
\includegraphics[width=6.in]{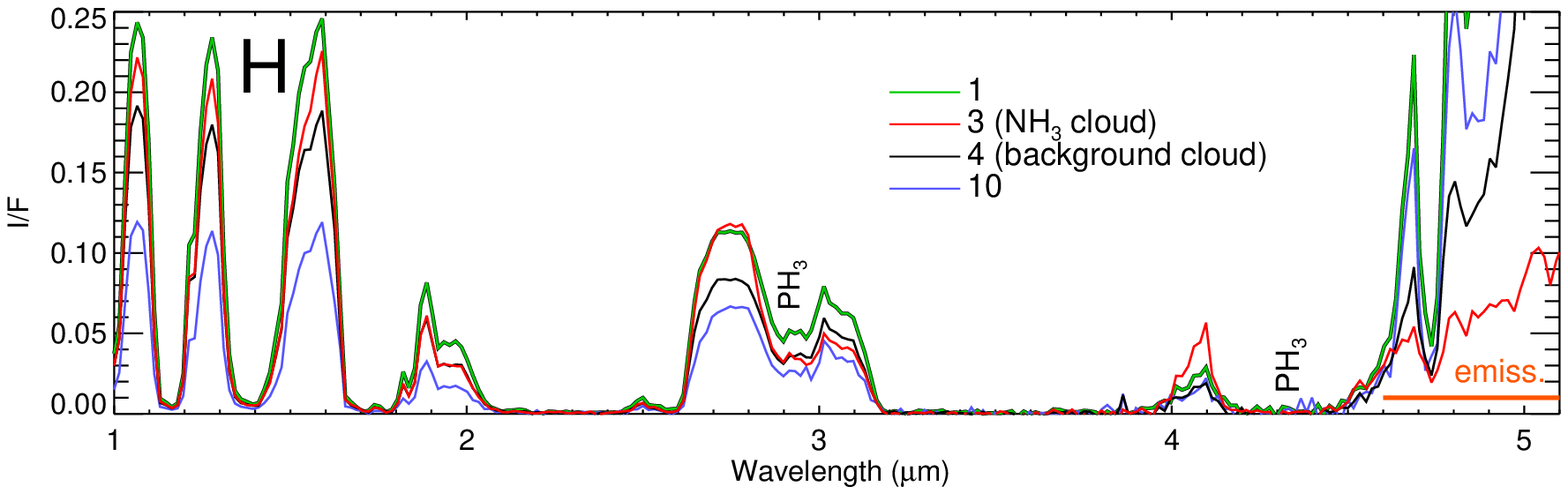}
\vspace{-0.2in}
\caption{These 11 October 2006 images from cube V1539290255.1, taken
  at 20:05 UT (Table\ \ref{Tbl:obslist}), cover the region roughly
  defined by the black square in Fig.\ \ref{Fig:polecontext}, using an
  exposure of 80 ms/pixel, yielding signals well below saturation
  levels. The top row of panels display image planes at key
  wavelengths indicated at the lower left of each panel.  Panels E
  (with target circles) and G (with latitude grids) display the same
  color composite that highlights \nht-signature clouds in
  magenta. Panel F highlights regions of high 5-\mum thermal emission.
  Panel H plots selected spectra of targets indicated by numbered
  circles in the image panels.  Note that the label ``\nht cloud''
  denotes that the cloud feature displays an ammonia spectral
  signature, not that all the aerosols in the column are made of
  ammonia.}\label{Fig:outereye}
\end{figure*}

Spectral images and spectra from just outside the inner eye to about
75\degx S are shown in Fig.\ \ref{Fig:outereye}.  Many small bright
and dark features are seen at pseudo continuum wavelengths of 1.59
\mum and 4.1 \mumx, while at 3.05 \mum these bright features appear
dark, a familiar characteristic of ammonia clouds seen in the major
storm clouds on Saturn (as in Fig.\ \ref{Fig:GWScase16}).  Spectra from a bright feature (3) and from a
nearby background region (4) are shown in the bottom panel of the figure,
using red and blue colors respectively.  Note that the bright feature
is only slightly darker than surrounding clouds at 3.05 \mumx, but is
20\% to 100\% brighter at the pseudo continuum wavelengths.  This
suggests that the underlying ammonia cloud is near its maximum reflectivity
for even an infinite optical depth, and that the  cloud layer overlying the ammonia layer is
sufficiently transparent to see some of the effects of ammonia absorption even in the
background region.  Also note that the background cloud
in this region is very transparent to thermal emission in the 4.6 -- 5.1 \mum region,
while the bright cloud feature has enough long wavelength absorption
to block most of that emission.

The color composite image in panels E and G of
Fig.\ \ref{Fig:outereye} displays a characteristic magenta color for
the bright features that are dark at 3.05 \mumx, which is essentially
the same color seen for storm clouds in similar color composites
displayed in Fig. \ref{Fig:GWScase16}. All of the small cloud features
that are bright at continuum wavelengths in Fig.\ \ref{Fig:outereye}
display a similar color. Note also that some of the dark features are
relatively dark at all the key wavelengths (feature 6 for example),
which might be a result of a local reduction in optical depth or, less
likely, the presence of a broadband absorber.  Also note that feature
5, while dark at 3.05 \mum and bright at 4.1 \mum is not brighter than
background clouds at 1.59 \mumx, which could be a result of larger
particles in the ammonia cloud layer that don't brighten as much at
shorter wavelengths.  It should be noted that we found no obvious
discrete features in any 2-\mum image (not shown), indicating that the
bright cloud features do not reach as high as the pressure level to
which that wavelength is sensitive (as will be shown later, its
two-way optical depth unity level is of the order of 300 mbar).

\begin{figure*}[!hbt]\centering
\includegraphics[width=6in]{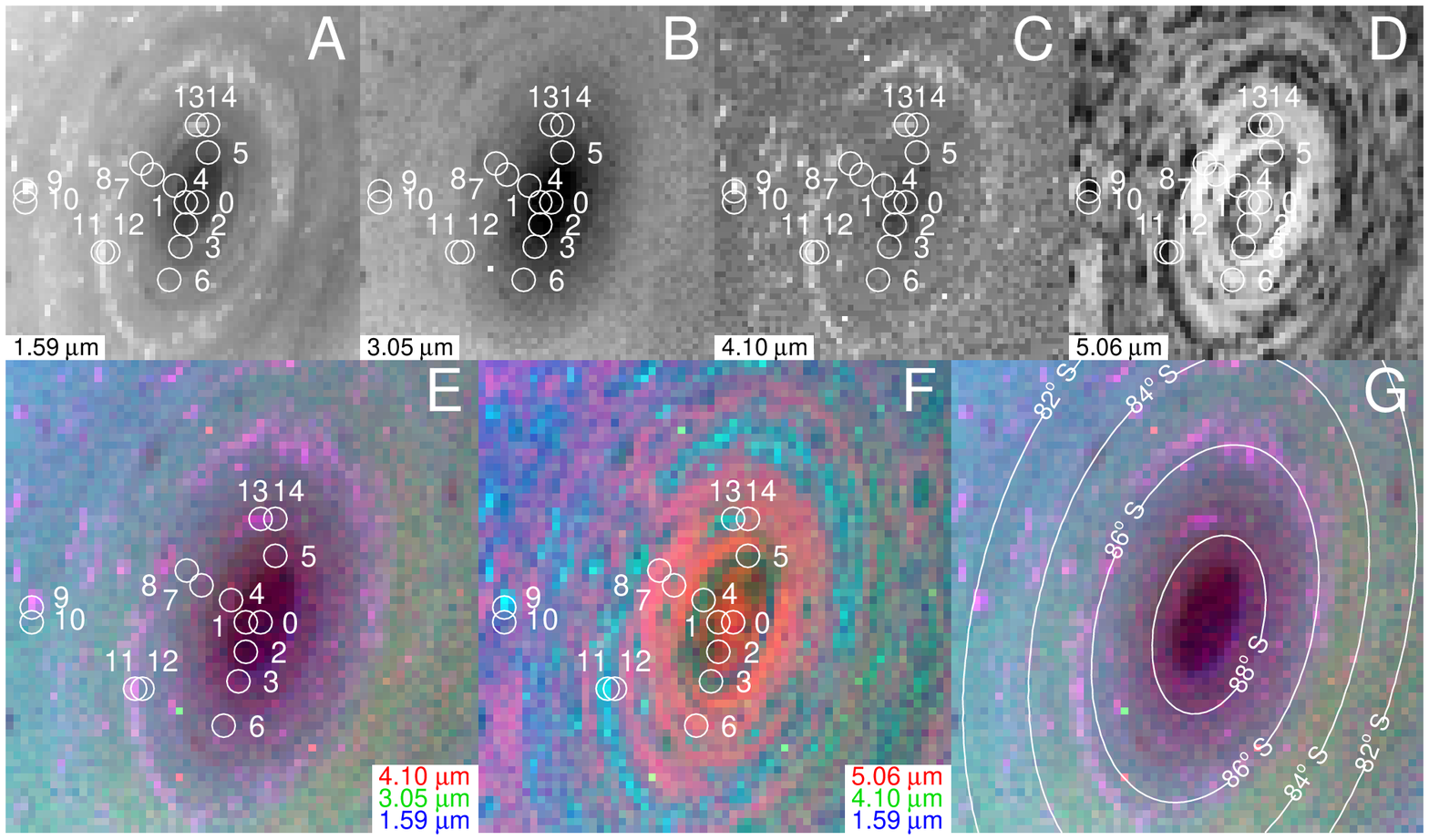}
\includegraphics[width=6in]{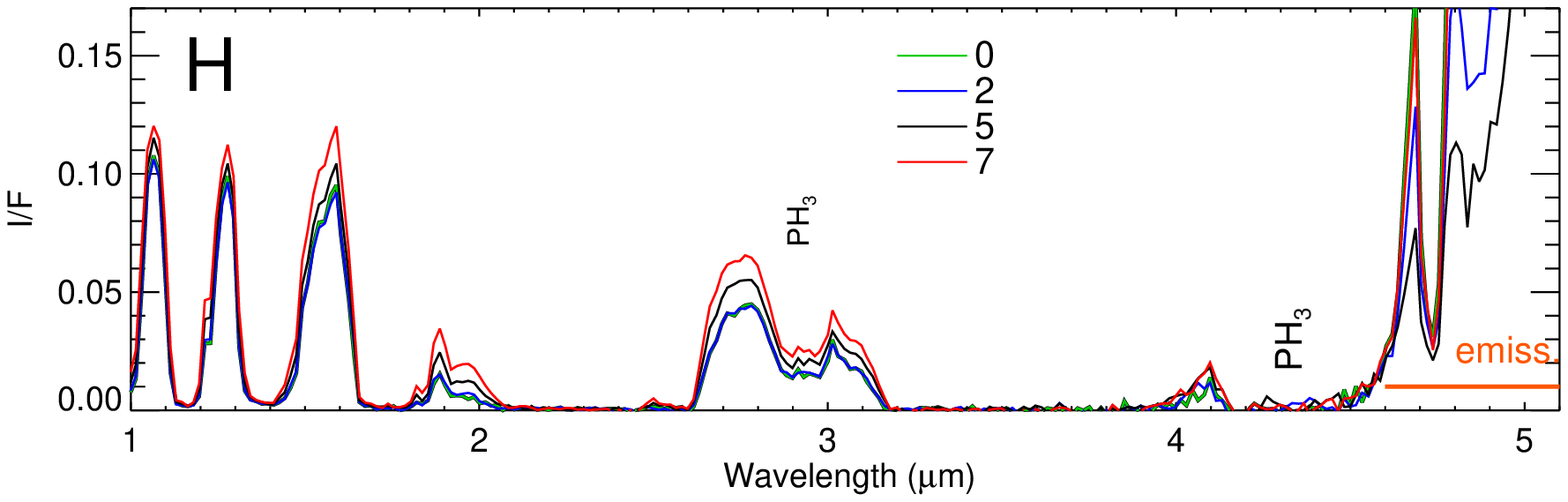}
\caption{As in Fig.\ \ref{Fig:outereye}, except that these 11 October 2006
  observations are  from cube V1539288419 taken at
  19:35 UT (Table\ \ref{Tbl:obslist}), which covers the inner eye region.}\label{Fig:innereye}
\end{figure*}

VIMS images and spectra from the inner eye region out to about
85\degx S are shown in Fig.\ \ref{Fig:innereye}.  These images also
contain many cloud features (9 and 13 for example) with the spectral
signatures of ammonia ice clouds similar to those seen in
Fig.\ \ref{Fig:outereye} and in the storm feature images in
Fig.\ \ref{Fig:GWScase16}.  This region has a more organized
distribution of ammonia-signature cloud features, with many
distributed along a bright ring with an inner edge at about 86\degx S.
This ring coincides with a dark ring seen at the thermal emission
wavelength of 5.06 \mum (in panel D), which indicates that the total
aerosol optical depth above the 4-bar level in this narrow band is quite large.  The portion
of the eye from about 89\degx S to the pole is relatively dark in
reflected sunlight compared to surrounding regions, which is a result
of reduced aerosol scattering. This is especially noticeable near 2
\mum (not shown in this figure, but illustrated later in
Fig.\ \ref{Fig:innershort}C in Section\ \ref{Sec:comp}), which is a
wavelength region sensitive to aerosols in the upper troposphere.  On
the other hand, the eye is quite bright at thermal emission
wavelengths, indicating a reduced level of aerosols at deeper levels.
However, the innermost region of the eye is not the brightest region
in emission.  Somewhat greater emission is seen near 88\degx S, in the
region between the inner and outer eye walls (location 8 for example).

\section{Radiative transfer modeling}

Before describing the model structure we infer from fitting the
observed VIMS spectra we first describe our approach to radiative
transfer modeling, i.e. how we parameterized the atmospheric
composition, how we model gas absorption and scattering properties of
aerosols, and how we handle multiple scattering and thermal emission.

\vspace{-0.1in} 
\subsection{Atmospheric structure and composition}

\begin{figure}[!hbt]\centering
\includegraphics[width=3.25in]{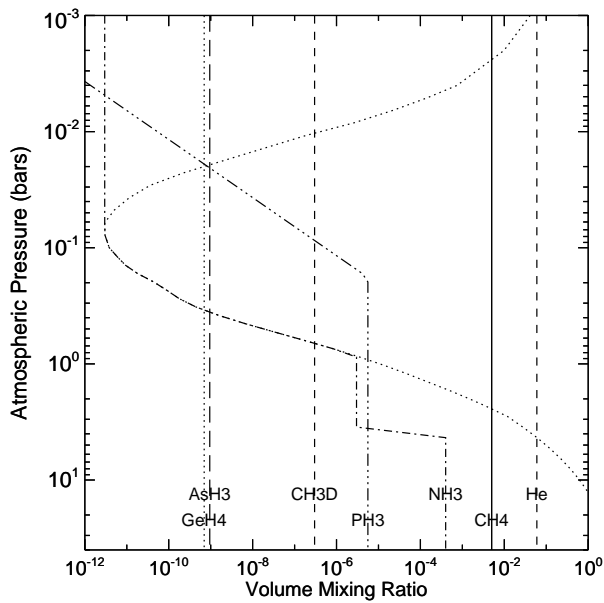}
\caption{Volume mixing ratios of spectroscopically important gases
in the atmosphere of Saturn.  These values were either assumed
or derived from fitting the spectrum from the background cloud at
location 2. See text for references. The dotted curve traces the
saturation vapor pressure profile of \nhtx.}
\label{Fig:gasmix}
\end{figure}

\vspace{-0.05in} We generally followed \cite{Sro2013gws,Sro2018dark}. We used
\cite{Lindal1985} to define the temperature structure between 0.2 mb
and 1.3 bars, and approximated the structure at deeper levels
using a dry adiabatic extrapolation.  The actual temperature structure in
the polar region, according to an analysis of Cassini Composite
Infrared Spectrometer (CIRS) observations by \cite{Fletcher2008}, is
somewhat warmer than our assumed profile (as shown in 
Fig.\ \ref{Fig:polecontext}C).  At the pole it is warmer by
5--15 K in the 500 -- 50 mbar region, but agreement continues to
improve with increasing pressures, and much better agreement is
obtained a few degrees from from the pole. Although these differences
seem large, test calculations for similar differences in the Storm Alley region
by \cite{Sro2018dark} produced spectral differences that were smaller than
the uncertainties.  We assumed
a He/H$_2$ volume mixing ratio (VMR) of 0.0638 \citep{Lindal1985}.  The assumed nominal
composition of the atmosphere as a function of pressure is displayed
in Fig.\ \ref{Fig:gasmix}.  For the \chf VMR we used the
\cite{Fletcher2009ch4saturn} value (4.7$\pm$0.2)$\times 10^{-3}$,
which corresponds to a \chfx/H$_2$ ratio of 5.3$\times 10^{-3}$.  For
\chtd we also used the \cite{Fletcher2009ch4saturn} VMR value of
3$\times 10^{-7}$.  Spectrally, \pht is the most
important variable gas (significant at 2.8-3.1 \mum and dominant at 4.1-5.1 \mumx).
 Its vertical distribution was adjusted to fit VIMS spectra using a
parameterization consisting of an adjustable deep mixing ratio
$\alpha_0$, a break-point pressure $p_b$, and an adjustable scale
 ratio $f$ of its scale height to the pressure scale height. At pressures less than $p_b$, the
mixing ratio can be written
as \begin{eqnarray} \alpha(p) = \alpha_0 (p/p_b)^{(1 - f)/f} \quad
  \mathrm{for} \quad p<p_b.\label{Eq:prof1}
\end{eqnarray}

\noindent
 Because we found a high correlation between spectral effects of $p_b$
 and $f$, only two of our three parameters could be well constrained by
 the VIMS observations. We chose $f$ = 0.1 and fit $p_b$ and the deep
 mixing ratio $\alpha_0$.

Arsine (\ashtx) has a
 noticeable effect on the VIMS spectra near 5 \mumx, which is where
 ammonia gas also plays a relatively minor role.
 \cite{Bezard1989} derived \asht mixing ratios of
 2.4$^{+1.4}_{-1.2}$ ppb (parts per billion) for the thermal component
 and 0.39$^{+0.21}_{-0.13}$ ppb for the reflected solar component. The
 latter is probably representative of the effective value in the 200
 -- 400 mbar range where they inferred a haze layer, while the former
 applies to the deep mixing ratio. \cite{Noll1989AsH3} inferred a
 value of 1.8$^{+1.8}_{-0.8}$ ppb.  \cite{Sro2016} inferred values in
 the 3.5-7 ppb range for a \herat of 0.064 from spectra in the cleared
 region following Saturn's Great Storm of
 2010-2011. \cite{Fletcher2011vims} inferred a value of 2-3 ppb. Our fits
 in the south polar region ranged from about 1 ppb to 3 ppb. 

 We initially tried to constrain the \nht mixing ratio, but we found
 that the results were unreliable because of the very low sensitivity
of the spectra to the ammonia profile parameters.  We eventually decided to fix the
 ammonia profile, using a deep value of 400 ppm, a pressure break point
 of 4 bars, at which point we decreased the \nht VMR to 3 ppm, finally
 setting it to the \nht saturation mixing ratio for pressures less
 than the saturation pressure. This profile was loosely guided by the
 analysis by \cite{Briggs1989} of Very Large Array (VLA) radio
 observations of Saturn, which yielded a deep \nht VMR of (4.8
 $\pm$1)$\times 10^{-4}$ and some evidence for depletion of \nht in
 the 2-4 bar region, which they interpret as evidence for an \nhfsh
 cloud and hydrogen sulfide (H$_2$S) vapor.  Our adopted profile
was also roughly consistent with the derived base pressure of the
cloud layer we assumed to consist of \nht ice.

\subsection{Gas absorption models}
\vspace{-0.05in} We used correlated-k gas absorption models for
methane, ammonia, phosphine, arsine, and collision-induced absorption
as described by \cite{Sro2018dark} and references cited therein.  For
methane we followed \cite{Sro2013gws} in using correlated-k models
based on line-by-line calculations down to 1.268 \mum
\citep{Sro2012LBL}, but for shorter wavelengths used correlated-k fits
to band models of \cite{Kark2010ch4} (we used P. Irwin's fits
available at http://users.ox.ac.uk/$\sim$atmp0035/ktables/ in files
ch4\_karkoschka\_IR.par.gz and ch4\_karkoschka\_vis.par.gz).  The \nht
absorption model fits are from \cite{Sro2010iso}, which are based primarily on band models of
\cite{Bowles2008}.
Absorption models for phosphine (\phtx) and arsine
(AsH$_3$) are the same as described by \cite{Sro2013gws}.
Collision-induced absorption (CIA) for H$_2$ and H$_2$-He was
calculated using programs downloaded from the Atmospheres Node of the
Planetary Data System, which are documented by \cite{Borysow1991h2h2f,
  Borysow1993errat} for the H$_2$-H$_2$ fundamental band,
\cite{Zheng1995h2h2o1} for the first H$_2$-H$_2$ overtone band, and by
\cite{Borysow1992h2he} for H$_2$-He bands.  The vertical penetration depths
permitted by these gases individually and in combination are shown in Fig. 7.

\begin{figure*}\centering
\includegraphics[width=6.in]{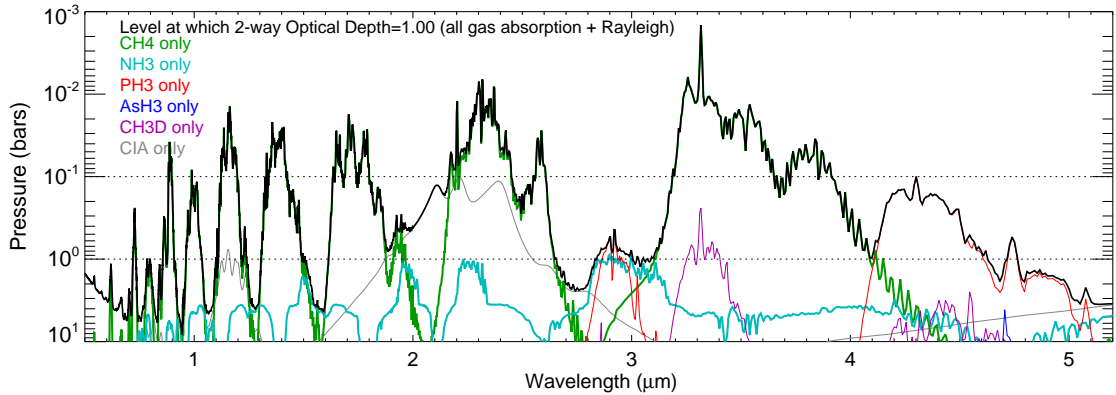}
\caption{Penetration depth in a clear atmosphere for various gases
singly (colored) and in combination (black), computed for gas mixing ratio profiles
are in Fig.\ \ref{Fig:gasmix}.}\label{Fig:pendepth}
\end{figure*}

\subsection{Cloud composition constraints}

\subsubsection{Chemical and photochemical constraints}

The Equilibrium Cloud Condensation Model (ECCM) provides 
constraints based on formation theories, chemistry, and
thermodynamic properties. An overview of those results was given
in Section\ \ref{Sec:intro}.  The main constituents inferred from
ECCM are  NH$_3$ ice, NH$_4$SH, and water ice.  However,
the vertical distribution suggested by vapor pressure curves can
be strongly modified by dynamical considerations, which modify
the mixing ratios and condensation levels from those suggested by
the simplest interpretation, i.e. that the gases are uniformly
mixed up to their condensation levels and above that follow their
saturated vapor pressure curves.  It is also possible for strong
convective events to transport materials upward and produce
clouds of mixed composition, with composite particles with a core
of one condensate, such as water ice, coated by other condensates
such as \nhfsh and \nhtx.  The Great Storm of 2010-2011 provided
one dramatic example of mixed composition being revealed by
spectral absorption signatures \citep{Sro2013gws}.
Besides ECCM materials, there are also candidate compounds produced
photochemically, the potentially most significant of which are
phosphorus (P$_4$ from photolysis of \phtx),  hydrazine (\nthfx), 
and diphosphine (\pthfx). 

\subsubsection{Spectral constraints}

\begin{figure*}[!htb]\centering
\includegraphics[width=5.2in]{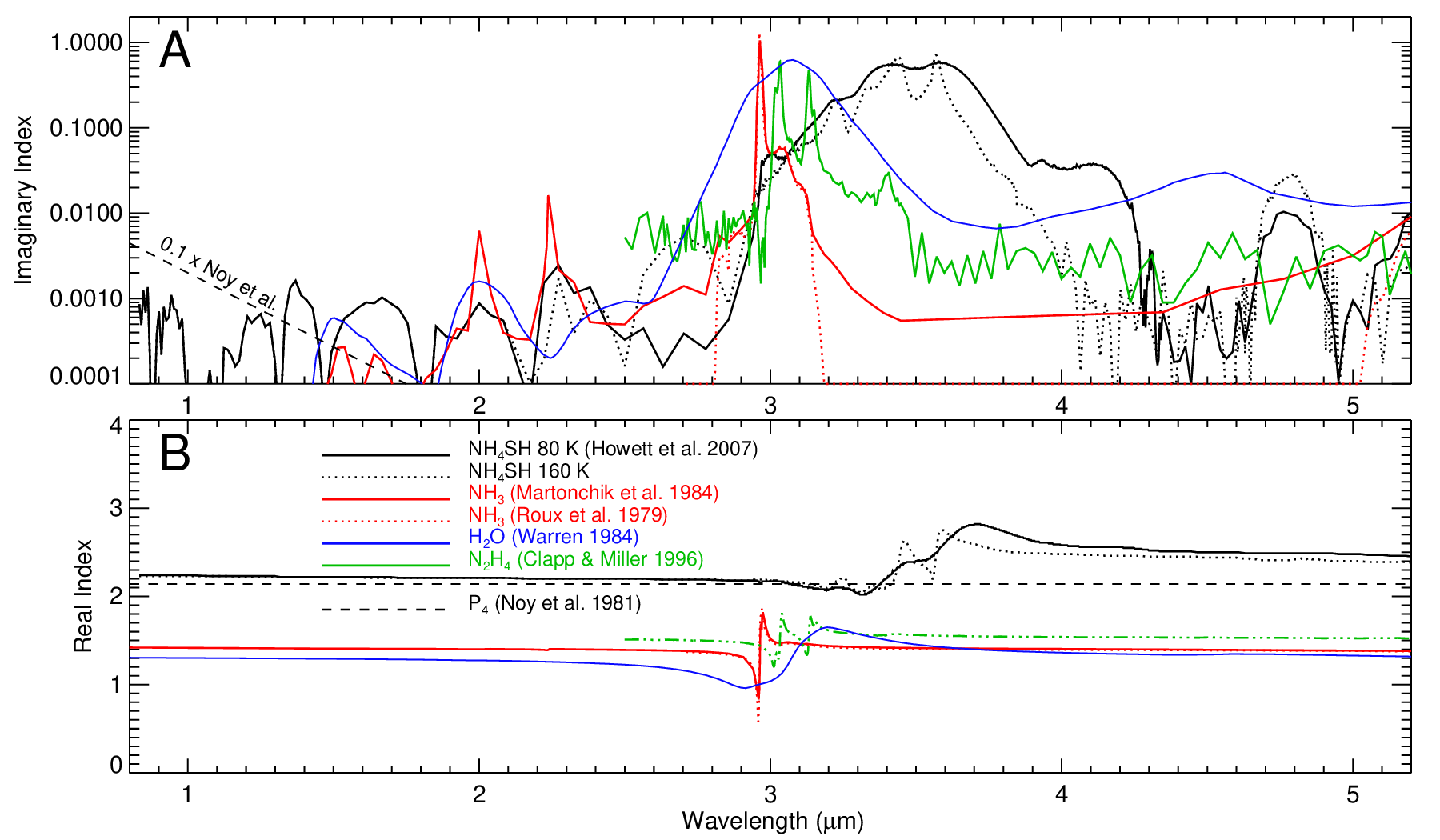}
\caption{Imaginary (A) and real (B) components of the refractive index
  vs. wavelength for candidate 3-\mum absorbers, including H$_2$O
  results of \cite{Warren1984}, \nht results of \cite{Roux1979} and
  \cite{Martonchik1984}, NH$_4$SH results of \cite{Howett2007}, and
  N$_2$H$_4$ results of \cite{Clapp1996}.}
\label{Fig:indexplot}
\end{figure*}

Spectral signatures provide a second important constraint.  
Refractive index plots for all of the previously discussed
materials (except diphosphine and phosphorus, which have not been
adequately characterized) are shown in Fig.\ \ref{Fig:indexplot}.
Among the more plausible
compounds, NH$_3$ has the strongest and sharpest absorption at 2
$\mu$m (the 2.25 $\mu$m absorption feature would not be visible due to
overlying gas absorption).  \nhfsh has an absorption that is roughly comparable
to that of NH$_3$ at 3 $\mu$m, but lacks a sharp spectral feature and
continues to increase beyond 3.1 $\mu$m while NH$_3$ absorption drops
significantly. Hydrazine, which is not expected to be very abundant on
Saturn, has a pair of strong absorption peaks between 3 and 3.2
$\mu$m, which would be very apparent unless cloud particles were very
large.  \pthf appears to have only very weak absorption in this region
\citep{Frankiss1968}.

\subsection{Parameterization of the vertical cloud structure}\label{Sec:param}

Based on the above considerations, we assumed a stack of four cloud
layers that is summarized in Fig.\ \ref{Fig:cloudmodel}B and Table \ref{Tbl:paramlist}.
  All layers were assumed to scatter like
spherical particles, even though they are most likely
composed of solid non-spherical particles.  This parameterization
scheme was a convenient way to incorporate composition-dependent and
wavelength-dependent scattering properties, and because
it allowed us to fit most spectra quite well, we were not
motivated to try more complex particle models or structures with more parameters to
be constrained. We also assumed a gamma size distribution with
variance 0.05 to smooth out phase function structure for the stratosphere
and a variance of 0.1 for the larger particles of the bottom three layers.  
The top three layers are essentially sheet clouds, which are
characterized by a base pressure, particle size, optical depth, and
refractive index.  All but the refractive index function are
adjustable parameters that are fitted by our retrieval code.  The
pressures at the top of these layers are large fractions of their base
pressures, 0.8 for the top layer, and 0.9 for the second and third
layers.  These fractions are arbitrary and not constrained by
observations.  Next we describe how we modeled each of the four layers.

\subsubsection{Stratospheric haze (layer 1)}

For this top model layer we assumed refractive index of 1.4, which is
comparable to the polar value of $\sim$1.45 inferred by
\cite{Perez-Hoyos2005}, who assumed that the haze extended from 10
mbar to 1 mbar, but were not able to constrain it any further from their
observations. In a subsequent paper \cite{Perez-Hoyos2016} constrained
only the integrated haze optical depth above 100 mbar.  Attempts to
define the vertical boundaries of this haze in Storm Alley latitudes
by \cite{Sro2018dark} yielded large uncertainties.  Nor is the
vertical distribution of this haze well constrained our by chosen south
polar VIMS observations.  We found that a wide range of base pressures
and scale heights could provide the same fit quality as a compact haze
as long as the optical depth and average pressure were the same.  Thus
we did not try to define its vertical distribution, but instead fit
its effective pressure under the assumption that it is a close to a
sheet cloud in vertical extent.  In fact, there is good evidence that
it might play a role in creating shadows, which would require that it
be a relatively compact structure instead of a vertically diffuse
structure. We show in the following that its effective pressure is
very well constrained by VIMS observations, as are its optical depth
and particle size.

\subsubsection{Upper tropospheric cloud (layer 2)}
The composition of this layer is unknown. As previously noted
\citep{Sro2018dark,Sro2016,Sro2013gws}, while absorbing light at short
visible wavelengths \citep{Perez-Hoyos2005}, this layer appears to
have no distinctive absorption features at near-IR wavelengths.
Possible compositions include some form of phosphorous or
diphosphine. According to a review by \cite{Fouchet2009}, diphosphine
is a leading candidate.  \cite{Visscher2009} predicted a much greater
abundance of diphosphine than phosphorus.  However, there is currently
no spectral evidence for or against diphosphine on Saturn (or on
Jupiter).  Although it has a distinctive double absorption peak near
4.3 \mum and another absorption feature at 6 \mum \citep{Nixon1956},
these are at least partially masked by phosphine gas absorption and
ammonia gas absorption respectively, The degree of masking is
uncertain because there is no quantitative characterization of the
diphosphine absorption strengths.  The white phosphorus (P$_4$)
molecule is a symmetric tetrahedron, has no dipole moment, and is thus
unlikely to have any significant near-IR absorption features.  We
decided to characterize the particles in this layer with a real
refractive index of 1.82 (the value for white phosphorus), which is
also close to the 1.74 value for \pthf at 195 K
\citep{LandoltBornstein2008}, the other candidate material for this
layer.  We chose an imaginary index with the spectral shape defined by
\cite{Noy1981} but reduced it by a factor of 10 in amplitude to avoid
excessive absorption at short wavelengths.  This layer cannot be an
ammonia condensate because it contains no 3-\mum absorption feature.

\subsubsection{Middle tropospheric cloud (layer 3)}
 
This layer we assumed to consist of ammonia
ice particles.  This is a plausible assumption because the layer appears
in a pressure range where \nht is likely to condense, and because
this layer clearly has an ammonia ice absorption signature. This
layer is found near 1 bar, well above the ECCM-suggested 1.7 bars, presumably
due to a depletion of \nht by formation of an underlying \nhfsh cloud \citep{Briggs1989}.
Its pressure is very well constrained by the VIMS observations.

\subsubsection{Deep tropospheric cloud (layer 4)}
Thermal emission in the 4.6-5.2 \mum spectral region originates from
the 2-10 bar level, with the contribution function at 5.12 \mum
peaking near 6 bars (Fig.\ 1), which is determined mainly by
collision-induced absorption by hydrogen.  However, the emission is
 attenuated, often significantly, by overlying aerosols in the
region between 1.7 and 5 bars where it is possible to have solid
condensates comprised of \nhfsh or \hto or both \citep{Atreya2005SSR}.
As these two materials have different spectral signatures
(Fig.\ \ref{Fig:indexplot}), there is a reasonable prospect of
constraining the composition of this cloud from 5-\mum emission
spectra.  In their analysis of VIMS night-time emission spectra,
\cite{Barstow2016} considered a variety of cloud models for the
emission-attenuating layer, obtaining their best fits with a
non-scattering cloud of \nhfsh particles, although completely gray
particles provided only slightly worse fits.  They expressed
confidence that whatever the bulk composition of the cloud, if \nht or
\nhfsh are present they must be contaminated with something that
darkens the individual particles and makes them more absorbing.  For
our model, we assumed the bottom layer to be a diffuse distribution
extending from the 5-bar level upwards to a fitted top pressure
determined by spectral constraints. We also assumed it to have
the same scale height as the atmosphere.  We initially used the
\cite{Howett2007} index for \nhfsh at 160 K.  However, preliminary
fits showed that we needed more absorption and spectrally flatter
properties than provided by pure \nhfshx.  We ultimately adopted a
refractive index of 2.0 + 0.03$i$ and a radius of 2 \mumx.  Thus, our
current analysis cannot confirm that this layer actually contains a significant
component of \nhfshx.  

\begin{table*}[!htb]\centering
\caption{Cloud and gas model parameters used in spectral
  calculations.}
\vspace{0.15in}
\begin{tabular}{r l l }
Parameter (unit) & Description & Value\\
\hline
$p_1$ (bar) & stratospheric haze base pressure & adjustable\\
$r_1$ (\mumx) & effective radius of stratospheric particles & adjustable\\
$n_1(\lambda)$ & refractive index of stratospheric particles & $n_1=1.4$\\
$\tau_1$ & stratospheric haze optical depth at 2 \mum & adjustable\\
\hline
$p_{1}$ (bar) & base of main visible cloud layer (P$_4$ or \pthf?)  & adjustable\\
$r_2$ (\mumx) & effective radius of main cloud particles & adjustable\\
$n_2(\lambda)$ & refractive index of main cloud & $n_2=1.82+i\times 0.1 n_{iN}$\\
$\tau_2$ & optical depth of upper cloud at 2 \mum & adjustable\\
\hline
$p_3$ (bar) & base pressure of putative \nht cloud & adjustable\\
$r_3$ (\mumx) & effective radius of \nht cloud particles & adjustable\\
$n_3(\lambda)$ & refractive index of \nht cloud particles & same as \nht \\
$\tau_3$ & optical depth of \nht cloud at 2 \mum & adjustable\\
\hline
$p_{4t}$ (bar) & pressure at top of deep cloud (\nhfsh + \hto?) & adjustable\\
$p_{4b}$ (bar) & bottom of deep cloud & fixed at 5 bars\\
$r_4$ (\mumx) & effective radius of deep cloud particles & fixed at 2 \mumx or adjusted\\
$n_4(\lambda)$ & refractive index of deep cloud particles &2 +$i\times$0.03  \\
$d\tau_4/dp$ (bar$^{-1}$) & optical density of deep cloud at 2 \mum & fixed at 20/bar\\
\hline
\hline
$p_b$ (bar) & \pht break-point pressure & adjustable\\ 
$\alpha_0$ & \pht VMR for $p > p_b$ & adjustable\\
$f$  & \pht to H$_2$ scale height ratio for $p < p_b$  & fixed at 0.1\\ 
$AsH_3v$ & arsine volume mixing ratio & adjustable\\
\hline\\[-0.1in]
\end{tabular}
\parbox{4.5in}{NOTE: Sheet clouds are assumed to have a top pressure
  that is 0.8 (stratosphere) or 0.9 times the bottom pressure and a particle to gas scale
  height ratio of 1.0; aerosol particles are assumed scatter like spheres with a gamma
  size distribution with variance parameter $b=0.1$, with distribution
  function $n(r) = \mbox{constant}\times r^{(1-3b)/b} e^{-r/ab}$,
  where with $a = r_\mathrm{i}$, for $i=0-3$, and $b=$ dimensionless variance, 
following
  \cite{Hansen1974}. Optical depths are given for a wavelength of 2
  \mumx.   $n_{iN}$ is the imaginary index determined by
  \cite{Noy1981}.}
\label{Tbl:paramlist}
\end{table*}

\subsection{Multiple scattering methods}

We used the same modified doubling and adding code with thermal source
capability that is described by \cite{Sro2013gws}.
We used a grid of 56 pressure levels from 0.5
mbar to 40 bars, distributed roughly in equal log increments, except
that additional layers are introduced where cloud boundaries are
inserted. 
To model the medium phase angle VIMS observations
we used 12 quadrature points per hemisphere in zenith angle and 12 in azimuth.
We approximated the line-spread function of the VIMS instrument as a
Gaussian with a wavelength-dependent full width at half maximum
(FWHM), then collected all the opacity values
within $\pm$FWHM of the sample wavelength, weighted those according to
the relative amplitude of the line-spread function, then sorted and
refit to ten terms again. (The near-IR FWHM values range from 0.0125 \mum to 0.0226 \mum
and are available from the Planetary Data System (PDS)
 at http://atmos.nmsu.edu/data\_and\_services/atmospheres\_data/
cassini/vims.html).
A special treatment is required in the 2.95-3.0 \mum
region where \nht cloud particles have very sharp absorption features.
In this region model calculations were made at 5-\icm intervals (the
maximum sampling frequency of our correlated-k models) and
subsequently smoothed to VIMS resolution.


\subsection{Estimation of uncertainties} We used a simplified model of uncertainties in measurement,
calibration, and gas absorption modeling, following that suggested by
\cite{Sro2016} after comparisons with more complex models.  The
simplified model expresses the I/F uncertainty as the square root of
the sum of two squared quantities. The first is an I/F uncertainty of
0.003 and the second is 10\% of the I/F signal.
These estimates are meant to capture combined effects of both
instrumental and modeling uncertainties, and allow us to fit most
observations with \chisq values close to the expected value, which is
the number of degrees of freedom N$_F$, so that \chisqx /N$_F$
$\approx$ 1, where N$_F$ = number of comparison points (usually 177)
less the number of adjusted parameters (13).

Our 1-$\sigma$ uncertainty range for each parameter individually is
based on the $\Delta\chi^2$ = 1 confidence interval for the \chisq
distribution obtained by adjusting all other parameters to
minimize \chisq for each value of the chosen parameter
\citep{Press1992}.  This is estimated by our Levenberg-Marquardt
algorithm under the assumption of normally distributed errors, which
is not entirely valid.  Further, parameter uncertainties we estimate
in this fashion are at best only locally valid, and do not guarantee
that there is not some other distant solution within this
13-dimensional parameter volume that provides a comparable or better
fit. In fact, we did find two distant solutions for the \nhtx
-signature cloud features, which will be described in
Section\ \ref{Sec:nh3sigfit}.  To fully explore this space to find the
absolute minimum in \chisq would take an impractically large time
because of the complex calculations that are involved and the high
dimensionality of the parameter space.

\section{Fit Results}

We fit observations by first selecting, with guidance from parameter
sensitivity calculations described in Sec.\ \ref{Sec:deriv}, an {\it a
  priori} model that provides usually only a very rough fit to the
observations, and then adjust parameters to minimize \chisq using a
form of the Levenberg-Marquardt algorithm as described by
\cite{Press1992}.  A typical fit takes about 2 hours of computation
time and usually converges to within 0.2 of the \chisq minimum in 6-15
iterations.  In the following, we first consider models of background
clouds, then proceed to models of discrete features, and follow that
with illustrations of how the model parameters affect the model spectra
and how initial values (first guesses) affect final solutions.

\subsection{Background cloud models}

We fit models to background clouds at locations identified in
Figs. \ref{Fig:outereye} and \ref{Fig:innereye}.  These were
segregated into two latitude regions for tabulation.  Results for the
outer region (from 71.1\degx S to 86.7\degx S) are given in
Table\ \ref{Tbl:backouter}, and for the inner region (from 87.3\degx S
to 89.9\degx S) in Table\ \ref{Tbl:backinner}.  The cloud structure
results for both regions are plotted in Fig.\ \ref{Fig:strucfits},
with the black symbols and lines referring to the background cloud
structures.  These structures do not show very much variation with
latitude over most of the outer region, especially within the 73\degx
S to 86\degx S latitude band, for which averages and standard
deviations are given in Table\ \ref{Tbl:meansd}.  In most cases the
standard deviations are smaller than the estimated uncertainties,
indicating that our retrieval error models were somewhat pessimistic.
Both layers 1 and 2 show an increased optical depth furthest from
the pole, accounting for the higher continuum I/F values seen in the
upper left corner of continuum images in Fig.\ \ref{Fig:outereye}.
Below we summarize fit results for each aerosol layer. 
The gas mixing ratio fits will be discussed in Section\ \ref{Sec:gas}.

\subsubsection{Stratospheric haze fit results}

Over the entire outer region, stratospheric base pressure ranged from
49$\pm$10 mbar to 70$\pm$10 mbar, optical depths ranged from about
0.027$\pm$0.002 to 0.019$\pm$0.005, and the particle radii ranged from
0.18$\pm$0.01 \mum to 0.20$\pm$0.01 \mumx, a remarkably small range.
Within the restricted outer region (73\degx
S to 86\degx S), the average stratospheric optical
depth is 0.023 with a standard deviation of only 0.002.  The
variation is much greater in the inner region, due to the sharp
decline in optical depth to 0.0010 as the pole is approached. The
transition begins around 86.5\degx S and is complete by 88.5\degx S.
There is also a bump in the stratospheric haze altitude centered around 88\degx S,
where there is a local pressure minimum near 33-40 mbar.

\begin{table*}[!hb]\centering
\caption{Fits to background cloud spectra from 71.1\degx S to 86.7\degx S}\label{Tbl:backouter}
\setlength{\tabcolsep}{0.03in}
\renewcommand{\arraystretch}{0.85} 
\vspace{0.1in}
\small
\begin{tabular}{l c c c c c c c c }
\hline\\[-0.1in]   
Locations:&                 4-1&                 4-2&                 4-4&                 4-7&                5-10&                 4-9&                5-12&                 5-8\\[0.05in]
PC Lat.: &          -71.1\degx  &          -73.9\degx  &          -77.2\degx  &          -80.1\degx  &          -82.1\degx  &          -82.2\degx  &          -85.8\degx  &          -86.7\degx \\[0.05in]
\hline\\[-0.15in]
              $p_1$, bar & 0.050$^{+0.01}_{-0.01}$ & 0.049$^{+0.01}_{-0.01}$ & 0.055$^{+0.01}_{-0.01}$ & 0.056$^{+0.01}_{-0.01}$ & 0.061$^{+0.01}_{-0.01}$ & 0.059$^{+0.01}_{-0.01}$ & 0.070$^{+0.01}_{-0.01}$ & 0.055$^{+0.01}_{-0.01}$\\[0.05in]
              $p_2$, bar &  0.35$^{+ 0.02}_{- 0.02}$ &  0.35$^{+0.02}_{-0.02}$ &  0.36$^{+0.02}_{-0.02}$ &  0.34$^{+0.02}_{-0.02}$ &  0.32$^{+0.02}_{-0.02}$ &  0.32$^{+0.02}_{-0.02}$ &  0.29$^{+0.02}_{-0.02}$ &  0.28$^{+0.02}_{-0.02}$\\[0.05in]
              $p_3$, bar &  0.96$^{+ 0.16}_{- 0.15}$ &  0.96$^{+0.17}_{-0.15}$ &  0.96$^{+0.25}_{-0.22}$ &  1.00$^{+0.27}_{-0.24}$ &  1.07$^{+0.34}_{-0.30}$ &  0.93$^{+0.23}_{-0.20}$ &  1.00$^{+0.23}_{-0.21}$ &  0.93$^{+0.18}_{-0.16}$\\[0.05in]
           $p_{4t}$, bar &  3.88$^{+ 0.07}_{- 0.07}$ &  2.65$^{+0.04}_{-0.04}$ &  2.98$^{+0.04}_{-0.04}$ &  3.63$^{+0.06}_{-0.06}$ &  3.10$^{+0.05}_{-0.05}$ &  3.54$^{+0.05}_{-0.05}$ &  3.45$^{+0.05}_{-0.06}$ &  4.15$^{+0.08}_{-0.08}$\\[0.05in]
   $\tau_1$$\times 10^2$ &  2.65$^{+ 0.59}_{- 0.49}$ &  2.46$^{+0.56}_{-0.46}$ &  2.53$^{+0.58}_{-0.47}$ &  2.34$^{+0.53}_{-0.44}$ &  2.19$^{+0.56}_{-0.45}$ &  2.17$^{+0.51}_{-0.42}$ &  2.31$^{+0.61}_{-0.48}$ &  1.85$^{+0.50}_{-0.39}$\\[0.05in]
                $\tau_2$ &  1.13$^{+ 0.09}_{- 0.10}$ &  0.86$^{+0.10}_{-0.10}$ &  0.87$^{+0.11}_{-0.11}$ &  0.83$^{+0.09}_{-0.09}$ &  0.75$^{+0.08}_{-0.08}$ &  0.71$^{+0.10}_{-0.09}$ &  0.59$^{+0.10}_{-0.09}$ &  0.57$^{+0.08}_{-0.08}$\\[0.05in]
                $\tau_3$ &  1.13$^{+ 0.35}_{- 0.27}$ &  0.92$^{+0.21}_{-0.17}$ &  0.56$^{+0.65}_{-0.31}$ &  0.76$^{+0.30}_{-0.22}$ &  0.78$^{+0.32}_{-0.23}$ &  0.68$^{+0.20}_{-0.16}$ &  0.94$^{+0.32}_{-0.24}$ &  0.77$^{+0.28}_{-0.21}$\\[0.05in]
             $r_1$, \mum & 0.189$^{+0.01}_{-0.01}$ & 0.194$^{+0.01}_{-0.01}$ & 0.197$^{+0.01}_{-0.01}$ & 0.191$^{+0.01}_{-0.01}$ & 0.182$^{+0.01}_{-0.01}$ & 0.183$^{+0.01}_{-0.01}$ & 0.181$^{+0.01}_{-0.01}$ & 0.186$^{+0.01}_{-0.01}$\\[0.05in]
             $r_2$, \mum &  0.50$^{+ 0.04}_{- 0.04}$ &  0.48$^{+0.04}_{-0.04}$ &  0.47$^{+0.04}_{-0.04}$ &  0.54$^{+0.06}_{-0.06}$ &  0.60$^{+0.08}_{-0.07}$ &  0.63$^{+0.08}_{-0.07}$ &  0.71$^{+0.08}_{-0.08}$ &  0.65$^{+0.08}_{-0.07}$\\[0.05in]
             $r_3$, \mum &  1.47$^{+ 0.92}_{- 0.55}$ &  1.10$^{+0.59}_{-0.36}$ &  1.92$^{+1.78}_{-0.91}$ &  1.51$^{+1.03}_{-0.59}$ &  1.60$^{+1.20}_{-0.66}$ &  1.32$^{+0.93}_{-0.52}$ &  1.74$^{+0.87}_{-0.57}$ &  1.75$^{+0.80}_{-0.54}$\\[0.05in]
            $p_b$, bar &  0.22$^{+ 0.05}_{- 0.04}$ &  0.18$^{+0.07}_{-0.05}$ &  0.15$^{+0.07}_{-0.05}$ &  0.18$^{+0.05}_{-0.04}$ &  0.15$^{+0.06}_{-0.04}$ &  0.17$^{+0.05}_{-0.04}$ &  0.18$^{+0.04}_{-0.03}$ &  0.21$^{+0.03}_{-0.03}$\\[0.05in]
     $\alpha_0$, ppm &   4.7$^{+  0.2}_{-  0.2}$ &   4.1$^{+ 0.4}_{- 0.4}$ &   4.4$^{+ 0.3}_{- 0.3}$ &   4.4$^{+ 0.2}_{- 0.2}$ &   4.6$^{+ 0.3}_{- 0.3}$ &   4.5$^{+ 0.2}_{- 0.2}$ &   4.3$^{+ 0.3}_{- 0.2}$ &   4.2$^{+ 0.2}_{- 0.2}$\\[0.05in]
    $AsH_3v$, ppb &  1.54$^{+ 0.33}_{- 0.27}$ &  1.98$^{+0.56}_{-0.45}$ &  1.77$^{+0.45}_{-0.38}$ &  1.65$^{+0.30}_{-0.25}$ &  1.92$^{+0.42}_{-0.36}$ &  1.20$^{+0.31}_{-0.26}$ &  1.99$^{+0.39}_{-0.33}$ &  1.65$^{+0.33}_{-0.28}$\\[0.05in]
$\chi^2$ & 204.18 & 174.85 & 179.99 & 184.53 & 138.96 & 170.10 & 148.85 & 146.10\\[0.05in]
$\chi^2/N_F$ &   1.25 &   1.07 &   1.10 &   1.13 &   0.85 &   1.04 &   0.91 &   0.89\\[0.05in]
\hline\\[-0.1in]
\end{tabular}
\normalsize
\parbox{5.4in}{NOTE: These fits cover the spectral range from 0.88 \mum to 5.12 \mumx, with exclusions for
order sorting filter joints and regions with very low S/N ratios.  These fits assumed fixed values of
   $p_4$ = 5 bars, $r_4$ = 2 \mumx, and $f$ = 0.1. }
\end{table*}

\subsubsection{Layer-2 (P$_4$ or P$_2$H$_4$?) fit results}
The putative diphosphine layer has even less variability than the
stratospheric haze.  Its base pressure smoothly decreases from about
350 mbar at 71\degx S to 290 mbar near 87.5\degx S, then increases to
400 mbar as the pole is approached.  Its optical depth declines
towards the pole, from near unity at 71\degx S, then slowly with
latitude until about 86\degx S, then more rapidly until it reaches a
low of 0.31$\pm$0.1 near the pole. Both vertical descent and
decreasing optical depth are consistent with downwelling motion as the
pole is approached.  The optical depth transition for this layer is
close to the edge of a sharp brightness decrease seen in ISS and VIMS
images that can sense scattering at the level of this layer, which is
also the boundary associated with shadows seen at continuum
wavelengths. A further drop from that point towards the pole seems to
be the main cause for the dark polar eye when viewed at solar
reflected wavelengths.  This decrease in the optical depth of the
putative diphosphine layer results in a partial unmasking of the
ammonia signature to such a degree that the entire region within 2\deg
of the pole can be considered an ammonia-signature cloud structure.
This is also evident from the magenta color it displays in
Fig.\ \ref{Fig:innereye}E/G.

\subsubsection{Ammonia layer fit results}

An even more muted variation is seen for the background \nht cloud layer.  Its pressure is almost
invariant with latitude all the way to the pole, although the uncertainty in our
constraints of this parameter results in a higher standard deviation than for the 
layer 2 pressures.  Its pressure averaged 979 mbar ($\sigma$ = 46 mbar) in the
restricted outer region (73\degx
S to 86\degx S), and declined to 810 mbar close to the pole, rising upward
slightly (in altitude) against the trend of the layers above it, which both move downward near the pole.
Its optical depth remains relatively close to its restricted outer mean of 1.05 ($\sigma$ = 0.13),
but also has a slight decline towards the pole, by about 20\%, a much smaller fraction
than for other layers.  Perhaps this indicates that the downwelling motion stops
near the level of the \nht cloud.  However, there is a decrease in particle radius near the
pole where it declines from near 2 \mum near 87\degx S by almost a factor of two near 90\degx S.

\subsubsection{Bottom layer fit results}

The top pressure of the deep cloud varies as needed to match the
rather patchy spatial variation in 5-\mum emissions, varying from 2.7
bars to 4.1 bars, with the inner region containing the larger values.
However, variations in this layer do not seem to be correlated with changes in the overlying layers,
with the one exception being close to the pole.  This is not necessarily a trend, however,
because there is quite a lot of spatial structure at thermal emission wavelengths that was
not adequately sampled by our small number of selected spectra.  For the background cloud
structures the upper three layers seem disconnected from the variations that occur in the
2.7-5 bar region.  In the following we show that this disconnection
 is not the case for the ammonia signature features.

\begin{table*}[!hbt]\centering
\caption{Fits to background cloud spectra at locations from 87.3\degx S to 89.8\degx S.}\label{Tbl:backinner}
\footnotesize
\setlength{\tabcolsep}{0.03in}
\renewcommand{\arraystretch}{0.85} 
\begin{tabular}{l c c c c c c c c c }
\hline\\[-0.05in]
Locations:&                 5-7&                 5-6&                5-14&                 5-5&                 5-4&                 5-3&                 5-2&                 5-1&                 5-0\\[0.05in]
PC Lat.: &          -87.3\degx  &          -87.5\degx  &          -87.6\degx  &          -88.4\degx  &          -88.5\degx  &          -88.6\degx  &          -89.2\degx  &          -89.3\degx  &          -89.8\degx \\
\hline\\
          $p_1$, bar & 0.054$^{+0.01}_{-0.01}$ & 0.040$^{+0.01}_{-0.01}$ & 0.041$^{+0.01}_{-0.01}$ & 0.040$^{+0.02}_{-0.01}$ & 0.033$^{+0.02}_{-0.01}$ & 0.044$^{+0.02}_{-0.01}$ & 0.058$^{+0.02}_{-0.02}$ & 0.056$^{+0.02}_{-0.02}$ & 0.061$^{+0.02}_{-0.02}$\\[0.05in]
          $p_2$, bar &  0.30$^{+ 0.02}_{- 0.02}$ &  0.29$^{+0.02}_{-0.02}$ &  0.27$^{+0.02}_{-0.02}$ &  0.32$^{+0.03}_{-0.03}$ &  0.30$^{+0.03}_{-0.02}$ &  0.33$^{+0.03}_{-0.03}$ &  0.41$^{+0.05}_{-0.05}$ &  0.41$^{+0.05}_{-0.04}$ &  0.43$^{+0.05}_{-0.04}$\\[0.05in]
          $p_3$, bar &  0.89$^{+ 0.16}_{- 0.14}$ &  0.87$^{+0.17}_{-0.15}$ &  0.82$^{+0.09}_{-0.08}$ &  0.84$^{+0.14}_{-0.13}$ &  0.82$^{+0.09}_{-0.09}$ &  0.84$^{+0.13}_{-0.12}$ &  0.85$^{+0.20}_{-0.18}$ &  0.81$^{+0.11}_{-0.10}$ &  0.81$^{+0.10}_{-0.09}$\\[0.05in]
       $p_{4t}$, bar &  3.80$^{+ 0.06}_{- 0.07}$ &  4.15$^{+0.08}_{-0.09}$ &  3.62$^{+0.06}_{-0.06}$ &  2.94$^{+0.04}_{-0.04}$ &  2.72$^{+0.03}_{-0.03}$ &  3.16$^{+0.05}_{-0.05}$ &  3.37$^{+0.05}_{-0.05}$ &  3.52$^{+0.04}_{-0.04}$ &  3.88$^{+0.07}_{-0.07}$\\[0.05in]
$\tau_1$$\times 10^2$ &  1.50$^{+ 0.43}_{- 0.33}$ &  1.14$^{+0.32}_{-0.25}$ &  1.13$^{+0.35}_{-0.27}$ &  0.90$^{+0.30}_{-0.23}$ &  0.76$^{+0.27}_{-0.20}$ &  0.96$^{+0.31}_{-0.23}$ &  0.91$^{+0.29}_{-0.22}$ &  0.97$^{+0.31}_{-0.23}$ &  1.03$^{+0.30}_{-0.23}$\\[0.05in]
           $\tau_2$ &  0.51$^{+ 0.08}_{- 0.08}$ &  0.52$^{+0.09}_{-0.08}$ &  0.44$^{+0.07}_{-0.06}$ &  0.44$^{+0.08}_{-0.07}$ &  0.43$^{+0.07}_{-0.06}$ &  0.43$^{+0.09}_{-0.08}$ &  0.37$^{+0.13}_{-0.10}$ &  0.32$^{+0.09}_{-0.08}$ &  0.31$^{+0.10}_{-0.08}$\\[0.05in]
           $\tau_3$ &  0.73$^{+ 0.29}_{- 0.21}$ &  0.81$^{+0.32}_{-0.23}$ &  0.93$^{+0.23}_{-0.19}$ &  0.70$^{+0.27}_{-0.20}$ &  0.86$^{+0.23}_{-0.18}$ &  0.99$^{+0.23}_{-0.19}$ &  0.64$^{+0.26}_{-0.18}$ &  0.70$^{+0.19}_{-0.15}$ &  0.79$^{+0.21}_{-0.17}$\\[0.05in]
        $r_1$, \mum & 0.182$^{+0.01}_{-0.01}$ & 0.181$^{+0.01}_{-0.01}$ & 0.181$^{+0.01}_{-0.01}$ & 0.179$^{+0.02}_{-0.01}$ & 0.170$^{+0.02}_{-0.02}$ & 0.178$^{+0.02}_{-0.01}$ & 0.173$^{+0.02}_{-0.02}$ & 0.180$^{+0.02}_{-0.01}$ & 0.183$^{+0.02}_{-0.01}$\\[0.05in]
        $r_2$, \mum &  0.65$^{+ 0.08}_{- 0.07}$ &  0.65$^{+0.08}_{-0.07}$ &  0.62$^{+0.08}_{-0.07}$ &  0.59$^{+0.07}_{-0.07}$ &  0.62$^{+0.07}_{-0.06}$ &  0.58$^{+0.08}_{-0.07}$ &  0.61$^{+0.11}_{-0.09}$ &  0.62$^{+0.12}_{-0.10}$ &  0.60$^{+0.13}_{-0.11}$\\[0.05in]
        $r_3$, \mum &  2.00$^{+ 0.71}_{- 0.52}$ &  2.04$^{+0.73}_{-0.54}$ &  1.70$^{+0.44}_{-0.35}$ &  1.75$^{+0.56}_{-0.42}$ &  1.78$^{+0.44}_{-0.35}$ &  1.34$^{+0.45}_{-0.33}$ &  1.24$^{+0.54}_{-0.36}$ &  1.28$^{+0.46}_{-0.33}$ &  1.05$^{+0.42}_{-0.29}$\\[0.05in]
         $p_b$, bar &  0.22$^{+ 0.05}_{- 0.04}$ &  0.21$^{+0.03}_{-0.02}$ &  0.23$^{+0.03}_{-0.03}$ &  0.25$^{+0.04}_{-0.03}$ &  0.26$^{+0.04}_{-0.03}$ &  0.28$^{+0.04}_{-0.03}$ &  0.33$^{+0.07}_{-0.06}$ &  0.39$^{+0.07}_{-0.06}$ &  0.40$^{+0.05}_{-0.04}$\\[0.05in]
    $\alpha_0$, ppm &   4.5$^{+  0.2}_{-  0.2}$ &   4.2$^{+ 0.2}_{- 0.2}$ &   4.3$^{+ 0.2}_{- 0.2}$ &   4.3$^{+ 0.3}_{- 0.3}$ &   4.1$^{+ 0.3}_{- 0.3}$ &   4.6$^{+ 0.3}_{- 0.3}$ &   4.2$^{+ 0.2}_{- 0.2}$ &   4.3$^{+ 0.2}_{- 0.2}$ &   4.1$^{+ 0.2}_{- 0.2}$\\[0.05in]
      $AsH_3v$, ppb &  1.37$^{+ 0.30}_{- 0.25}$ &  1.63$^{+0.30}_{-0.25}$ &  1.32$^{+0.28}_{-0.23}$ &  1.24$^{+0.36}_{-0.29}$ &  1.15$^{+0.36}_{-0.28}$ &  1.10$^{+0.24}_{-0.19}$ &  1.08$^{+0.24}_{-0.20}$ &  1.24$^{+0.26}_{-0.21}$ &  1.14$^{+0.24}_{-0.20}$\\[0.05in]
$\chi^2$ & 134.05 & 150.34 & 148.52 & 128.59 & 125.36 & 128.33 & 133.19 & 153.03 & 152.91\\[0.05in]
$\chi^2/N_F$ &   0.82 &   0.92 &   0.91 &   0.78 &   0.76 &   0.78 &   0.81 &   0.93 &   0.93\\[0.05in]
\hline\\[-0.1in]
\end{tabular}
\normalsize
\parbox{5.6in}{NOTE: These fits cover the spectral range from 0.88 \mum to 5.12 \mumx, with exclusions for
order sorting filter joints and regions with very low S/N ratios.  These fits assumed fixed values of
   $p_4$ = 5 bars, $r_4$ = 2 \mumx, $f$ = 0.1, and $d\tau_4/dp$ = 20/bar. }
\end{table*}

\begin{figure*}[!ht]\centering
\hspace{-0.125in}\includegraphics[width=3.5in]{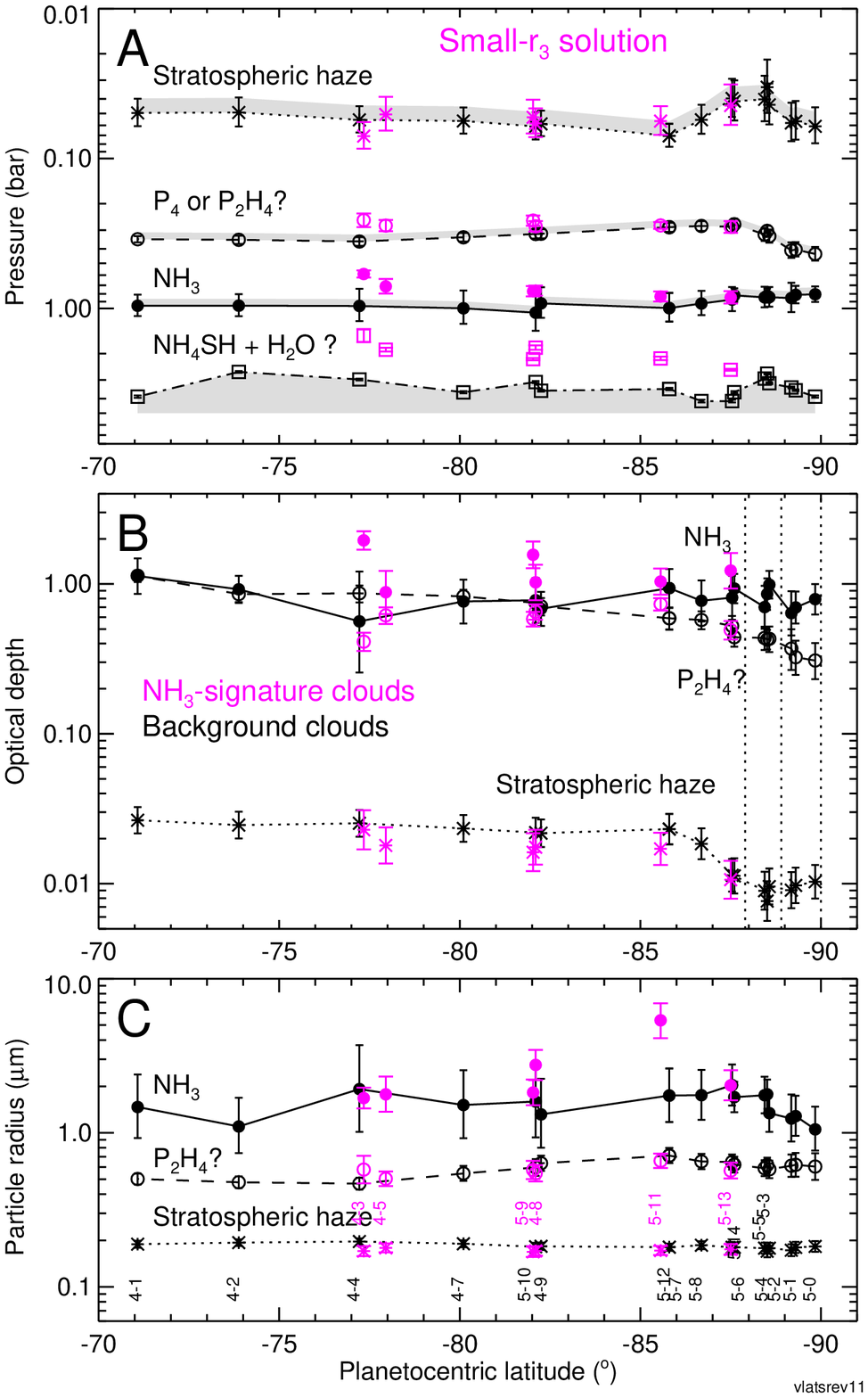}
\hspace{-0.125in}\includegraphics[width=3.5in]{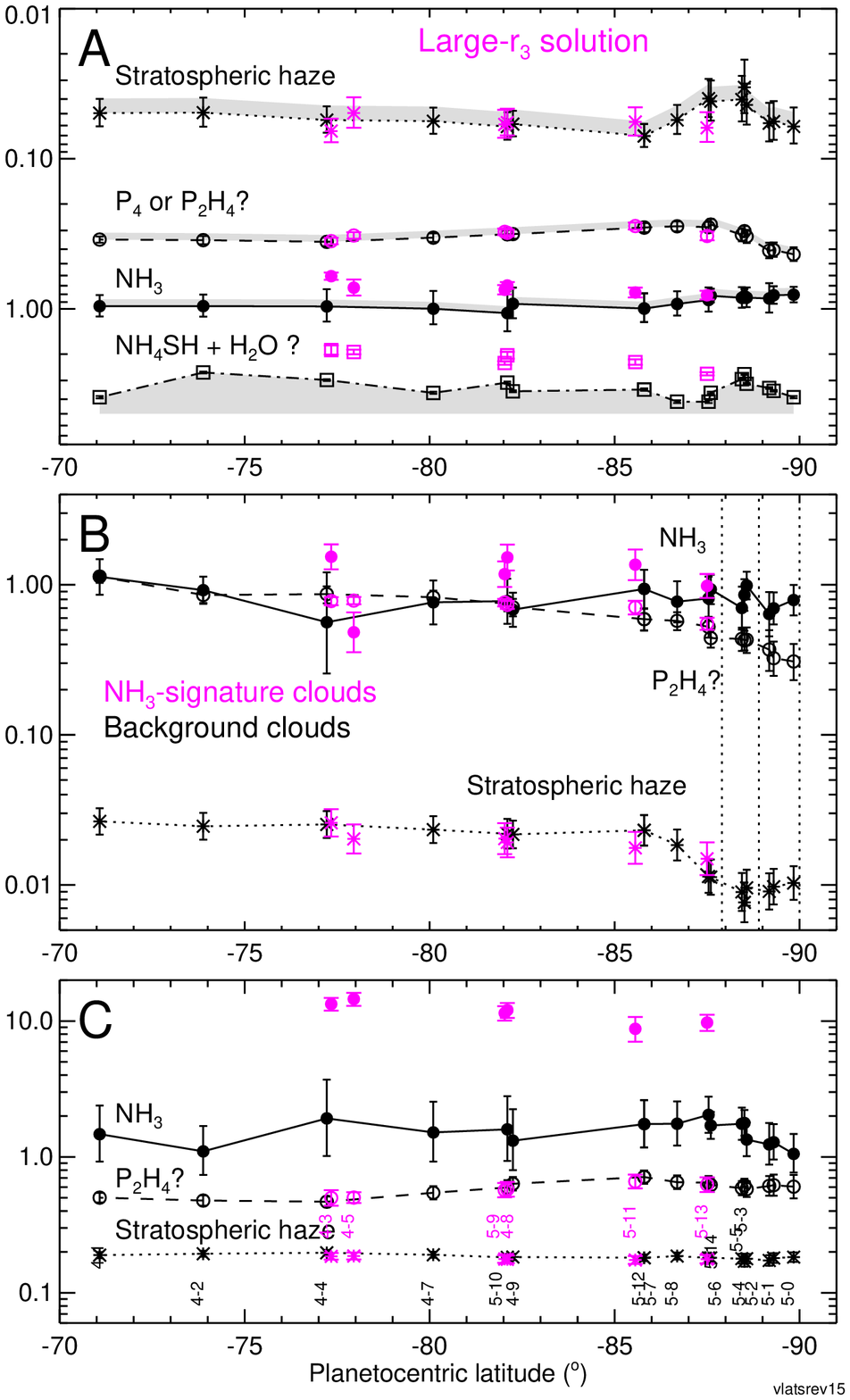}
\caption{Best-fit model parameters for background clouds (black) and
  \nhtx -signature clouds (magenta). The left column displays the
  small-$r_3$ solutions for the \nhtx-signature clouds, and the right
  column the large-$r_3$ solutions.  In each column, panel A displays
  pressures versus latitude for the putative diphosphine layer (open
  circles), the \nht layer (filled circles), and the top of the bottom
  layer (squares). Optical depths (B) and particle radii (C) are
  plotted for the top three layers.  The location numbers referred to in
  Figs.\ \ref{Fig:outereye} and \ref{Fig:innereye} are noted above the
  x-axis in panel C. The ammonia-signature clouds have larger
  particles and larger optical depths than the background clouds, and
  deep cloud tops that extend to higher altitudes. Dotted vertical
  lines at 87.9\degx S and 88.9\degx S mark locations of brightness
  step changes in ISS 752-nm images (discussed in
  Sec.\ \ref{Sec:comp}.)}\label{Fig:strucfits}
\end{figure*}

\begin{table*}\centering
\caption{Mean, standard deviation of fitted values about their mean, and uncertainty of the mean,
 for fits to background and
large-$r_3$ and small-$r_3$ fits to ammonia-signature clouds within the latitude
band from 73\degx S to 86\degx S, where other parameters are relatively independent
of latitude.}\label{Tbl:meansd}
\setlength{\tabcolsep}{0.04in}
\renewcommand{\arraystretch}{0.75} 
\begin{tabular}{l c c c| c c c| c c c }
                &\multicolumn{3}{c}{Background (N=6)} & \multicolumn{3}{c}{\nhtx -sig., large $r_3$ (N=4)} & \multicolumn{3}{c}{\nhtx -sig., small $r_3$ (N=4)}\\[0.05in]
Par., unit & mean & $\sigma$ & $\sigma/\sqrt{N-1}$ & mean &  $\sigma$ &  $\sigma/\sqrt{N-1}$ & mean &  $\sigma$ &  $\sigma/\sqrt{N-1}$ \\
\hline
p$_1$, bar & 0.060 & 0.008 & 0.003 & 0.058 & 0.006 & 0.003 & 0.058 & 0.008 & 0.003\\[0.05in]
p$_2$, bar & 0.317 & 0.029 & 0.010 & 0.315 & 0.027 & 0.012 & 0.272 & 0.012 & 0.006\\[0.05in]
p$_3$, bar & 0.979 & 0.046 & 0.016 & 0.710 & 0.066 & 0.029 & 0.733 & 0.091 & 0.041\\[0.05in]
p$_{4t}$, bar & 3.369 & 0.458 & 0.162 & 2.084 & 0.188 & 0.084 & 1.915 & 0.271 & 0.121\\[0.05in]
$\tau_1$ x 100 & 2.270 & 0.209 & 0.074 & 2.068 & 0.313 & 0.140 & 1.834 & 0.263 & 0.118\\[0.05in]
$\tau_2$ & 0.720 & 0.123 & 0.044 & 0.754 & 0.032 & 0.014 & 0.600 & 0.119 & 0.053\\[0.05in]
$\tau_3$ & 0.795 & 0.134 & 0.047 & 1.216 & 0.435 & 0.195 & 1.292 & 0.452 & 0.202\\[0.05in]
r$_1$, \mum & 0.187 & 0.006 & 0.002 & 0.180 & 0.006 & 0.003 & 0.172 & 0.004 & 0.002\\[0.05in]
r$_2$, \mum & 0.599 & 0.096 & 0.034 & 0.562 & 0.066 & 0.029 & 0.572 & 0.057 & 0.025\\[0.05in]
r$_3$, \mum & 1.586 & 0.269 & 0.095 &12.019 & 2.175 & 0.973 & 2.682 & 1.565 & 0.700\\[0.05in]
p$_b$, bar & 0.177 & 0.019 & 0.007 & 0.245 & 0.045 & 0.020 & 0.179 & 0.035 & 0.016\\[0.05in]
$\alpha_0$, ppm & 4.343 & 0.181 & 0.064 & 9.428 & 3.186 & 1.425 & 5.330 & 0.841 & 0.376\\[0.05in]
$AsH_3v$, ppb & 1.769 & 0.272 & 0.096 & 2.542 & 0.554 & 0.248 & 1.322 & 0.325 & 0.145\\[0.05in]
\chisqx/$N_F$ & 0.985 & 0.108 & 0.038 & 1.089 & 0.111 & 0.050 & 1.265 & 0.304 & 0.136\\[0.05in]
\hline
\end{tabular}
\parbox{4.5in}{NOTE: The averages are unweighted, $\sigma$ is the standard deviation
of fits about the mean and  $\sigma/\sqrt{N-1}$ is the estimated uncertainty of the mean.}
\end{table*}

\subsection{Models of ammonia-signature clouds}\label{Sec:nh3sigfit}

\begin{figure*}[!htb]\centering
\includegraphics[width=6.25in]{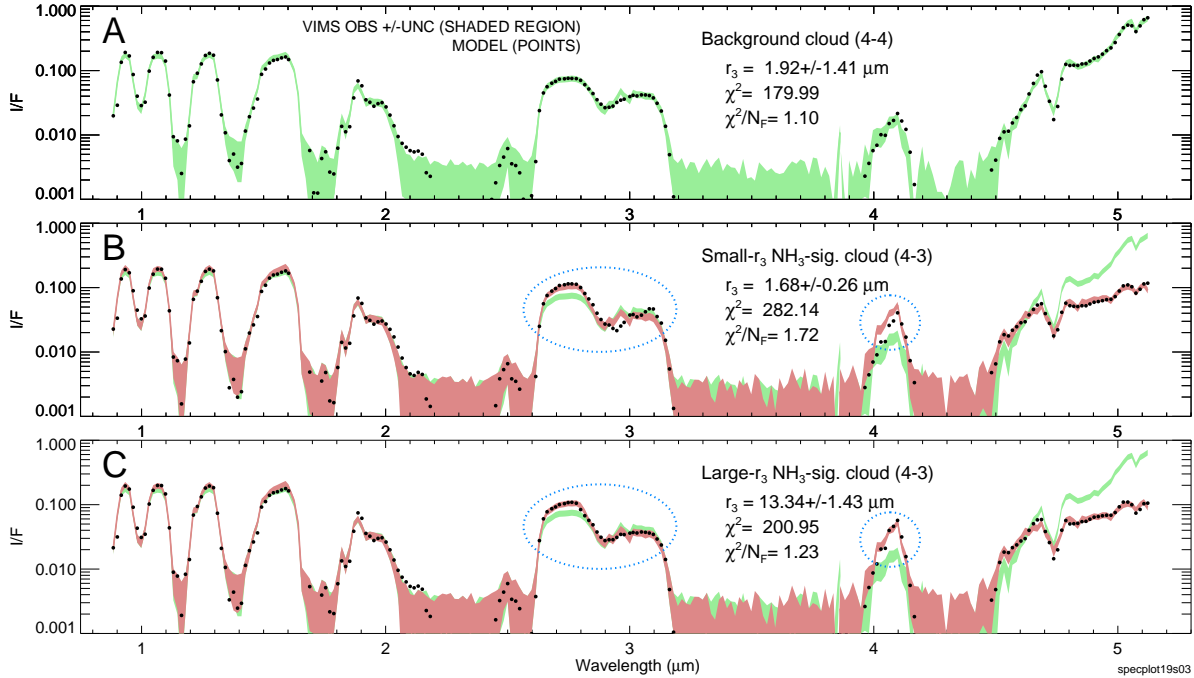}
\caption{{\bf (A:)} Model spectrum (points) compared to observed background spectrum from location 4-4 (light
green shaded region, with shading indicating combined uncertainty of measurement and modeling).
{\bf (B:)} As in A except the model is the small-$r_3$ fit to the \nht signature spectrum from location 4-3, with
shading in red, and background spectrum from A shown with green shading. {\bf (C:)}   
As in B, except the model spectrum is for the large-$r_3$ solution, which provides a
significantly better fit in this case, especially in regions within dotted blue ovals. }\label{Fig:spectralfits}
\end{figure*}

The ammonia signature clouds are taken to be those that appear
brighter than background clouds at pseudo-continuum wavelengths of
1.59 \mum and 4.1 \mumx, while being darker than or at least no
brighter than background clouds at 3.05 \mumx, a wavelength at which
\nht ice absorbs strongly.  These are the cloud features that have a
magenta color in panels E and G of Figs.\ \ref{Fig:outereye} and
\ref{Fig:innereye}.  Two different solutions were found for the
ammonia-signature features: a solution with a small (1-2 \mumx)
particle radius in the ammonia layer (layer 3), and a second solution
with a much larger (10-13 \mumx) particle radius in the ammonia layer,
comparable to the 14-\mum ammonia particle radius found by \cite{Baines2018GeoRL}
for a bright feature in the north polar eye.
The best-fit model results are plotted in Fig.\ \ref{Fig:strucfits}
using a magenta color, with parameter values and uncertainties given
in Table\ \ref{Tbl:nh3sig}.  Here the left column is for the
small-$r_3$ solutions and the right column is for the large-$r_3$
solutions.  The same background structure is plotted in both columns
for reference.  Average characteristics over the restricted outer
region are also given for both solutions in Table\ \ref{Tbl:meansd},
along with standard deviations, and uncertainty estimates for the
averages.  The larger particle solution produces slightly better
overall fit qualities and in some cases makes significant improvements
in local regions where ammonia spectral signatures are more evident.
This is illustrated in Fig.\ \ref{Fig:spectralfits}, which provides a
comparison of measured spectra with corresponding model spectra at
background location 4-4 and \nht -signature spectra at location 4-3.
Here the improvement provided by the large-$r_3$ solution is
significant and is obvious near 3 \mum and 4.1 \mumx.  But there are
other cases in which the alternate solutions are of comparable
quality, which can be seen by comparing \chisq values in
Tables\ \ref{Tbl:nh3sig} and \ref{Tbl:nh3sigbigr3}.  Although there is
no fit for which the small-$r_3$ solution is better than the
large-$r_3$ solution, we kept both because in several cases the
differences are insignificant, and in all cases the large-$r_3$
solutions are accompanied by significantly different gas mixing ratios
compared to background fits, while the small-$r_3$ solutions are in
better agreement with the background model gas parameters (discussed
in Sec.\ \ref{Sec:gas}). On the other hand, the large-$r_3$ solutions
are more consistent with background optical depths and particle sizes
for the stratospheric haze, and with particle sizes and pressures for
the putative diphosphine layer.

\begin{table*}[!hb]\centering
\caption{Fits to \nhtx-signature cloud spectra, small $r_3$ solution from all regions.}\label{Tbl:nh3sig}
\setlength{\tabcolsep}{0.08in}
\renewcommand{\arraystretch}{0.85} 
\begin{tabular}{l c c c c c c c c }
Locations:&                 4-3&                 4-5&                 5-9&                 4-8&                5-11&                5-13\\[0.05in]
PC. Lat. &          -77.3\degx  &          -77.9\degx  &          -82.0\degx  &          -82.1\degx  &          -85.6\degx  &          -87.5\degx \\[0.05in]
\hline\\
              $p_1$, bar & 0.071$^{+0.015}_{-0.014}$ & 0.051$^{+0.014}_{-0.012}$ & 0.053$^{+0.015}_{-0.013}$ & 0.058$^{+0.014}_{-0.012}$ & 0.056$^{+0.013}_{-0.012}$ & 0.044$^{+0.015}_{-0.012}$\\[0.05in]
              $p_2$, bar &  0.26$^{+ 0.03}_{- 0.03}$ &  0.28$^{+0.03}_{-0.02}$ &  0.26$^{+0.02}_{-0.02}$ &  0.28$^{+0.02}_{-0.02}$ &  0.28$^{+0.02}_{-0.01}$ &  0.29$^{+0.03}_{-0.02}$\\[0.05in]
              $p_3$, bar &  0.59$^{+ 0.03}_{- 0.03}$ &  0.71$^{+0.08}_{-0.08}$ &  0.77$^{+0.06}_{-0.06}$ &  0.76$^{+0.06}_{-0.06}$ &  0.83$^{+0.07}_{-0.06}$ &  0.85$^{+0.08}_{-0.08}$\\[0.05in]
           $p_{4t}$, bar &  1.52$^{+ 0.16}_{- 0.15}$ &  1.89$^{+0.06}_{-0.05}$ &  2.18$^{+0.04}_{-0.04}$ &  1.83$^{+0.06}_{-0.06}$ &  2.16$^{+0.06}_{-0.06}$ &  2.57$^{+0.04}_{-0.04}$\\[0.05in]
   $\tau_1$$\times 10^2$ &  2.29$^{+ 0.80}_{- 0.60}$ &  1.80$^{+0.58}_{-0.44}$ &  1.62$^{+0.55}_{-0.41}$ &  1.75$^{+0.54}_{-0.41}$ &  1.71$^{+0.48}_{-0.37}$ &  1.06$^{+0.36}_{-0.27}$\\[0.05in]
                $\tau_2$ &  0.41$^{+ 0.06}_{- 0.06}$ &  0.62$^{+0.08}_{-0.08}$ &  0.58$^{+0.07}_{-0.06}$ &  0.65$^{+0.08}_{-0.07}$ &  0.73$^{+0.07}_{-0.07}$ &  0.49$^{+0.07}_{-0.07}$\\[0.05in]
                $\tau_3$ &  1.95$^{+ 0.29}_{- 0.26}$ &  0.88$^{+0.34}_{-0.25}$ &  1.57$^{+0.35}_{-0.29}$ &  1.03$^{+0.32}_{-0.25}$ &  1.04$^{+0.23}_{-0.19}$ &  1.23$^{+0.38}_{-0.30}$\\[0.05in]
             $r_1$, \mum & 0.171$^{+0.013}_{-0.012}$ & 0.179$^{+0.012}_{-0.011}$ & 0.169$^{+0.013}_{-0.012}$ & 0.171$^{+0.013}_{-0.012}$ & 0.172$^{+0.012}_{-0.011}$ & 0.175$^{+0.014}_{-0.013}$\\[0.05in]
             $r_2$, \mum &  0.58$^{+ 0.13}_{- 0.11}$ &  0.50$^{+0.06}_{-0.05}$ &  0.57$^{+0.08}_{-0.07}$ &  0.55$^{+0.07}_{-0.06}$ &  0.66$^{+0.08}_{-0.07}$ &  0.57$^{+0.07}_{-0.06}$\\[0.05in]
             $r_3$, \mum &  1.68$^{+ 0.28}_{- 0.24}$ &  1.78$^{+0.54}_{-0.41}$ &  1.83$^{+0.39}_{-0.32}$ &  2.76$^{+0.70}_{-0.56}$ &  5.37$^{+1.54}_{-1.26}$ &  2.03$^{+0.51}_{-0.41}$\\[0.05in]
            $p_b$, bar &  0.21$^{+ 0.04}_{- 0.03}$ &  0.21$^{+0.03}_{-0.03}$ &  0.17$^{+0.06}_{-0.04}$ &  0.13$^{+0.07}_{-0.05}$ &  0.18$^{+0.04}_{-0.03}$ &  0.18$^{+0.05}_{-0.04}$\\[0.05in]
     $\alpha_0$, ppm &   5.3$^{+  0.7}_{-  0.6}$ &   4.9$^{+ 0.7}_{- 0.6}$ &   4.3$^{+ 0.6}_{- 0.5}$ &   5.6$^{+ 1.1}_{- 0.9}$ &   6.6$^{+ 1.2}_{- 1.0}$ &   4.1$^{+ 0.4}_{- 0.4}$\\[0.05in]
    $AsH_3v$, ppb &  1.15$^{+ 0.46}_{- 0.33}$ &  1.06$^{+0.44}_{-0.31}$ &  1.12$^{+0.42}_{-0.31}$ &  1.44$^{+0.55}_{-0.40}$ &  1.84$^{+0.62}_{-0.49}$ &  0.99$^{+0.37}_{-0.27}$\\[0.05in]
$\chi^2$ & 282.14 & 193.41 & 178.86 & 229.21 & 153.61 & 154.14\\[0.05in]
$\chi^2/N_F$ &   1.72 &   1.18 &   1.09 &   1.40 &   0.94 &   0.94 \\[0.05in]
\hline\\[-0.05in]
\end{tabular}
\normalsize
\parbox{5in}{NOTE: These fits cover the spectral range from 0.88 \mum to 5.12 \mumx, with exclusions for
order sorting filter joints and regions with very low S/N ratios.  These fits assumed fixed values of
   $p_4$ = 5 bars, $r_4$ = 2 \mumx, $f$ = 0.1, and $d\tau_4/dp$ = 20/bar. }
\end{table*}

\begin{table*}[!hb]\centering
\caption{Fits to \nhtx-signature cloud spectra, large $r_3$ solution from all regions.}\label{Tbl:nh3sigbigr3}
\setlength{\tabcolsep}{0.08in}
\renewcommand{\arraystretch}{0.85} 
\begin{tabular}{l c c c c c c c c }
\small
Locations:&                 4-3&                 4-5&                 5-9&                 4-8&                5-11&                5-13\\
Planetocent. Lat. &          -77.3\degx  &          -77.9\degx  &          -82.0\degx  &          -82.1\degx  &          -85.6\degx  &          -87.5\degx \\[0.05in]
\hline\\
              $p_1$, bar & 0.065$^{+0.012}_{-0.011}$ & 0.050$^{+0.013}_{-0.011}$ & 0.059$^{+0.013}_{-0.012}$ & 0.057$^{+0.012}_{-0.011}$ & 0.057$^{+0.013}_{-0.011}$ & 0.062$^{+0.015}_{-0.013}$\\[0.05in]
              $p_2$, bar &  0.35$^{+ 0.02}_{- 0.02}$ &  0.32$^{+0.02}_{-0.02}$ &  0.31$^{+0.02}_{-0.02}$ &  0.31$^{+0.02}_{-0.02}$ &  0.28$^{+0.02}_{-0.01}$ &  0.33$^{+0.02}_{-0.02}$\\[0.05in]
              $p_3$, bar &  0.61$^{+ 0.03}_{- 0.03}$ &  0.72$^{+0.09}_{-0.09}$ &  0.75$^{+0.05}_{-0.05}$ &  0.70$^{+0.04}_{-0.04}$ &  0.78$^{+0.06}_{-0.06}$ &  0.81$^{+0.06}_{-0.05}$\\[0.05in]
           $p_{4t}$, bar &  1.88$^{+ 0.13}_{- 0.13}$ &  1.94$^{+0.07}_{-0.07}$ &  2.30$^{+0.06}_{-0.06}$ &  2.04$^{+0.11}_{-0.11}$ &  2.26$^{+0.09}_{-0.09}$ &  2.71$^{+0.07}_{-0.07}$\\[0.05in]
   $\tau_1$$\times 10^2$ &  2.59$^{+ 0.60}_{- 0.49}$ &  2.03$^{+0.50}_{-0.41}$ &  2.03$^{+0.55}_{-0.43}$ &  1.92$^{+0.49}_{-0.39}$ &  1.77$^{+0.49}_{-0.38}$ &  1.49$^{+0.43}_{-0.33}$\\[0.05in]
                $\tau_2$ &  0.78$^{+ 0.05}_{- 0.05}$ &  0.78$^{+0.04}_{-0.04}$ &  0.76$^{+0.04}_{-0.04}$ &  0.73$^{+0.04}_{-0.04}$ &  0.71$^{+0.07}_{-0.07}$ &  0.55$^{+0.05}_{-0.05}$\\[0.05in]
                $\tau_3$ &  1.54$^{+ 0.32}_{- 0.27}$ &  0.48$^{+0.17}_{-0.13}$ &  1.18$^{+0.25}_{-0.21}$ &  1.52$^{+0.33}_{-0.28}$ &  1.36$^{+0.36}_{-0.29}$ &  0.98$^{+0.20}_{-0.17}$\\[0.05in]
             $r_1$, \mum & 0.185$^{+0.011}_{-0.010}$ & 0.187$^{+0.011}_{-0.011}$ & 0.177$^{+0.012}_{-0.011}$ & 0.176$^{+0.012}_{-0.011}$ & 0.174$^{+0.012}_{-0.011}$ & 0.177$^{+0.013}_{-0.012}$\\[0.05in]
             $r_2$, \mum &  0.50$^{+ 0.07}_{- 0.06}$ &  0.50$^{+0.05}_{-0.04}$ &  0.57$^{+0.07}_{-0.07}$ &  0.58$^{+0.08}_{-0.07}$ &  0.66$^{+0.08}_{-0.07}$ &  0.63$^{+0.08}_{-0.07}$\\[0.05in]
             $r_3$, \mum & 13.34$^{+ 1.43}_{- 1.43}$ & 14.52$^{+1.58}_{-1.61}$ & 11.44$^{+1.43}_{-1.38}$ & 12.03$^{+1.53}_{-1.49}$ &  8.76$^{+1.93}_{-1.73}$ &  9.74$^{+1.37}_{-1.29}$\\[0.05in]
            $p_b$, bar &  0.32$^{+ 0.04}_{- 0.04}$ &  0.23$^{+0.06}_{-0.05}$ &  0.21$^{+0.04}_{-0.03}$ &  0.26$^{+0.03}_{-0.03}$ &  0.21$^{+0.03}_{-0.02}$ &  0.20$^{+0.05}_{-0.04}$\\[0.05in]
     $\alpha_0$, ppm &  12.8$^{+  2.5}_{-  2.2}$ &   6.7$^{+ 1.3}_{- 1.1}$ &   6.6$^{+ 0.9}_{- 0.8}$ &  12.9$^{+ 1.9}_{- 1.7}$ &   8.2$^{+ 1.1}_{- 1.0}$ &   5.6$^{+ 0.5}_{- 0.5}$\\[0.05in]
    $AsH_3v$, ppb &  3.24$^{+ 0.96}_{- 0.83}$ &  2.07$^{+0.80}_{-0.61}$ &  2.95$^{+0.83}_{-0.70}$ &  2.50$^{+0.76}_{-0.63}$ &  1.95$^{+0.67}_{-0.52}$ &  1.66$^{+0.51}_{-0.40}$\\[0.05in]
$\chi^2$ & 200.95 & 181.96 & 174.57 & 184.57 & 150.86 & 154.94\\[0.05in]
$\chi^2/N_F$ &   1.23 &   1.11 &   1.06 &   1.13 &   0.92 &   0.94\\[0.05in]
\hline\\[-0.05in]
\end{tabular}
\normalsize
\parbox{5.2in}{NOTE: These fits cover the spectral range from 0.88 \mum to 5.12 \mumx, with exclusions for
order sorting filter joints and regions with very low S/N ratios.  These fits assumed fixed values of
   $p_4$ = 5 bars, $r_4$ = 2 \mumx, $f$ = 0.1, and $d\tau_4/dp$ = 20/bar. }
\end{table*}

The most consistent characteristic of the \nhtx-signature cloud structures is that they provide
a strong blocking of Saturn's thermal emission.  But in those structures the cloud layer that provides
almost all the blocking is the deep cloud, not the ammonia cloud.  The deep cloud layer extends upwards
to much lower pressures than is typical for the background clouds, averaging 1.9$\pm$0.12 bars compared
to the background average of 3.24$\pm$0.16 bars, both in the outer region, where
the uncertainties of the unweighted averages are computed as $\sigma$/$\sqrt{(N-1)}$.  The effect
of these deep clouds is mainly restricted to the thermal emission spectral region, although
they do provide a small boost to the continuum spectra at wavelengths less than 1.65 \mum (as
evident from derivative spectra shown in Section\ \ref{Sec:deriv}). We chose a model in which the base
of the deep cloud is fixed at 5 bars and has a fixed optical density, but an adjustable (fitted)
cloud top.  This choice is consistent with the idea that
 vertical convection elevates cloud particles to pressures low enough to block thermal emission.
It also would have been possible to attenuate thermal emission with a suspended sheet cloud, although
that structure is less appealing on dynamical grounds.

For the \nhtx-signature structures in the outer restricted region, the
elevation of the deep cloud is accompanied by a smaller local
elevation of the \nht cloud layer, from 979$\pm$16 mbar to 710$\pm$29 mbar
or 733$\pm$41 mbar for large and small $r_3$ solutions respectively,
 as well as an increase in its optical depth from 0.80$\pm$0.05
to an average of 1.22$\pm$0.20 or 1.29$\pm$0.20.  The \nhtx --signature cloud
structures have significant variability in the ammonia layer, and very
little variability in the putative diphosphine layer. The \nht
 spectral signature for the discrete feature we analyzed closest to the pole
(5-13 at 87.5\degx S) is made more obvious by the small 0.49 optical
depth of the diphosphine layer.

\subsection{Comparison of background and \nhtx -signature cloud structures}

\begin{figure*}[!ht]\centering
\hspace{-0.15in}\includegraphics[width=5.5in]{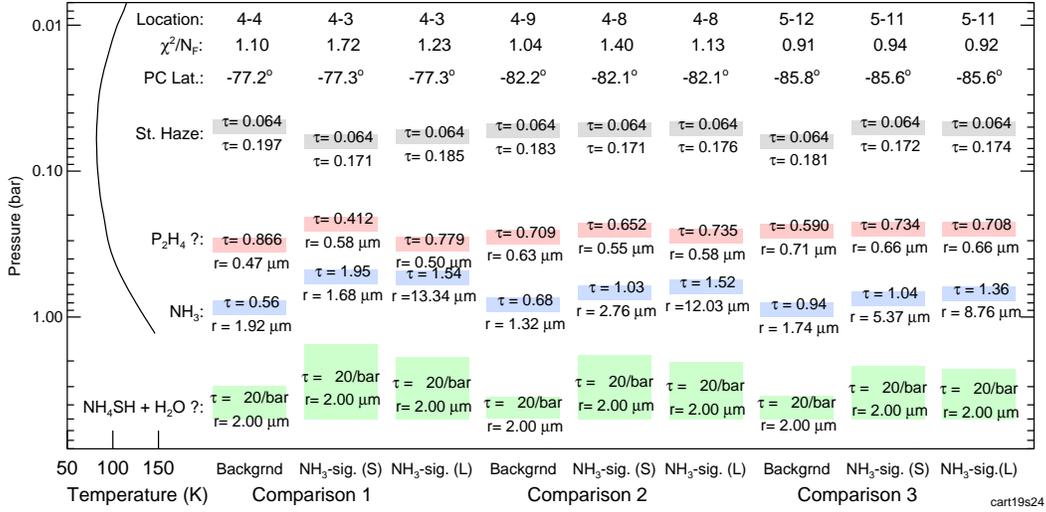}
\caption{Comparison of inferred structures of three \nhtx -signature
  clouds with the structures of the nearby background clouds.  We show
  both the small-$r_3$ and large-$r_3$ solutions for the
  \nhtx-signature clouds, identified in the x-axis labels by (S) and
  (L) respectively.  At the top are ID codes that refer to locations
  marked in Figs. \ref{Fig:outereye} and
  \ref{Fig:innereye}.}\label{Fig:nh3backexamples}
\end{figure*}

The most consistent feature of the \nhtx-signature clouds, compared to
background clouds, is the relatively low top pressure of their deep
\nhfsh+\hto layer, which makes that layer more effective in blocking
Saturn's thermal emission.  All the \nhtx -signature clouds have this
feature.  Most of the background clouds have the top of this layer
about 1 bar deeper, although there are a few background clouds that
have thicker bottom layers.  Three examples of \nhtx -signature cloud
structures in comparison with nearby background cloud structures are
shown in Fig.\ \ref{Fig:nh3backexamples}.  For the left hand pair,
from near 77\degx S, the main differences of the \nhtx -signature
structure from the background structure are (1) all the cloud layers
are elevated relative to the background, most dramatically for the top
of the deep layer, which moved from 3 bars to 1.5 bars, (2) the
optical depth of the ammonia layer is up by more than a factor of two,
and (3) the optical depth of the putative diphosphine layer is
slightly lower.  The elevated top of the deep layer suggests that deep
convection is responsible for initiating the other observed effects.

For the \nhtx-signature structures, the elevation of the \nht layer
and its increased opacity is a plausible consequence of the
Taylor-Proudman theorem, in which Coriolis forces prevent motion
perpendicular to the rotation axis \citep{Pedlosky1982}.  Thus a pulse
of deep convection can push higher cloud layers upward, without
convective penetration of those layers, as if the column above the convective pulse
were confined by vertical walls confining the motion of gases within
it.  The fact that the putative diphosphine layer does not exhibit an
optical depth increase, or much elevation for the large-$r_3$
solution, seems to contradict this idea, perhaps because the column
confinement does not quite extend to that pressure level.  While the
typical assumptions used to prove the Taylor-Proudman theorem, such as
the fluid being incompressible, are not strictly satisfied on Saturn,
\cite{Hide1966PSS} argued that the planet's rotation was sufficiently
rapid to make confining effects (tending to force motions to be
constant along columns parallel to the rotation axis) potentially
important on both Jupiter and Saturn.  Near the poles these confining
columns become close to vertical.

The source of the unique spectral signature is in the upper
troposphere, mainly controlled by the optical depth and pressure
levels of the diphosphine and ammonia layers. The \nhtx -signature clouds have optically
thicker and higher \nht layers, but generally somewhat less optical depth
 in the diphosphine layer. The alternate solutions to the \nhtx -signature
cloud structures are compared with each other and with the nearby background
structures in Fig.\ \ref{Fig:nh3backexamples}. We see that the large-$r_3$ solutions
suggest less perturbation of the putative diphosphine layer by the presumed
convective pulse that generates the \nhtx -signature structure.

\subsection{Gas profile results}\label{Sec:gas}

Besides the aerosol parameters,  we also
included adjustable parameters to constrain the vertical gas profiles
for  \pht and \ashtx.  For \phtx, we made use of three \pht
absorption bands of differing strengths (near 2.9 \mumx, 4.3 \mumx,
and 4.75 \mumx) to try to constrain the main parameters defining its
vertical profile: the deep mixing ratio $\alpha_0$, the pressure break
point $p_b$, and the scale height ratio $f$ above the break point.
What we found in general was a very loosely constrained pressure break
point near the top of the main cloud layer (near 200 mbar) rather than
the value of 550 mbar used by \cite{Fletcher2009ph3}, or the 1.3 bar value
inferred from VIMS 5-\mum emission spectra by \cite{Fletcher2011vims}.
More recent analysis of the VIMS 5-\mum emission spectra by
\cite{Barstow2016} found that a break point at 1.1 bars was the
best-fit value for nadir-only spectral fits, but was much lower when
limb darkened observations were included and might be even lower than
500 mbar, which was the lowest pressure they considered.  However,
even with our use of daytime observations that allow us to sample
three spectral bands of \phtx, there remains such a strong correlation
between the spectral effects of a scale height change and those due to
a change in pressure break point (as shown in Section\ \ref{Sec:deriv} in
Figs. \ref{Fig:logder} and \ref{Fig:logder2}), that these two cannot
be independently well constrained.  Thus,  we chose to
fix the scale height ratio at a small value of 0.1 and fit only the
other two profile parameters (the deep VMR, and the pressure break
point).  This relatively sharp drop in \pht above the cloud tops is
similar to that found by \cite{Prinn1975} in their models of \pht
photolysis in the weakly convective regions on Jupiter.

\begin{figure*}[!htb]\centering
\includegraphics[width=4.5in]{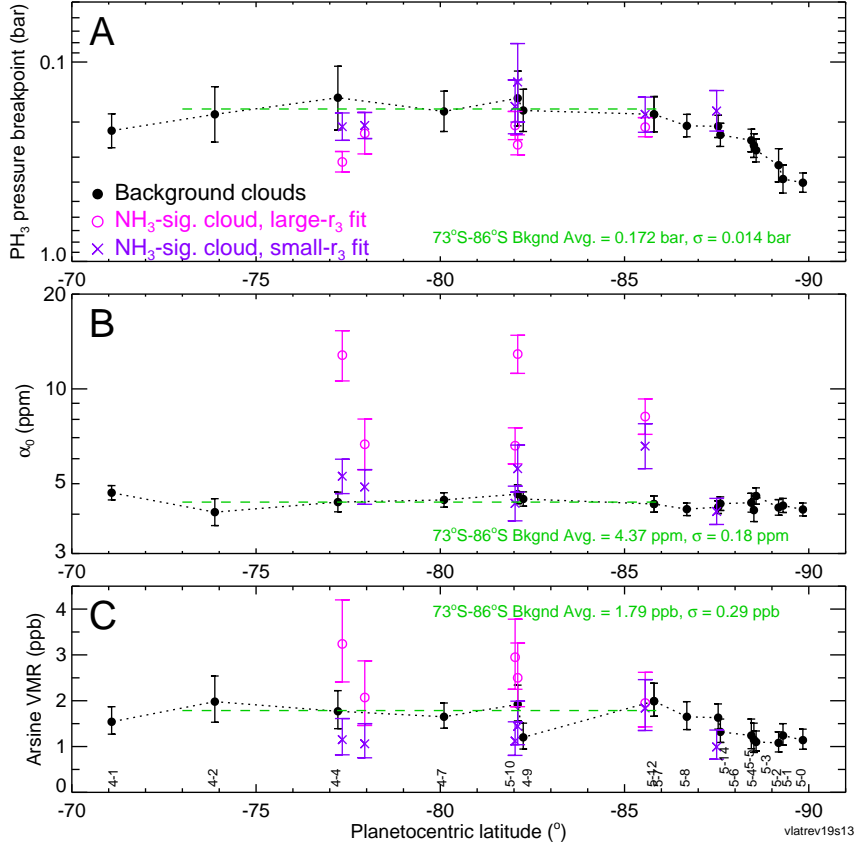}
\caption{Retrieved \pht break point pressures (A), deep \pht mixing ratios (B),
and arsine deep mixing ratios (C), all versus planetocentric latitude.
Filled circles (black) indicate fits to background cloud models. 
Fits to spectra of \nhtx --signature cloud features
are plotted as open circles (magenta) for large-$r_3$ solutions and as $\times$'s (purple)
for small-$r_3$ solutions.  
}\label{Fig:gasres}
\end{figure*}

Our gas parameter fit results for background cloud models given in
Tables\ \ref{Tbl:backouter} and \ref{Tbl:backinner}, as well as \nhtx
-signature cloud fits given in Tables \ref{Tbl:nh3sig} and \ref{Tbl:nh3sigbigr3}, are all plotted
in Fig.\ \ref{Fig:gasres} as a function of latitude.  The parameters
derived from background spectra have remarkably smooth variations with
latitude, all remaining fairly constant between about 73\degx S and
86\degx S, with averages over that region of 172 mb (with standard
deviation of $\sigma$ = 14 mbar) for the \pht pressure breakpoint, 4.4
ppm ($\sigma$ = 0.2 ppm) for the deep \pht mixing ratio (which remains
at essentially the same value to the pole), and 1.8 ppb ($\sigma$ =
0.3 ppb) for the \asht mixing ratio. However, between 86\degx S and
the pole, the \pht breakpoint pressure increased to 400 mb, while the
arsine mixing ratio decreased to about 1.1 ppb, both of which suggest
downward motions in this latitude band, as well as that \asht probably
is not uniformly mixed, but declines with altitude.  In general the break-point is
found slightly above the the putative diphosphine cloud, which is
consistent with the idea that at large incident angles the
photochemical destruction of \pht is strongly inhibited below the top
cloud level. 
  
Our deep \pht VMR values in the polar region are
generally at the lower end of the 4-9 ppm range of
\cite{Fletcher2008}, averaging about 4.4 ppm instead of roughly 6 ppm,
and do not show any evidence of their large 8 ppm peak near 82\degx
S. Perhaps this should not be too surprising given their different
modeling assumptions, namely that of a fixed pressure knee at 550 mbar
and an adjustable scale height, as well as the fact that we can make
use of the very strong \pht band centered at 4.3 \mumx, which appears
in reflected sunlight.

From about 76\degx S to 86\degx S, our mean arsine VMR
(1.79$\pm$0.13 ppb) is between the two values derived by
\cite{Bezard1989} of 2.4$^{+1.4}_{-1.2}$ ppb for the thermal component
and 0.39$^{+0.21}_{-0.13}$ ppb for the reflected solar component.
These two values are consistent with a mixing ratio that declines with
altitude.  Our results are also consistent with the deep value of
1.8$^{+1.8}_{-0.9}$ ppb inferred by \cite{Noll1989AsH3}, but somewhat
lower than the more global value of 2.2$\pm$0.3 ppb obtained by
\cite{Fletcher2011vims} from VIMS nighttime observations assuming
scattering clouds.  Given evidence for descending motions in the south
polar region, it is not too surprising to find a somewhat reduced
level of arsine relative to global mean values.  It is also the case
that we are probably characterizing cloud properties better by
combining thermal and solar reflected light and including scattering
by cloud layers, and that may also be a factor in retrieval of
different arsine abundances.

The \pht results for the \nhtx --signature cloud structures are
generally more uncertain than those obtained for the background cloud
structures, perhaps because the \nht absorption feature near 3 \mum
interferes somewhat with the 2.9-\mum \pht absorption feature and the
extra cloud opacity in these structures reduce the visibility of gas
absorption features.  The gas parameters in these regions also have
some consistent deviations, most notably the greatly increased deep
mixing ratio of \pht found for the large-$r_3$ solutions
(Fig.\ \ref{Fig:gasres}B), which is accompanied by an increased
pressure breakpoint (Fig.\ \ref{Fig:gasres}A). \cite{Baines2018GeoRL} also
found comparably enhanced \pht mixing ratios associated
with the \nhtx--signature clouds in the north polar region.
 The arsine VMR results
for the \nhtx--signature structures are also elevated for the
large-$r_3$ solutions, but somewhat depressed for the small-$r_3$
solutions.  Much smaller deviations from the background gas parameter
solutions are found for the small-$r_3$ solutions for the \pht
parameters.  The somewhat better fits obtained from the large-$r_3$
solutons suggest that there may be more vertical variation in the \pht
and \asht gas profiles than are simplified models have assumed.
This issue warrants further investigation.

\begin{figure*}[!htb]\centering
\includegraphics[width=4.1in]{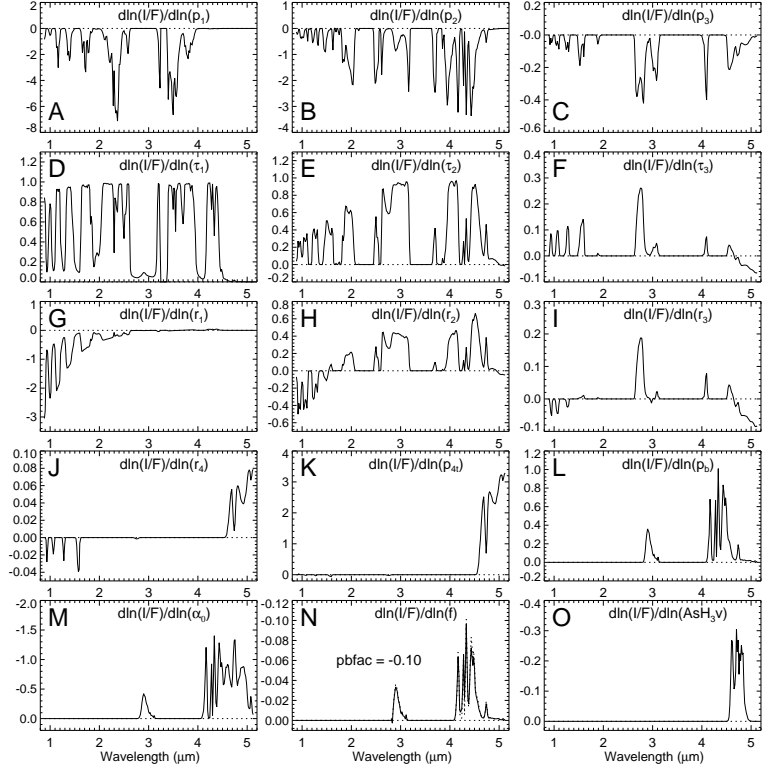}
\caption{Logarithmic derivatives of I/F
  with respect to aerosol model parameters (A-K) and gas parameters
  (L-O), for the background model (5-10) taken at best fit
values given in Table \ref{Tbl:backouter}.
 Derivatives are taken with respect to $p_1$ (A), $p_2$ (B), $p_3$ (C),
  $\tau_1$ (D), $\tau_2$ (E), $\tau_3$ (F), $r_1$ (G), $r_2$ (H), $r_3$
   (I), $r_4$ (J), $p_\mathrm{4t}$ (K), $p_b$ (L), $\alpha_0$ (M),
  $f$ (N), and $AsH_3v$ (O).   The dotted curve in N is the curve in L scaled
  by -0.16.
}
\label{Fig:logder}
\end{figure*}

\subsection{Sensitivity of model spectra to model parameters}\label{Sec:deriv}

The strength of influence on
the spectrum of various model parameters, and their correlations, are perhaps most easily
understood with the help of logarithmic derivatives, of the
form \begin{eqnarray} \frac{\partial \log (I(\lambda; x_1, \cdots,
    x_N))}{\partial \log (x_i)} & = & \frac{(1/I)\partial I(\lambda; x_1,
    \cdots, x_N)}{(1/x_i)\partial x_i} \nonumber \\
&  = & \frac{x_i \partial I(\lambda; x_1, \cdots, x_N)}{I\partial x_i},  \label{Eq:lnder}
\end{eqnarray}
where $x_i$ is any of the model parameters $x_1, \cdots, x_N$, and $I$ is the
model spectral radiance for the given set of parameters.  The relation also holds
if radiance $I$ replaced by reflectivity $I(\lambda; x_1, \cdots, x_N)/F(\lambda)$.
The middle expression in the above equation is easiest to interpret.  It
states that the logarithmic derivative is the ratio of the fractional change in $I$ (or $I/F$)
to the fractional change in parameter $x_i$ that produced it.  If the ratio
is large, then the parameter will be well constrained by the observed spectrum,
unless the spectral ratio for that parameter has a shape similar to that for one
of the other parameters, in which case their effects may be difficult to distinguish.

\begin{figure*}[!htb]\centering
\includegraphics[width=4.1in]{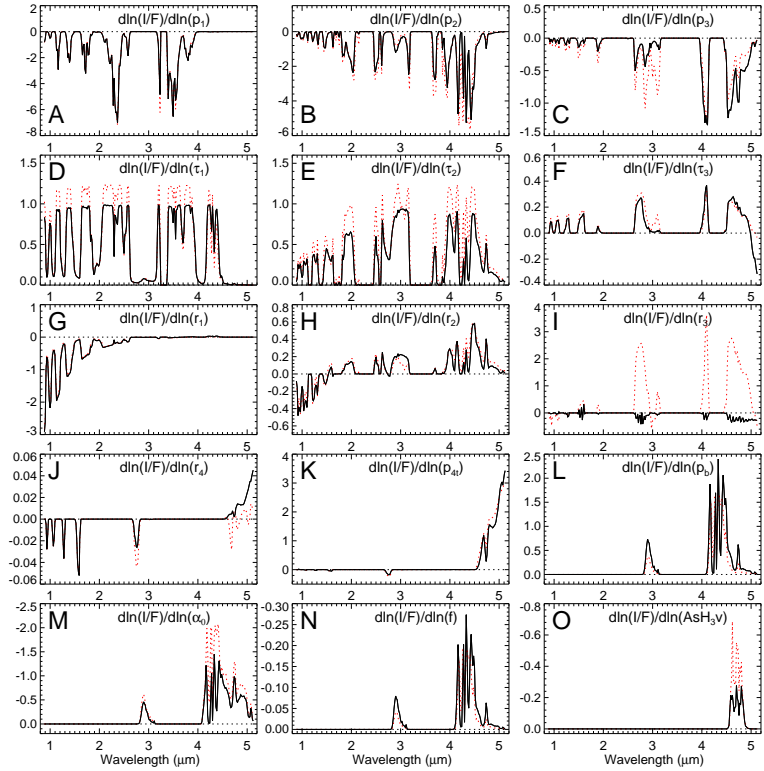}
\caption{As in Fig.\ \ref{Fig:logder} except that derivatives are
  taken for the \nhtx-signature model (5-9) at best fit parameter
  values given in Table\ \ref{Tbl:nh3sig}, with results for the large-$r_3$ solution in solid black,
  and for the small-$r_3$ solution in dotted red. }\label{Fig:logder2}
\end{figure*}


Logarithmic derivatives are displayed for the location 5-10 background
model structure in Fig.\ \ref{Fig:logder} and for the location 5-9
\nhtx-signature structures (both large-$r_3$ and large-$r_3$
solutions) in Fig.\ \ref{Fig:logder2}.  Both figures show derivatives
for both aerosol parameters (panels A-K) and gas parameters (panels
L-O).  Note that
most derivative spectra are distinctly different from each other,
suggesting that most parameters can be well constrained by the
observations, if they have sufficient influence on the spectrum.
 In the aerosol group for \nhtx -signature
models, the derivatives for $\tau_3$ and $r_3$ (in panels F and I) are quite similar at
wavelengths greater than 2 \mumx.  However, their influence at shorter
wavelengths is very different.  Long-exposure VIMS spectra for which
the shorter wavelengths are saturated cannot take advantage of this difference,
which is one reason we chose not to use them. 

A more serious ambiguity appears in the gas parameter group,
where panels L and N show extremely similar spectral shapes for
derivatives with respect to the phosphine pressure break point (L) and
with respect to the phosphine-to-pressure scale height ratio
(N). Plotting a scaled version of the former (dotted curve in panel N)
on top of the latter's plot shows that their spectral shapes are
nearly indistinguishable.  Thus, out of the three parameters used to
characterize the vertical distribution of \phtx, only one (the deep
phosphine mixing ratio, $\alpha_0$) can be constrained well at this model
point.  This does not mean that there are no boundaries for the other
two parameters, only that at the local point where the derivatives are
taken, one cannot distinguish between a small fractional increase in
the break-point pressure from a small fractional decrease in the phosphine
to pressure scale height ratio.

Neither structure allows much sensitivity either to the upper
tropospheric VMR of ammonia or its relative humidity.  A 100\% change
in these parameters produces only a few percent fractional change in
I/F, which is why we did not try to fit these parameters (or show them
in the derivative plots).  Also note (in panel K) that increasing the
deep layer cloud top pressure has the main effect of increasing the
thermal emission (seen at wavelengths beyond 4.5 \mumx), although the
spectral shapes are different for the two models, due to the much
higher pressure for the background model, which provides significantly
more visibility of the arsine absorption (evident in panel O). In
Fig.\ \ref{Fig:logder2}, the different mixing ratio profiles for
phosphine and arsine for the large-$r_3$ solution makes their
derivatives have less influence on the spectrum.

\subsection{Sensitivity of fits to initial guesses}

\begin{figure}[!htb]
\hspace{-0.25in}\includegraphics[width=3.75in]{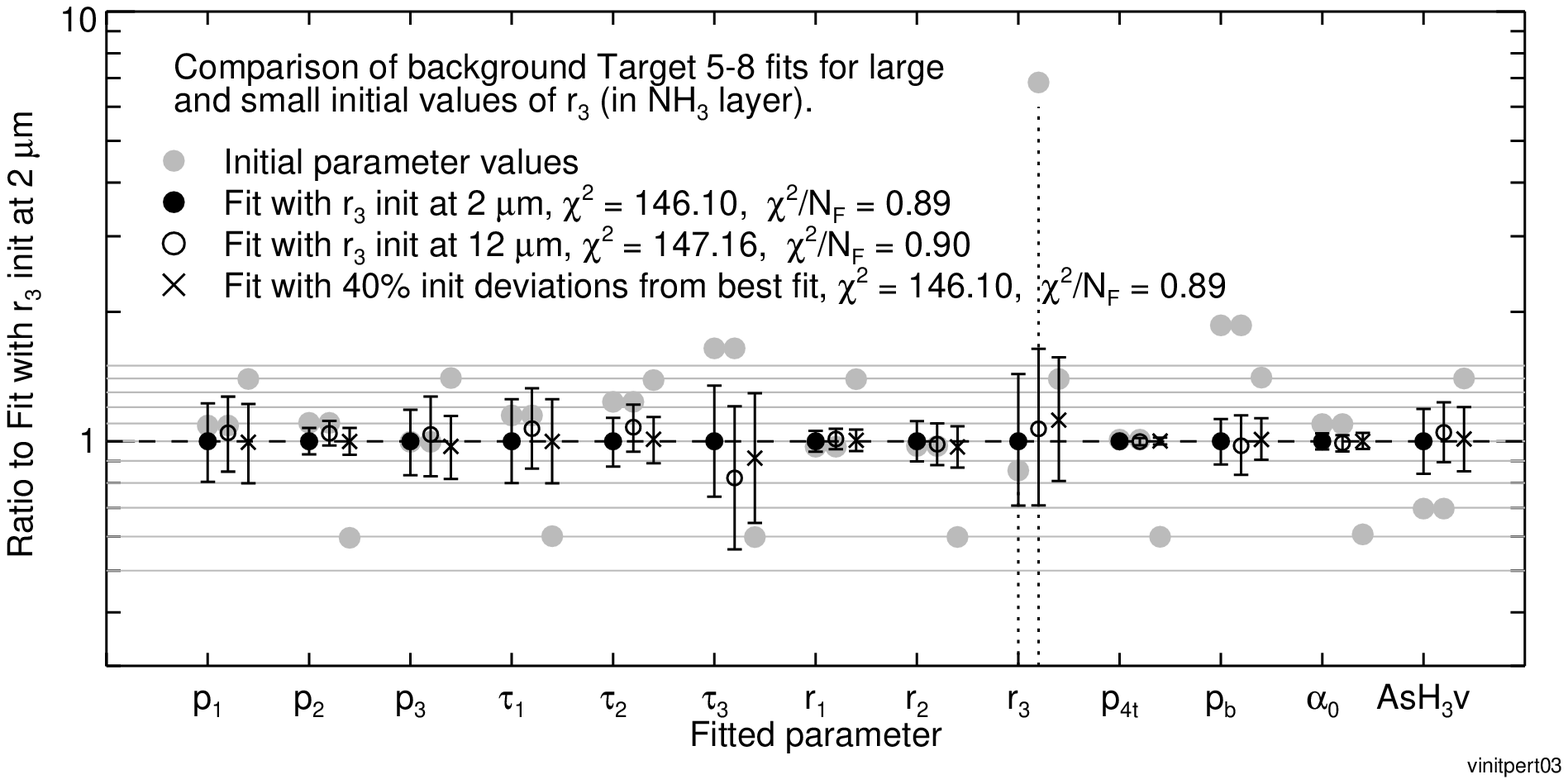}\par
\hspace{-0.25in}\includegraphics[width=3.75in]{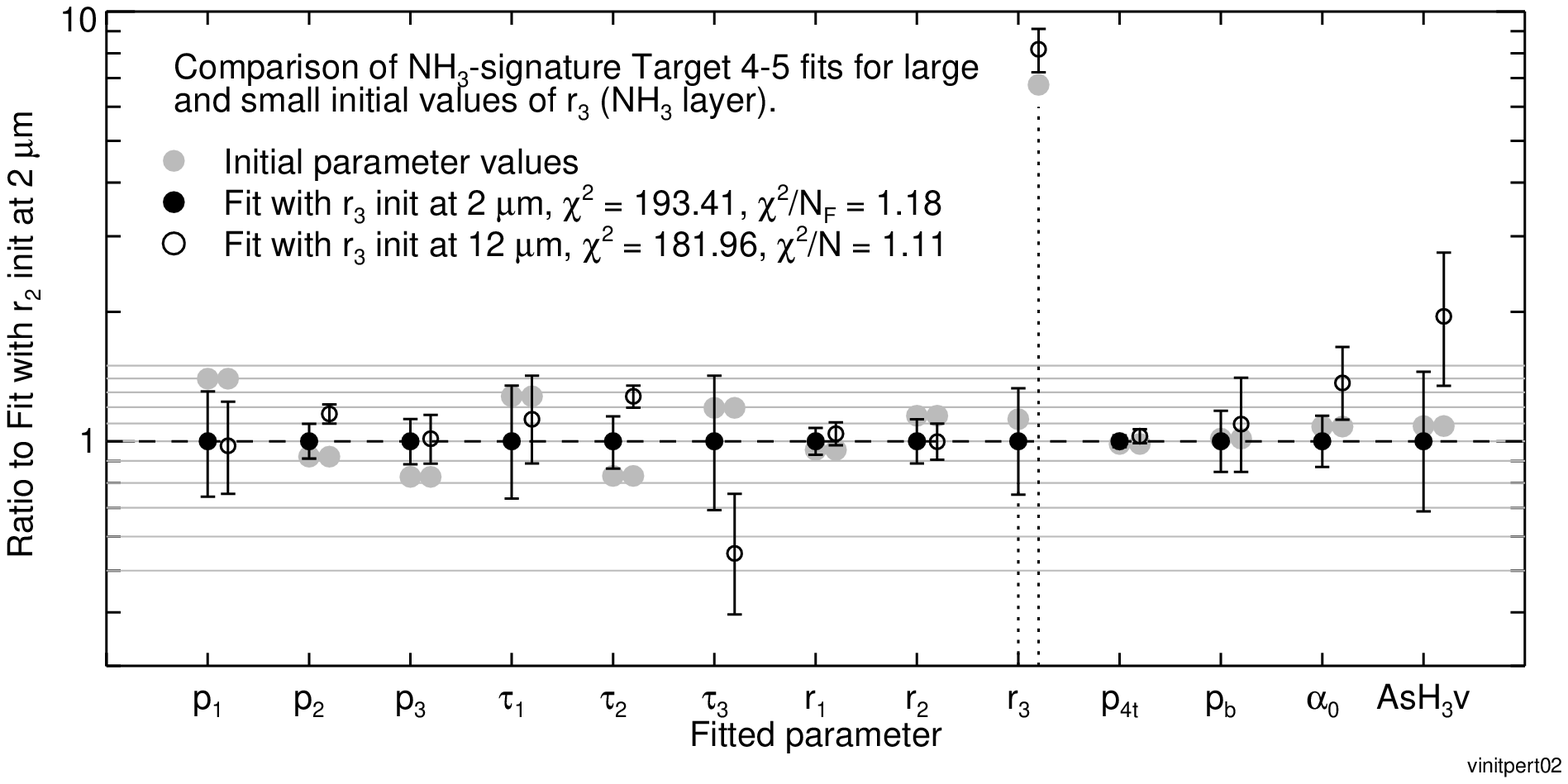}
\caption{Effects of different initial values of $r_3$ on best fit parameters for 
 a background cloud (top) and an ammonia-signature cloud (bottom). In each panel the grey
dots show the initial parameter values, the filled circles the fit that results from
a small value for the inital $r_3$ value, and the open circles the result for a large guess for
the initial $r_3$ value.  Each parameter is plotted as a ratio to that parameter for the
small $r_3$ initial value (that is why the first fit results are all on the unit line).
  For the background example (top) all fitted parameter values for the large-$r_3$ guess are nearly identical
to the results from the first small-particle fit, well within estimated uncertainties.  
For the \nhtx-signature example (bottom) the ammonia layer parameters and
gas mixing ratios are perturbed the most by the new solution obtained. }\label{Fig:perturb}
\end{figure}

Physical insight, guided
by the logarithmic derivative spectra, was used to formulate manual initial
guesses that provide at least crude fits to the observed spectra,
which were then refined by our L-M algorithm.  To see how sensitive the
final results were to the initial crude estimates, we did some trial perturbed
calculations, samples of which are illustrated in Fig.\ \ref{Fig:perturb}.  The
top panel illustrates various fits to the spectrum from location 5-8,
which samples a background cloud.  Three fits are shown here, with
initial guess and best-fit parameter values plotted for each case,
shown as ratios to the parameter values obtained from the first fit.
The initial guess values for each case are plotted using gray filled
circles.  The best fit values for the first case are plotted as black dots with
error bars.  These all have unit central values because they have a
unit ratio to themselves.  The most uncertain parameter is seen to be
the particle radius of the \nht cloud layer ($r_3$).  For the next
case, the initial value of that parameter was changed from 2 \mum to 12
\mumx. The resulting best fit using that guess, shown by open circles,
was very close to the initial fit, with parameter values well within the
fitting uncertainties.  The next case used an initial guess for each
fitted parameter that was either 40\% larger or 40\% smaller than the
 best fit value for the first case. 
Again, the results were almost identical to the initial
fit.  We conclude that the background fits are not very easily
perturbed.  It is also worth noting that the parameters 
best constrained by these fits are r$_1$, PH$_3$v, p$_{4t}$, p$_2$, and r$_2$.

Somewhat different results were obtained from perturbations of fits
to the \nhtx -signature spectra. In the bottom panel of Fig.\ \ref{Fig:perturb} we show fit results
for such a cloud at location 4-5. The first case uses
an inital guess of 2 \mum for $r_3$, which settles on $r_3$ = 1.75$^{+0.8}_{-0.5}$ \mum
as the best fit, and the second uses an inital guess
of 12 \mum for that parameter, yielding the very different best fit
value of $r_3$ = 14.5$\pm$1.5 \mumx.  Choosing intermediate first guess values
between 2 \mum and 12 \mum resulted in best fit solutions that were
close to the original small-$r_3$ or large-$r_3$ solutions, not to some
intermediate value.   For the location 4-5 spectrum the large-$r_3$
 solution actually fits better than the original solution, which is what led us to
carry out and present fits with both initial guesses for the \nhtx -signature features. 
This second solution
is also seen to result in a somewhat smaller optical depth for the ammonia
layer, a larger optical depth for the putative diphosphine layer, and larger 
values for the two gas mixing ratios.
The remaining parameters are within uncertainties of the initial fit, so that
the main features of the vertical structure are very similar for both solutions.


\section{Discussion}

\subsection{Vertical structure in the ``eyewall'' regions.}\label{Sec:comp}

Shadow measurements by \cite{Dyudina2009} led to the
concept of inner and outer eyewalls, each casting shadows on interior
cloud regions. One interpretation of these results is illustrated in
the right panel of Fig.\ \ref{Fig:polecontext}, which displays a
stair-step of cloud levels, with each step matching the height of the
obstruction needed to cast a shadow of the observed length on a flat
region interior to the obstruction.  However, as can be seen from
cloud structure models displayed in Fig.\ \ref{Fig:strucfits}, our
radiative transfer modeling provides no corroborating evidence for such
large changes of cloud elevation with latitude, nor of any structures
with large optical depths that would be expected for eyewall clouds.
Thus, it at first appears perplexing that the observed shadows
even exist. However, a detailed investigation of the structure of these
layers suggests an explanation: small step changes in optical depth versus latitude
in the somewhat translucent layers above the ammonia layer can produce shadows on the underlying layer.
This mechanism and others are evaluated in a companion paper \citep{Sro2019shadows},
which concludes that it is the only one plausibly consistent with the observations.
Here we present evidence for the existence of sharp optical depth
transisions that can make that mechanism work.

\begin{figure}[!hb]\centering
\hspace{-0.1in}\includegraphics[width=3.5in]{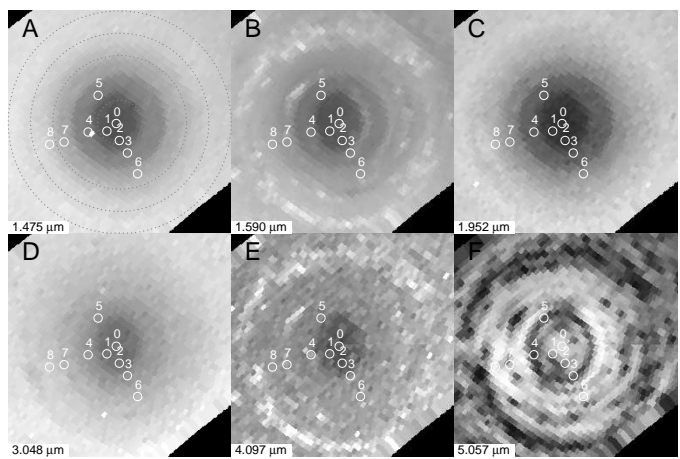}
\caption{Polar projection images identifying locations of background
  spectral samples obtained from VIMS cube V1539288419. For these
  images the subsolar and sub-spacecraft longitudes are at 178.9\degx
  E and 149.9\degx E respectively, which are clockwise angles from a
  horizontal line extending from the pole to the right.}\label{Fig:innershort}
\end{figure} 

To model the different latitude bands in the inner and outer eye regions,
 we selected nine 
spectral samples: three in the darkest inner region, three in the next
region between the two shadow boundaries, two just outside the outer
shadow, and one in the outer region.  The locations of these spectral
samples on selected polar projection images are shown in Fig.\ \ref{Fig:innershort}.  
Note that they all avoid the \nhtx-signature features.  These sample
 locations are the same as shown
in Fig.\ \ref{Fig:innereye}, and labeled with the same numbers. 
Our model fits for the locations shown in Fig.\ \ref{Fig:innershort}
 were already plotted in Fig.\ \ref{Fig:strucfits} and parameter
 values and uncertainties given in
 Table\ \ref{Tbl:backinner}. However, optical depths presented earlier
 were given at a wavelength of 2 \mumx.  It is also useful to convert
 these optical depths to the 752-nm wavelength at which shadows are
 observed in ISS images.

\begin{figure}[!htb]\centering
\includegraphics[width=3.15in]{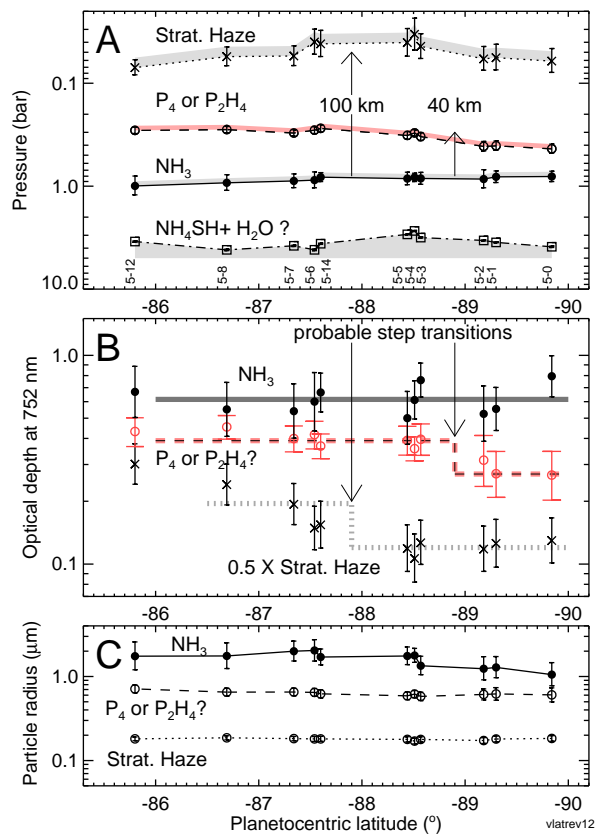}
\caption{Best-fit parameter values versus latitude for fits to inner polar region, 
including pressures (A), optical depths at
  752 nm (B), and particle sizes (C).  Recall that optical depths in
all other figures (and tables) are for a reference wavelength of 2 \mumx. Vertical
arrows are placed at latitudes indicated by step changes in brightness
of ISS images displayed in Fig.\ \ref{Fig:isstriple}.  The possible
step changes are indicated by dashed and dotted lines for upper tropospheric
and stratospheric layers respectively. The stratospheric haze optical depths
 are scaled by a factor of 1/2 to avoid overlap.}\label{Fig:odconvert}
\end{figure}

\begin{figure*}[!htb]\centering
\includegraphics[width=5.in]{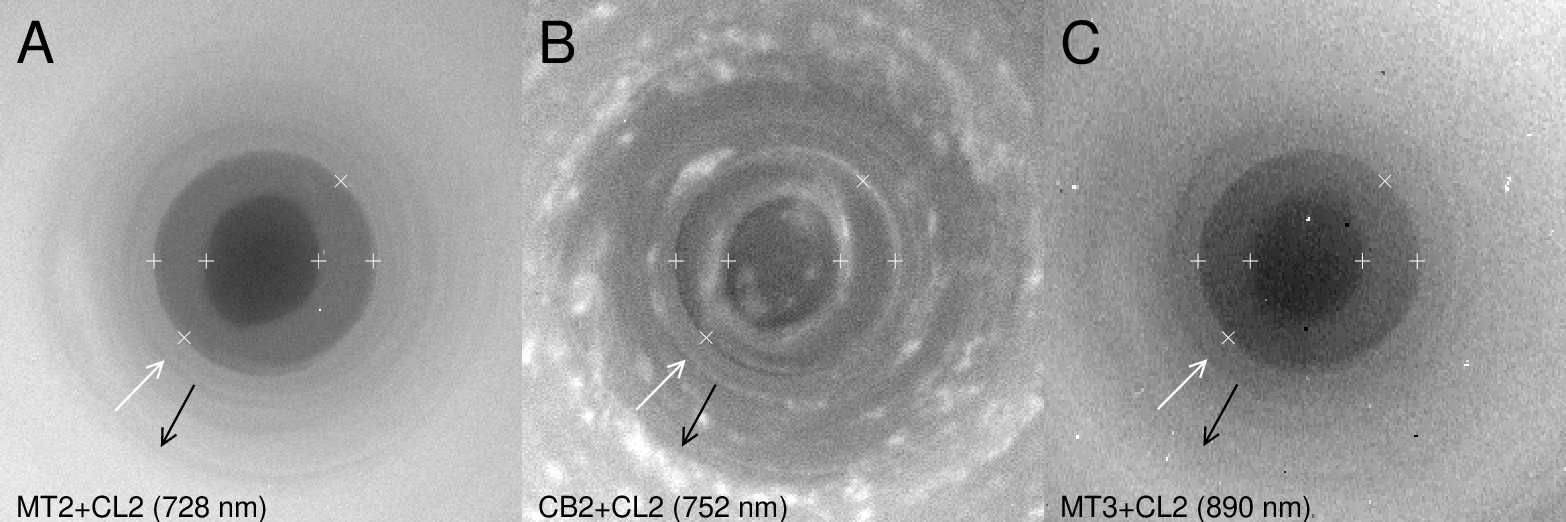}
\includegraphics[width=2.5in]{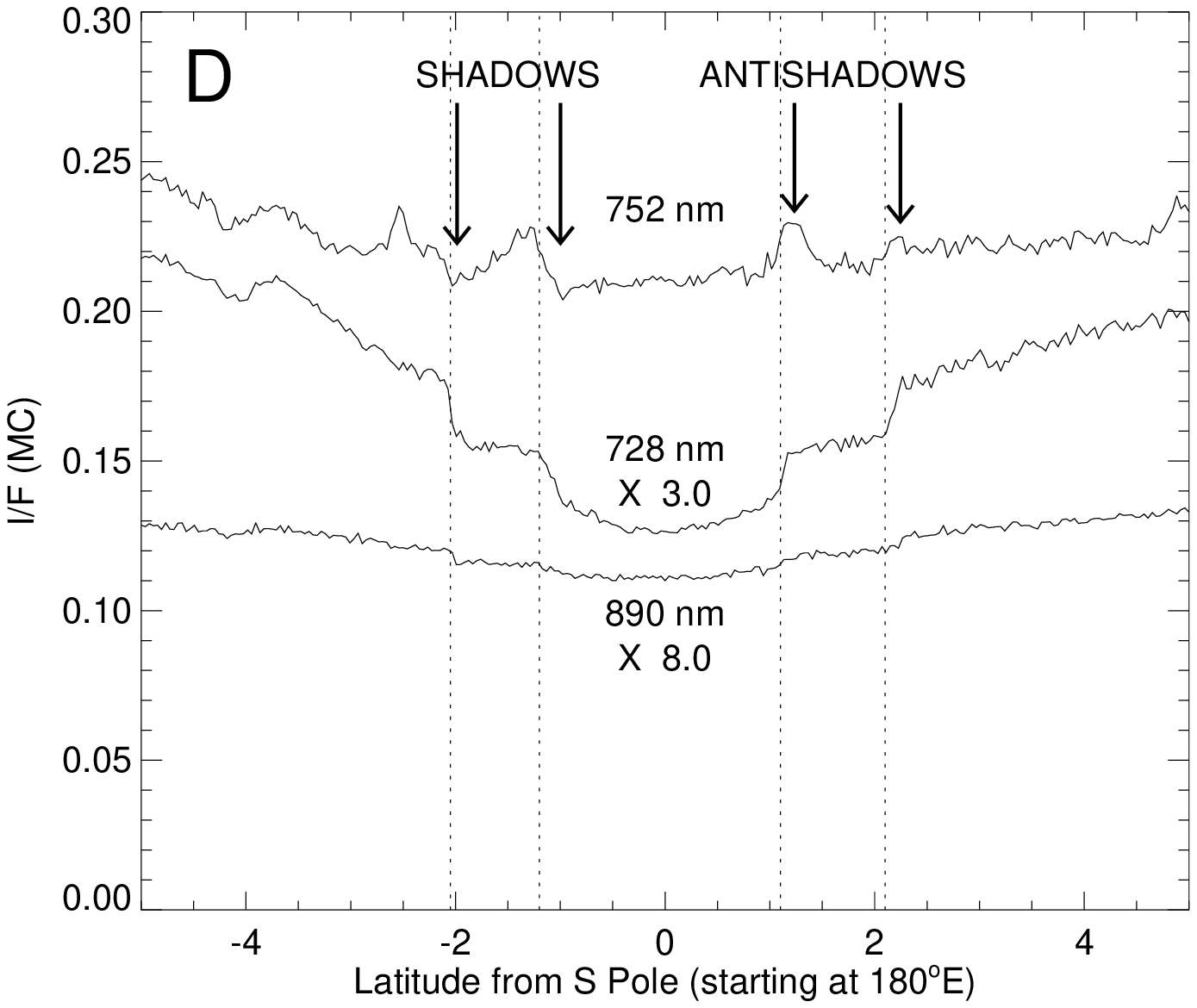}
\includegraphics[width=2.5in]{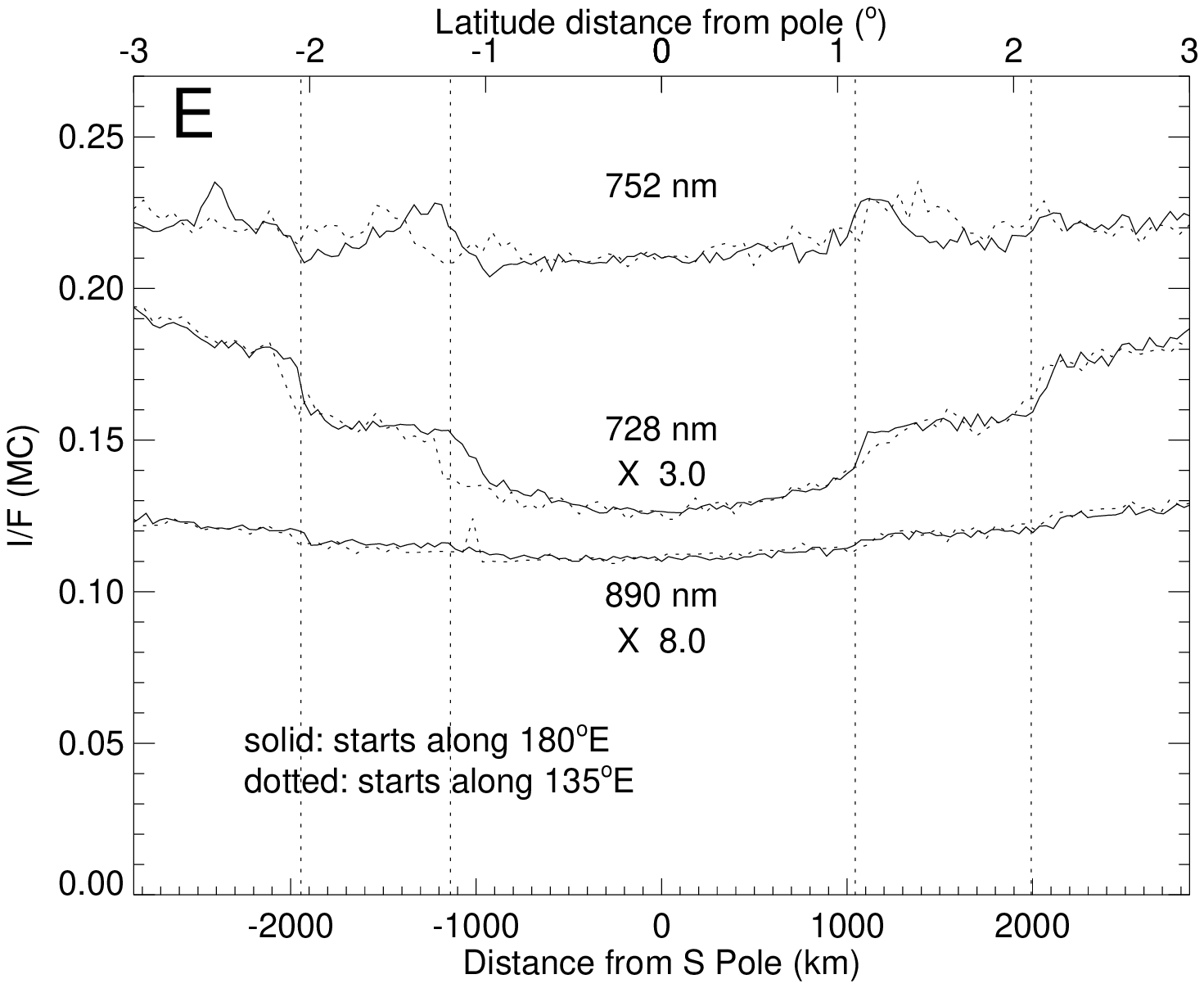}
\caption{Polar projections of 2006 ISS images for wavelengths with
  intermediate (A), low (B), and strong (C) methane gas
  absorption.  Plus signs provide reference marks at identical
  locations in each image. The direction of incoming sunlight
is indicated by white arrows, and direction toward the Cassini
spacecraft by the black arrows.  Images have been Minnaert corrected to
  approximately remove limb darkening. Bottom: plots of meridional
  scans in each image in two directions: horizontal (D) and
  sun-aligned (E, dotted). Vertical dotted lines are plotted at the
  same latitudes as the fiducial marks.  Arrows in D mark locations of
shadow and antishadow features.
}\label{Fig:isstriple}
\end{figure*}

The 752-nm optical depths are displayed versus latitude in
Fig.\ \ref{Fig:odconvert}, which indicate a possible step decrease
versus latitude in the optical depth of the stratospheric haze layer
might be close to the latitude of the outer ``eyewall'' shadow and a
similar stepped decrease in the optical depth of the putative
diphosphine layer may be located near the shadow related to the inner
``eyewall''.  These step changes, if real, are of roughly the
magnitude needed to create shadows on the underlying layer of \nht
aerosols, i.e. 0.1-0.2 optical depths according to \cite{Sro2019shadows}.  However,
given the relatively large error bars on the fitted optical depth
values, the steps are not fully established by our retrievals.
Although our radiative transfer analysis shows a decline in optical
depth towards the pole, the existence of step changes cannot be firmly
inferred from that analysis for two reasons: first, we cannot properly
model radiative transfer in close proximity to such a step change
because our model relies on the assumption of horizontal homogeneity
over a reasonable length scale, and second, the uncertainty in the
value of the derived parameters is too large to define a step change
over a small spatial region. Our radiative transfer results are consistent with step changes
and do constrain the size of the steps, if not their precise shapes.
 
\begin{figure}[!t]\centering
\includegraphics[width=3.in]{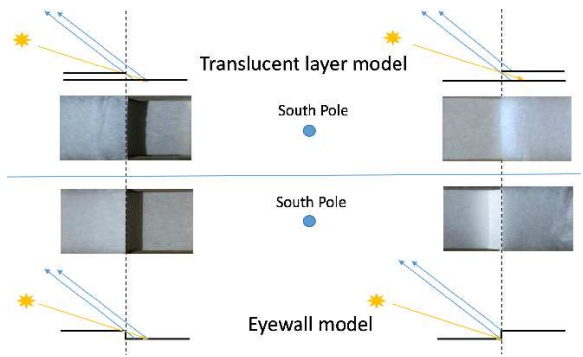}
\caption{Photographs of tracing-paper physical models illustrating two
  alternative shadow production mechanisms: a sharp change in optical
  depth of a translucent layer (top) or a step change in cloud
  pressure at an eyewall edge (bottom). Light from the sun is
  indicated by orange rays, and light to the observer is indicated by
  blue rays.  Each model produces a shadow on the lower layer when the
  transition is located on the sunward side (left) of the pole.  But
  on the opposite side (right) the translucent layer model displays a
  moderately bright ``antishadow'' produced by light shining
  underneath the top layer and providing extra illumination from
below, while the alternate model displays a
  brightly illuminated eyewall.  When illuminated and observed at the angles
  illustrated here, which are comparable to those for the ISS images
  in Fig.\ \ref{Fig:isstriple}, the antishadow is seen to extend from
  the upper layer boundary (marked by dashed lines) away from the
  pole, which is also the direction seen in the ISS images, while the
  bright eyewall is seen to extend towards the pole, in conflict with
  ISS observations.}
\vspace{0.2in}
\label{Fig:physmod}
\end{figure}

To confirm the sharpness of these transitions, we used ISS images with up to 8.6 times
the spatial resolution of VIMS. The ISS images shown in Fig.\ \ref{Fig:isstriple} illustrate a
correlation between features in the upper altitude cloud layers sensed
by 728-nm and 890-nm images, and the shadow features that appear in
the more deeply sensing 752-nm image.  Figure\ \ref{Fig:isstriple} also shows scans
along horizontal and sun-aligned meridians in the lower two panels.
 The ISS 728-nm image senses deeply enough
to register sharp changes in both stratospheric and \nht layers, but
the ISS 890-nm image sees a highly attenuated view due to overlying
methane absorption. The scans make clear that the shadows are relatively subtle features
of the order of 10\% or smaller. The scans also show brightening on the opposite
side of the pole, labeled in the figure as antishadows.  These arise from
extra illumination underneath the layers that cast shadows when they are on the
sunward side of the pole, as illustrated by the photographs of a toy physical model displayed
in Fig.\ \ref{Fig:physmod}.  These bright features cannot be due to light
reflected by eyewalls because they are on the wrong side of the
boundary that produces shadows when it is on the opposite side of the pole.  
 A more complete and quantitative treatment of this topic is
provided by \cite{Sro2019shadows}.

\subsection{Making and masking of the \nht spectral signature}\label{Sec:masking}

\begin{figure*}[!htb]\centering
\includegraphics[width=6in]{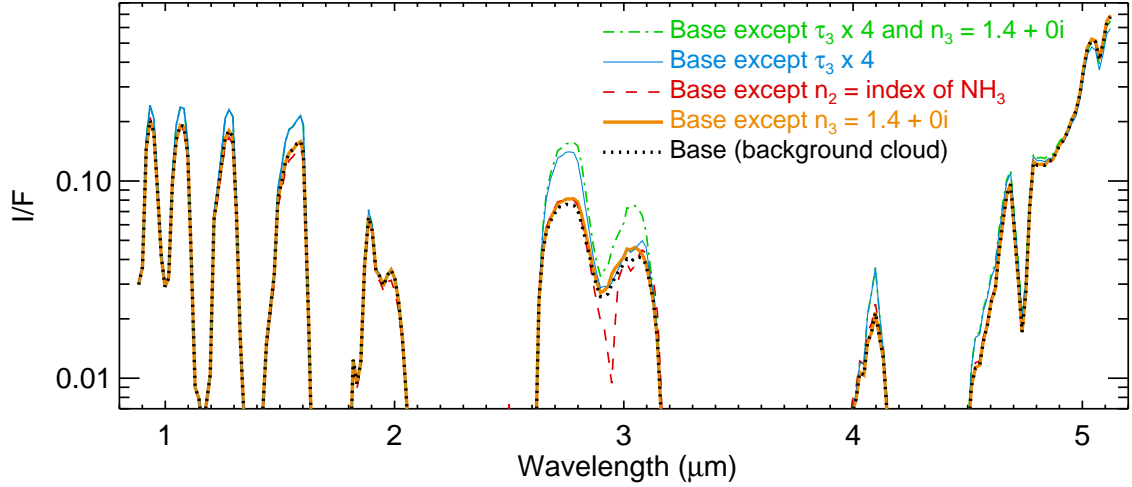}
\caption{Model calculations illustrating how the spectral signature of
  \nht is altered by aerosol vertical location and optical depth
  changes.  The base model is background model 5-4 in
  Table\ \ref{Tbl:backouter}.  Other curves show the effects of
  changing layer-3 index (orange) from the normal ammonia index
to one with the same real value but without absorption, changing the layer 2 index (red)
from its normal real value to the ammonia index, increasing
  the layer 3 (ammonia layer) optical depth (blue), and increasing the optical depth
  of layer 3 but removing its ammonia absorption (green).
}\label{Fig:mask}
\end{figure*}

Figure\ \ref{Fig:mask} illustrates how the ammonia signature varies
with the vertical level and optical depth of the ammonia layer.  The
base model in this figure (shown in black) is for the background cloud
spectrum from location 4 in Fig.\ \ref{Fig:outereye}.  If the top
tropospheric layer (the putative diphosphine layer) were instead
composed of \nhtx, the result would be a sharp and significant
spectral feature near 2.986 \mumx, as illustrated by the red model
spectrum in Fig.\ \ref{Fig:mask}.  There are two reasons that feature
is not seen in the base model.  First, the \nht cloud layer is
underneath 0.87 optical depths of the ``diphosphine'' layer, which is
helped in obscuring the feature by the large zenith angles of sun and
observer.  Second, the \nht layer is deep enough that it is affected
by overlying phosphine gas absorption, which also obscures the
absorption features produced by the cloud particles.  The
effectiveness of these masking effects is shown by the very small
difference between the base model and the model in which the \nht
layer absorption is turned off by setting n$_3$ = 1.4 + 0$i$ (orange curve).
   
The effect of \nht absorption is more apparent when the overlying
layer optical depth is reduced, or when the ammonia layer optical
depth is increased.  This is shown in Fig.\ \ref{Fig:mask} by the blue
and green curves. The blue spectrum is computed for the base model 
except for a factor of 4 increase in the
ammonia layer optical depth.  The green spectrum is for the same model
except that the ammonia absorption is turned off by setting n$_3$ = 1.4 +0$i$.
Without ammonia absorption, the I/F values at 2.7 \mum and 3.1 \mum increase
by comparable fractional amounts relative to the base model (black).  But with ammonia
absorption included, the I/F increase at 3.1 \mum becomes negligible, while the increase
at 2.7 \mum remains nearly the same as without ammonia absorption. This
shows that the reflectivity of the ammonia cloud at 3.1 \mum becomes
saturated because of its low single-scattering albedo.

\begin{figure*}[!htb]\centering
\includegraphics[width=6.in]{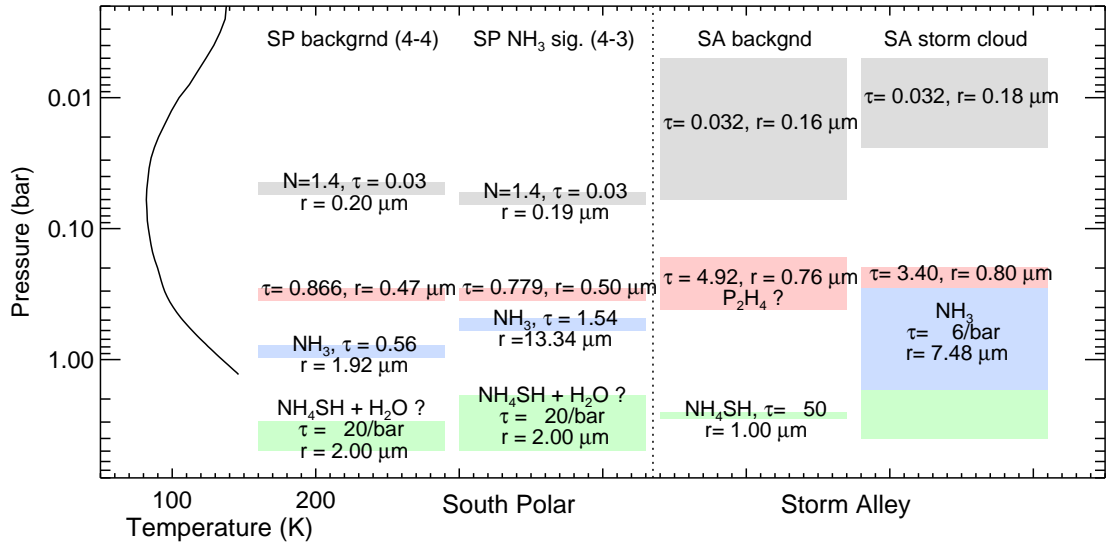}
\caption{Vertical cloud structure inferred from fitting background and
  \nht signature cloud spectra (large-$r_3$ solution) in comparison
  with models of the Storm Alley background and storm cloud structures
   for locations L and H in Fig. 22 of
  \cite{Sro2018dark}. Note that the derived bottom boundary of the
  stratospheric haze for the Storm Alley fits is highly uncertain and
  differences shown here are roughly within
  uncertainties.}\label{Fig:alleycomp}
\end{figure*}

\subsection{Comparison of polar clouds to Storm Alley clouds}

Here we compare our polar cloud structure models with typical models
derived by \cite{Sro2018dark} for the Storm Alley cloud features.  In
the left two columns of Fig.\ \ref{Fig:alleycomp} is a pair of south
polar model fits for an ammonia-signature cloud (at location 4-3) and
a background cloud (at location 4-4), and in the right is a pair of
Storm Alley fits for background and ammonia-signature storm clouds
from \cite{Sro2018dark}.  The most striking difference between these
models is the large optical depth of the upper tropospheric (pink)
layer in the Storm Alley case, both for the background and for the
storm cloud. The optical depth for Storm Alley background clouds is nearly
six times larger than for the background cloud in the south polar region.
For the storm cloud, although the optical depth of that layer is
reduced, its ratio to ammonia-signature polar clouds is more than a factor
 of four.  In Storm Alley, producing an ammonia signature requires
significant penetration of ammonia ice particles into the overlying
putative diphosphine layer, while in the south polar region the upper
tropospheric layer has such a low optical depth that relatively small
increases in the ammonia layer opacity or decreases in the upper layer
opacity can produce the signature.  There is a caveat regarding the
optical depth of the putative diphosphine layer in the Storm Alley
background cloud structure.  The \cite{Sro2018dark} fit to that
structure did not include a potential \nht ice particle layer.  It is
likely the case that the 4.9 optical depths of the diphosphine layer
could be split roughly equally between an upper layer of diphosphine
particles and a lower layer of \nht particles.  With that much optical
depth, it would be hard to distinguish the two configurations.

\subsection{The structure of polar dark spots}

There are two prominent dark features in Fig.\ \ref{Fig:outereye} that
are darker than background clouds over a wide range of continuum and
pseudo-continuum wavelengths.  One appears at the upper left and a
second can be seen near the bottom, intersecting the left edge of the
images. Although we did not conduct a systematic investigation of dark
features, we did fit one spectrum from the dark feature near location
4-2.  The immediate background structure surrounding this feature is
similar to prior fits at similar latitudes.  The main difference in
the structure of the dark feature itself was found in the ammonia
layer, which had a low optical depth of 0.54$\pm$0.18, while the
nearby background cloud had a rather normal ammonia layer optical
depth of 0.98$\pm$0.17.  These dark features merit further
investigation to search for possible clues about the dynamical
mechanisms that might create them.

\section{Summary and Conclusions}

 We were able to accurately model VIMS near-IR spectra of the south
 polar region, sampling background clouds from 71\degx S to nearly
 90\degx S, and \nhtx--signature features from 77.3\degx S to
 87.5\degx S, using a four-layer aerosol model, each with different
 scattering properties and composition: a stratospheric haze, assumed
 conservative at n = 1.4 at near-IR wavelengths, a putative diphosine
 cloud, an ammonia ice cloud layer, and a deep diffuse cloud assumed to extend from
 5 bars upward to an adjustable pressure, which is in the region where
 \nhfsh and \hto are likely condensibles. All layers were
 characterized as spherical particles.  The top three layers were
 characterized by fitted particle radius, pressure, and optical depth,
 and given a small physical thickness, making them essentially sheet
 clouds.  We also fit \pht and \asht gas mixing ratios.  Our analysis of
 Cassini/VIMS spectra of the south polar region of Saturn have led to
 the following conclusions.

{\it Background structure. }  The top three layers
have typical fitted pressures of 50 mbar, 320 mbar, and 980 mbar, with corresponding
particle radii of 0.19 \mumx, 0.6 \mumx, and 1.6 \mumx, and 2-\mum optical depths
of 0.023, 0.72, and 0.80, respectively.  The optical depth increases of the stratospheric haze
increases are a factor of 20 larger at 752 nm,  making the stratospheric haze as impactful as the
upper tropospheric layer in possible creation of shadows 
on the underlying \nht layer, the composition of which is confirmed by the
detection of its 2.9-3.1 \mum absorption feature. As the pole is approached, the stratospheric haze (layer 1)
optical depth decreases by a factor of 2 near 88\degx N and the putative
diphosphine haze (layer 2) decreases by about 30\% near 89\degx N.

{\it  \nhtx-signature structures. } In these structures the tops of the deep layer
  are all elevated by about 1 bar (higher in altitude) relative to the
  same layer in the typical background cloud structure, and the
  ammonia layer relative to its properties in background regions is
  typically elevated by about 200 mb, increased in optical depth by as
  much as 100\%, and increased in particle size (significantly for
the large-$r_3$ solutions).  This suggests that
  deep convection, perhaps in the water layer, is involved in
  generating the observed differences in the
  \nht layers, although there is no evidence of vertical mass
  transport from deep levels to the visible cloud deck, as there is
  for the Great Storm of 2010-2011. However, this might be an
  approximate realization of the Taylor-Proudman theorem, in which an
  upward convective pulse at the 2-3 bar level could force the entire
  column above that pulse to move upward, inducing a pulse of
  condensation at lower pressures, more evident in the ammonia layer
  than in the putative diphosphine layer.

{\it Why penetrating convection is not needed to see \nht signatures. }
The 3-\mum absorption is more apparent in the polar region due to the
reduced optical depth of the upper cloud layer (the putative
 diphosphine layer), which can be an order of
magnitude smaller than in other regions on Saturn, perhaps because of
polar downwelling or reduced rates of photochemical production,
or some combination of both.

{\it Lack of evidence for eyewalls. } The cloud structures inferred from our radiative transfer
  modeling are not consistent with the south polar cloud structure suggested by the results
 of \cite{Dyudina2009}.  None of the modeled cloud layers disappear,
  and their optical depths are relatively small and do not change
  dramatically with latitude. There is no evidence for an optically
  thick vertical wall of clouds as found in hurricanes on earth, and
no detection of lightning.

{\it Alternative shadow mechanism: step changes in optical depths
  of translucent layers. } From ISS imaging at 728 nm, which can
detect the top two layers of our model, but not the ammonia layer, it
is clear that the optical depth changes noted above probably occur in
sharp steps, one at 2\deg from the pole, and a second near 1.2\deg
from the pole. As confirmed by a simplified Monte Carlo analysis
presented in our companion paper \citep{Sro2019shadows}, these optical
depth transitions are probably large enough and sharp enough to
produce shadows and antishadows of roughly the observed magnitudes.

{\it Existence of antishadows supports the alternative mechanism. }
The observations of bright features emanating from the same horizontal
aerosol boundaries as shadows, but seen on opposite sides of the pole,
can arise from extra sunlight shining underneath a layer due to lower
optical depth on the sunward side of the layer, producing what we call
antishadows.  For the VIMS and ISS observing geometry, these features
extend away from the pole, while illuminated eyewalls would be seen as
bright features extending toward the pole, the opposite of what is
observed on Saturn. This provides additional strong evidence that step
changes in physically thin overlying layers are the source of shadows.

{\it Phosphine. }  From fitting models to background clouds, there seems to be
 a fairly latitude
  independent deep mixing ratio, with an average of 4.33$\pm$0.05 ppm,
  and a peak-to-peak range of 4.1-4.7 ppm.  However, we found a significant increase in
  the break-point pressure from about 200 mbar near 86\degx S to 400
  mbar at 89.8\degx S.  This result seems at least qualitatively
  consistent with the polar decline within 2\deg of the south pole
  that was found by \cite{Fletcher2008}, although our different
  modeling approach makes direct comparisons uncertain. Fitting of \nhtx-signature
structures suggests that the \pht mixing ratios are locally increased by factors of 2-3
over background values and that pressure break points in those regions
also seem to be generally increased.

{\it Arsine. }  From background fits it appears that south polar arsine volume
  mixing ratios are relatively latitude independent between
 77\degx S and 86\degx S, over which the average value
was found to be 1.71$\pm$0.13 ppb.
From 86\degx to the pole the mixing ratio declines to 1.1$\pm$0.2 ppb.
This suggests that arsine's mixing ratio declines with altitude, as suggested
previously by results of \cite{Bezard1989}, such that
descending motions near the pole produce a lower effective mixing ratio.
For \nhtx-signature fits we find somewhat decreased mixing ratios for the
small-$r_3$ solutions, but the opposite for the large-$r_3$ solutions.

There are more than a few questions to be resolved about the south polar cloud
structure of Saturn.  The composition of the ubiquitous upper tropospheric layer seems
likely to be either diphosphine or phosphorus. But to test which
better fits spectral observations, we need better measurements of the
optical properties of both of these materials under Saturnian
conditions, especially of diphosphine about which very little is
known.  There is also a caveat that must be recognized in the modeling of
the discrete bright features.  These are probably affected to at least
a small degree by the violation of the assumption of horizontal homogeneity,
as many of these features are of the same horizontal scale as the vertical
distance between cloud layers.  Furthermore, it must be noted that these
results are based on one snapshot taken in 2006.  Temporal evolution of
the south polar cloud structure is also uncertain because of the scarcity
of observations of adequate spatial resolution.  However, based on analysis of
HST observations by \cite{Perez-Hoyos2005} there is some evidence that optical
depths at 80\degx S and 86\degx S may have declined significantly in the several
years prior to the 2003 southern solstice.  If and when they begin to increase
remains to be determined. 

\section*{Acknowledgments.} \addcontentsline{toc}{section}{Acknowledgments}

Support for this work was provided by
NASA through its Cassini Data Analysis and
Participating Scientists Program via grant NNX15AL10G and by
the subsequent Cassini Data Analysis Program via grant 80NSSC18K0966.
This research has made use of the USGS Integrated Software for Imagers and 
Spectrometers (ISIS).  We thank an anonymous reviewer for detailed comments
that led us to make significant improvements in the paper.
The archived data associated with the paper can be obtained from
Planetary Data System (PDS) Atmospheres Node at
https://pds-atmospheres.nmsu.edu/data\_and\_services/atmospheres\_data/ catalog.htm,
and includes calibrated Cassini VIMS and ISS datasets, along with a 
description of the calibration, and tabular data for all tables and 
figures contained in the paper as well as a public domain 
version of the paper.

\vspace{-0.15in}

\begin{thebibliography}{53}
\expandafter\ifx\csname natexlab\endcsname\relax\def\natexlab#1{#1}\fi
\expandafter\ifx\csname url\endcsname\relax
  \def\url#1{\texttt{#1}}\fi
\expandafter\ifx\csname urlprefix\endcsname\relax\def\urlprefix{URL }\fi

\bibitem[{{Acton}(1996)}]{Acton1996}
{Acton}, C.~H., 1996. {Ancillary data services of NASA's Navigation and
  Ancillary Information Facility}. Planet. and Space Sci. 44, 65--70.

\bibitem[{{Anderson} et~al.(2004){Anderson}, {Sides}, {Soltesz}, {Sucharski},
  and {Becker}}]{Anderson2004}
{Anderson}, J.~A., {Sides}, S.~C., {Soltesz}, D.~L., {Sucharski}, T.~L.,
  {Becker}, K.~J., 2004. {Modernization of the Integrated Software for Imagers
  and Spectrometers}. In: {Mackwell}, S., {Stansbery}, E. (Eds.), Lunar and
  Planetary Science Conference. Vol.~35 of Lunar and Planetary Science
  Conference. p. 2039.

\bibitem[{{Atreya} and {Wong}(2005)}]{Atreya2005SSR}
{Atreya}, S.~K., {Wong}, A., 2005. {Coupled Clouds and Chemistry of the Giant
  Planets - A Case for Multiprobes}. Space Sci. Rev. 116, 121--136.

\bibitem[{{Baines} et~al.(2009){Baines}, {Delitsky}, {Momary}, {Brown},
  {Buratti}, {Clark}, and {Nicholson}}]{Baines2009stormclouds}
{Baines}, K.~H., {Delitsky}, M.~L., {Momary}, T.~W., {Brown}, R.~H., {Buratti},
  B.~J., {Clark}, R.~N., {Nicholson}, P.~D., 2009. {Storm clouds on Saturn:
  Lightning-induced chemistry and associated materials consistent with
  Cassini/VIMS spectra}. \planss 57, 1650--1658.

\bibitem[{{Baines} et~al.(2008){Baines}, {Momary}, {Kim}, {Roos-Serote},
  {Showman}, {Atreya}, {Brown}, {Buratti}, {Clark}, and
  {Nicholson}}]{Baines2008}
{Baines}, K.~H., {Momary}, T.~W., {Kim}, J.~H., {Roos-Serote}, M., {Showman},
  A.~P., {Atreya}, S.~K., {Brown}, R.~H., {Buratti}, B.~J., {Clark}, R.~N.,
  {Nicholson}, P.~D., 2008. {Saturn's dynamic atmosphere at depth: Physical
  characteristics, winds, and spatial constraints on trace gas variability near
  the 3-bar level and their dynamical implications from Cassini-Hygens/VIMS}.
  In: Poster Presented at Saturn after Cassini Huygens Conference, Imperial
  College, London.

\bibitem[{{Baines} et~al.(2018){Baines}, {Sromovsky}, {Fry}, {Momary}, {Brown},
  {Buratti}, {Clark}, {Nicholson}, and {Sotin}}]{Baines2018GeoRL}
{Baines}, K.~H., {Sromovsky}, L.~A., {Fry}, P.~M., {Momary}, T.~W., {Brown},
  R.~H., {Buratti}, B.~J., {Clark}, R.~N., {Nicholson}, P.~D., {Sotin}, C.,
  2018. {The Eye of Saturn's North Polar Vortex: Unexpected Cloud Structures
  Observed at High Spatial Resolution by Cassini/VIMS}. \grl 45, 5867--5875.

\bibitem[{{Barstow} et~al.(2016){Barstow}, {Irwin}, {Fletcher}, {Giles}, and
  {Merlet}}]{Barstow2016}
{Barstow}, J.~K., {Irwin}, P.~G.~J., {Fletcher}, L.~N., {Giles}, R.~S.,
  {Merlet}, C., 2016. {Probing Saturn's tropospheric cloud with Cassini/VIMS}.
  Icarus 271, 400--417.

\bibitem[{{B\'ezard} et~al.(1989){B\'ezard}, {Drossart}, {Lellouch}, {Tarrago},
  and {Maillard}}]{Bezard1989}
{B\'ezard}, B., {Drossart}, P., {Lellouch}, E., {Tarrago}, G., {Maillard},
  J.~P., 1989. {Detection of arsine in Saturn}. \apj 346, 509--513.

\bibitem[{{Borysow}(1991)}]{Borysow1991h2h2f}
{Borysow}, A., 1991. {Modeling of collision-induced infrared absorption spectra
  of H$_2$-H$_2$ pairs in the fundamental band at temperatures from 20 to 300
  K}. Icarus 92, 273--279.

\bibitem[{{Borysow}(1992)}]{Borysow1992h2he}
{Borysow}, A., 1992. {New model of collision-induced infrared absorption
  spectra of H$_2$-He pairs in the 2-2.5 micron range at temperatures from 20
  to 300 K - an update}. Icarus 96, 169--175.

\bibitem[{{Borysow}(1993)}]{Borysow1993errat}
{Borysow}, A., 1993. {Erratum}. Icarus 106, 614.

\bibitem[{{Bowles} et~al.(2008){Bowles}, {Calcutt}, {Irwin}, and
  {Temple}}]{Bowles2008}
{Bowles}, N., {Calcutt}, S., {Irwin}, P., {Temple}, J., 2008. {Band parameters
  for self-broadened ammonia gas in the range 0.74 to 5.24 {$\mu$}m to support
  measurements of the atmosphere of the planet Jupiter}. Icarus 196, 612--624.

\bibitem[{{Briggs} and {Sackett}(1989)}]{Briggs1989}
{Briggs}, F.~H., {Sackett}, P.~D., 1989. {Radio observations of Saturn as a
  probe of its atmosphere and cloud structure}. Icarus 80, 77--103.

\bibitem[{{Brown} et~al.(2004){Brown}, {Baines}, {Bellucci}, {Bibring},
  {Buratti}, {Capaccioni}, {Cerroni}, {Clark}, {Coradini}, {Cruikshank},
  {Drossart}, {Formisano}, {Jaumann}, {Langevin}, {Matson}, {McCord},
  {Mennella}, {Miller}, {Nelson}, {Nicholson}, {Sicardy}, and
  {Sotin}}]{Brown2004}
{Brown}, R.~H., {Baines}, K.~H., {Bellucci}, G., {Bibring}, J.-P., {Buratti},
  B.~J., {Capaccioni}, F., {Cerroni}, P., {Clark}, R.~N., {Coradini}, A.,
  {Cruikshank}, D.~P., {Drossart}, P., {Formisano}, V., {Jaumann}, R.,
  {Langevin}, Y., {Matson}, D.~L., {McCord}, T.~B., {Mennella}, V., {Miller},
  E., {Nelson}, R.~M., {Nicholson}, P.~D., {Sicardy}, B., {Sotin}, C., 2004.
  {The Cassini Visual and Infrared Mapping Spectrometer (VIMS) Investigation}.
  Space Sci. Rev. 115, 111--168.

\bibitem[{{Clapp} and {Miller}(1996)}]{Clapp1996}
{Clapp}, M.~L., {Miller}, R.~E., 1996. {Complex Refractive Indices of
  Crystalline Hydrazine from Aerosol Extinction Spectra}. Icarus 123, 396--403.

\bibitem[{{Clark} et~al.(2012){Clark}, {Cruikshank}, {Jaumann}, {Brown},
  {Stephan}, {Dalle Ore}, {Eric Livo}, {Pearson}, {Curchin}, {Hoefen},
  {Buratti}, {Filacchione}, {Baines}, and {Nicholson}}]{Clark2012}
{Clark}, R.~N., {Cruikshank}, D.~P., {Jaumann}, R., {Brown}, R.~H., {Stephan},
  K., {Dalle Ore}, C.~M., {Eric Livo}, K., {Pearson}, N., {Curchin}, J.~M.,
  {Hoefen}, T.~M., {Buratti}, B.~J., {Filacchione}, G., {Baines}, K.~H.,
  {Nicholson}, P.~D., 2012. {The surface composition of Iapetus: Mapping
  results from Cassini VIMS}. Icarus 218, 831--860.

\bibitem[{Drummond and Thekaekara(1973)}]{Drummond1973}
Drummond, A., Thekaekara, M., 1973. The extraterrestrial solar spectrum.
  Institute of Environmental Sciences, Mt. Prospect Illinois.

\bibitem[{{Dyudina} et~al.(2009){Dyudina}, {Ingersoll}, {Ewald}, {Vasavada},
  {West}, {Baines}, {Momary}, {Del Genio}, {Barbara}, {Porco}, {Achterberg},
  {Flasar}, {Simon-Miller}, and {Fletcher}}]{Dyudina2009}
{Dyudina}, U.~A., {Ingersoll}, A.~P., {Ewald}, S.~P., {Vasavada}, A.~R.,
  {West}, R.~A., {Baines}, K.~H., {Momary}, T.~W., {Del Genio}, A.~D.,
  {Barbara}, J.~M., {Porco}, C.~C., {Achterberg}, R.~K., {Flasar}, F.~M.,
  {Simon-Miller}, A.~A., {Fletcher}, L.~N., 2009. {Saturn's south polar vortex
  compared to other large vortices in the Solar System}. Icarus 202, 240--248.

\bibitem[{{Dyudina} et~al.(2008){Dyudina}, {Ingersoll}, {Ewald}, {Vasavada},
  {West}, {Del Genio}, {Barbara}, {Porco}, {Achterberg}, {Flasar},
  {Simon-Miller}, and {Fletcher}}]{Dyudina2008Sci}
{Dyudina}, U.~A., {Ingersoll}, A.~P., {Ewald}, S.~P., {Vasavada}, A.~R.,
  {West}, R.~A., {Del Genio}, A.~D., {Barbara}, J.~M., {Porco}, C.~C.,
  {Achterberg}, R.~K., {Flasar}, F.~M., {Simon-Miller}, A.~A., {Fletcher},
  L.~N., 2008. {Dynamics of Saturn's South Polar Vortex}. Science 319, 1801.

\bibitem[{{Fletcher} et~al.(2011){Fletcher}, {Baines}, {Momary}, {Showman},
  {Irwin}, {Orton}, {Roose-Serote}, and {Merlit}}]{Fletcher2011vims}
{Fletcher}, L.~N., {Baines}, K.~H., {Momary}, T.~M., {Showman}, A.~S., {Irwin},
  P.~G.~J., {Orton}, G.~S., {Roose-Serote}, M.~R., {Merlit}, C., 2011. {Saturn's tropospheric
  composition and clouds from Cassini/VIMS 4.6 -- 5.1 $\mu$m nightside
  spectroscopy}. Icarus 214, 510--533.

\bibitem[{{Fletcher} et~al.(2008){Fletcher}, {Irwin}, {Orton}, {Teanby},
  {Achterberg}, {Bjoraker}, {Read}, {Simon-Miller}, {Howett}, {de Kok},
  {Bowles}, {Calcutt}, {Hesman}, and {Flasar}}]{Fletcher2008}
{Fletcher}, L.~N., {Irwin}, P.~G.~J., {Orton}, G.~S., {Teanby}, N.~A.,
  {Achterberg}, R.~K., {Bjoraker}, G.~L., {Read}, P.~L., {Simon-Miller}, A.~A.,
  {Howett}, C., {de Kok}, R., {Bowles}, N., {Calcutt}, S.~B., {Hesman}, B.,
  {Flasar}, F.~M., 2008. {Temperature and Composition of Saturn's Polar Hot
  Spots and Hexagon}. Science 319, 79.

\bibitem[{{Fletcher} et~al.(2009{\natexlab{a}}){Fletcher}, {Orton}, {Teanby},
  and {Irwin}}]{Fletcher2009ph3}
{Fletcher}, L.~N., {Orton}, G.~S., {Teanby}, N.~A., {Irwin}, P.~G.~J.,
  2009{\natexlab{a}}. {Phosphine on Jupiter and Saturn from Cassini/CIRS}.
  Icarus 202, 543--564.

\bibitem[{{Fletcher} et~al.(2009{\natexlab{b}}){Fletcher}, {Orton}, {Teanby},
  {Irwin}, and {Bjoraker}}]{Fletcher2009ch4saturn}
{Fletcher}, L.~N., {Orton}, G.~S., {Teanby}, N.~A., {Irwin}, P.~G.~J.,
  {Bjoraker}, G.~L., 2009{\natexlab{b}}. {Methane and its isotopologues on
  Saturn from Cassini/CIRS observations}. Icarus 199, 351--367.

\bibitem[{{Fouchet} et~al.(2009){Fouchet}, {Moses}, and
  {Conrath}}]{Fouchet2009}
{Fouchet}, T., {Moses}, J.~I., {Conrath}, B.~J., 2009. {Saturn: Composition and
  Chemistry}. In: {Dougherty}, M.~K., {Esposito}, L.~W., {Krimigis}, S.~M.
  (Eds.), Saturn from Cassini-Huygens. Springer Dordrecht Heidelberg London New
  York, pp. 83--112.

\bibitem[{{Frankiss}(1968)}]{Frankiss1968}
{Frankiss}, S.~G., 1968. {Vibrational Spectrum and Structure of Solid
  Diphosphine}. Inorg. Chem. 7, 1931--1933.

\bibitem[{{Hansen} and {Travis}(1974)}]{Hansen1974}
{Hansen}, J.~E., {Travis}, L.~D., 1974. {Light scattering in planetary
  atmospheres}. Space Sci. Rev. 16, 527--610.

\bibitem[{{Hide}(1966)}]{Hide1966PSS}
{Hide}, R., 1966. {On the circulation of the atmospheres of Jupiter and
  Saturn}. \planss 14, 669.

\bibitem[{{Howett} et~al.(2007){Howett}, {Carlson}, {Irwin}, and
  {Calcutt}}]{Howett2007}
{Howett}, C.~J.~A., {Carlson}, R.~W., {Irwin}, P.~G.~J., {Calcutt}, S.~B.,
  2007. {Optical constants of ammonium hydrosulfide ice and ammonia ice}.
  Journal of the Optical Society of America B Optical Physics 24, 126--136.

\bibitem[{{Karkoschka} and {Tomasko}(2010)}]{Kark2010ch4}
{Karkoschka}, E., {Tomasko}, M.~G., 2010. {Methane absorption coefficients for
  the jovian planets from laboratory, Huygens, and HST data}. Icarus 205,
  674--694.

\bibitem[{{Lindal} et~al.(1985){Lindal}, {Sweetnam}, and
  {Eshleman}}]{Lindal1985}
{Lindal}, G.~F., {Sweetnam}, D.~N., {Eshleman}, V.~R., 1985. {The atmosphere of
  Saturn - an analysis of the Voyager radio occultation measurements}. \aj 90,
  1136--1146.

\bibitem[{{Martonchik} et~al.(1984){Martonchik}, {Orton}, and
  {Appleby}}]{Martonchik1984}
{Martonchik}, J.~V., {Orton}, G.~S., {Appleby}, J.~F., 1984. {Optical
  properties of NH$_3$ ice from the far infrared to the near ultraviolet}.
  Appl. Optics 23, 541--547.

\bibitem[{{Miller} et~al.(1996){Miller}, {Klein}, {Juergens}, {Mehaffey},
  {Oseas}, {Garcia}, {Giandomenico}, {Irigoyen}, {Hickok}, {Rosing}, {Sobel},
  {Bruce}, {Flamini}, {Devidi}, {Reininger}, {Dami}, {Soufflot}, {Langevin},
  and {Huntzinger}}]{Miller1996SPIE}
{Miller}, E.~A., {Klein}, G., {Juergens}, D.~W., {Mehaffey}, K., {Oseas},
  J.~M., {Garcia}, R.~A., {Giandomenico}, A., {Irigoyen}, R.~E., {Hickok}, R.,
  {Rosing}, D., {Sobel}, H.~R., {Bruce}, C.~F., {Flamini}, E., {Devidi}, R.,
  {Reininger}, F.~M., {Dami}, M., {Soufflot}, A., {Langevin}, Y., {Huntzinger},
  G., 1996. {The Visual and Infrared Mapping Spectrometer for Cassini}. In:
  {Horn}, L. (Ed.), Society of Photo-Optical Instrumentation Engineers (SPIE)
  Conference Series. Vol. 2803. pp. 206--220.

\bibitem[{{Nixon}(1956)}]{Nixon1956}
{Nixon}, E.~R., 1956. {The Infrared Spectrum of Biphosphine}. J. Phys. Chem.
  60, 1054--1059.

\bibitem[{{Noll} et~al.(1989){Noll}, {Geballe}, and {Knacke}}]{Noll1989AsH3}
{Noll}, K.~S., {Geballe}, T.~R., {Knacke}, R.~F., 1989. {Arsine in Saturn and
  Jupiter}. \apjl 338, L71--L74.

\bibitem[{{Noy} et~al.(1981){Noy}, {Podolak}, and {Bar-Nun}}]{Noy1981}
{Noy}, N., {Podolak}, M., {Bar-Nun}, A., 1981. {Photochemistry of phosphine and
  Jupiter's Great Red Spot}. \jgr 86, 11985--11988.

\bibitem[{{Pedlosky}(1982)}]{Pedlosky1982}
{Pedlosky}, J., 1982. {Geophysical fluid dynamics}. New York and Berlin,
  Springer-Verlag, 1982.~636 p.

\bibitem[{{P{\'e}rez-Hoyos} et~al.(2005){P{\'e}rez-Hoyos},
  {S{\'a}nchez-Lavega}, {French}, and {Rojas}}]{Perez-Hoyos2005}
{P{\'e}rez-Hoyos}, S., {S{\'a}nchez-Lavega}, A., {French}, R.~G., {Rojas},
  J.~F., 2005. {Saturn's cloud structure and temporal evolution from ten years
  of Hubble Space Telescope images (1994 -- 2003)}. Icarus 176, 155--174.

\bibitem[{{P{\'e}rez-Hoyos} et~al.(2016){P{\'e}rez-Hoyos}, {Sanz-Requena},
  {S{\'a}nchez-Lavega}, {Irwin}, and {Smith}}]{Perez-Hoyos2016}
{P{\'e}rez-Hoyos}, S., {Sanz-Requena}, J.~F., {S{\'a}nchez-Lavega}, A.,
  {Irwin}, P.~G.~J., {Smith}, A., 2016. {Saturn's tropospheric particles phase
  function and spatial distribution from Cassini ISS 2010-11 observations}.
  Icarus 277, 1--18.

\bibitem[{{Porco} et~al.(2004){Porco}, {West}, {Squyres}, {McEwen}, {Thomas},
  {Murray}, {Delgenio}, {Ingersoll}, {Johnson}, {Neukum}, {Veverka}, {Dones},
  {Brahic}, {Burns}, {Haemmerle}, {Knowles}, {Dawson}, {Roatsch}, {Beurle}, and
  {Owen}}]{Porco2004SSR}
{Porco}, C.~C., {West}, R.~A., {Squyres}, S., {McEwen}, A., {Thomas}, P.,
  {Murray}, C.~D., {Delgenio}, A., {Ingersoll}, A.~P., {Johnson}, T.~V.,
  {Neukum}, G., {Veverka}, J., {Dones}, L., {Brahic}, A., {Burns}, J.~A.,
  {Haemmerle}, V., {Knowles}, B., {Dawson}, D., {Roatsch}, T., {Beurle}, K.,
  {Owen}, W., 2004. {Cassini Imaging Science: Instrument Characteristics and
  Anticipated Scientific Investigations at Saturn}. Space Science Reviews 115,
  363--497.

\bibitem[{{Press} et~al.(1992){Press}, {Teukolsky}, {Vetterling}, and
  {Flannery}}]{Press1992}
{Press}, W.~H., {Teukolsky}, S.~A., {Vetterling}, W.~T., {Flannery}, B.~P.,
  1992. {Numerical recipes in FORTRAN. The art of scientific computing, 2nd
  ed.} Cambridge: University Press.

\bibitem[{{Prinn} and {Lewis}(1975)}]{Prinn1975}
{Prinn}, R.~G., {Lewis}, J.~S., 1975. {Phosphine on Jupiter and implications
  for the Great Red Spot}. Science 190, 274--276.

\bibitem[{{Roux} et~al.(1979){Roux}, {Wood}, and {Smith}}]{Roux1979}
{Roux}, J.~A., {Wood}, B.~E., {Smith}, A.~M., 1979. {Optical properties of thin
  H$_2$0, NH$_3$, and CO$_2$ cryofilms. AEDC-TR-79-57}. Arnold Engineering
  Development Center, Tennessee.

\bibitem[{{Sromovsky} et~al.(2013){Sromovsky}, {Baines}, and
  {Fry}}]{Sro2013gws}
{Sromovsky}, L.~A., {Baines}, K.~H., {Fry}, P.~M., 2013. {Saturn's Great Storm
  of 2010-2011: Evidence for ammonia and water ices from analysis of VIMS
  spectra}. Icarus 226, 402--418.

\bibitem[{{Sromovsky} et~al.(2018){Sromovsky}, {Baines}, and
  {Fry}}]{Sro2018dark}
{Sromovsky}, L.~A., {Baines}, K.~H., {Fry}, P.~M., 2018. {Models of bright
  storm clouds and related dark ovals in Saturn's Storm Alley as constrained by
  2008 Cassini/VIMS spectra}. Icarus 302, 360--385.

\bibitem[{{Sromovsky} et~al.(2016){Sromovsky}, {Baines}, {Fry}, and
  {Momary}}]{Sro2016}
{Sromovsky}, L.~A., {Baines}, K.~H., {Fry}, P.~M., {Momary}, T.~W., 2016.
  {Cloud clearing in the wake of Saturn's Great Storm of 2010-2011 and
  suggested new constraints on Saturn's He/H$_{2}$ ratio}. Icarus 276,
  141--162.

\bibitem[{{Sromovsky} and {Fry}(2010)}]{Sro2010iso}
{Sromovsky}, L.~A., {Fry}, P.~M., 2010. {The source of 3-{$\mu$}m absorption in
  Jupiter's clouds: Reanalysis of ISO observations using new NH$_{3}$
  absorption models}. Icarus 210, 211--229.

\bibitem[{{Sromovsky} et~al.(2019){Sromovsky}, {Fry}, and
  {Baines}}]{Sro2019shadows}
{Sromovsky}, L.~A., {Fry}, P.~M., {Baines}, K.~H., 2019. {Interpretation of
  south polar cloud shadows and antishadows on Saturn}. Icarus, submitted.

\bibitem[{{Sromovsky} et~al.(2012){Sromovsky}, {Fry}, {Boudon}, {Campargue},
  and {Nikitin}}]{Sro2012LBL}
{Sromovsky}, L.~A., {Fry}, P.~M., {Boudon}, V., {Campargue}, A., {Nikitin}, A.,
  2012. {Comparison of line-by-line and band models of near-IR methane
  absorption applied to outer planet atmospheres}. Icarus 218, 1--23.

\bibitem[{{Visscher} et~al.(2009){Visscher}, {Sperier}, {Moses}, and
  {Keane}}]{Visscher2009}
{Visscher}, C., {Sperier}, A.~D., {Moses}, J.~I., {Keane}, T.~C., 2009.
  {Phosphine and Ammonia Photochemistry in Jupiter's Troposphere}. In: Lunar
  and Planetary Science Conference. Vol.~40 of Lunar and Planetary Science
  Conference. p. 1201.

\bibitem[{{Warren}(1984)}]{Warren1984}
{Warren}, S.~G., 1984. {Optical constants of ice from the ultraviolet to the
  microwave}. Appl. Optics 23, 1206--1225.

\bibitem[{{Weidenschilling} and {Lewis}(1973)}]{Weidenschilling1973}
{Weidenschilling}, S.~J., {Lewis}, J.~S., 1973. {Atmospheric and cloud
  structures of the jovian planets}. Icarus 20, 465--476.

\bibitem[{Wohlfarth(2008)}]{LandoltBornstein2008}
Wohlfarth, C., 2008. Refractive index of diphosphine: Datasheet from
  landolt-b{\"o}rnstein - group iii condensed matter {\textperiodcentered}
  volume 47: ``refractive indices of pure liquids and binary liquid mixtures
  (supplement to iii/38)'' in springermaterials). Copyright 2008
  Springer-Verlag Berlin Heidelberg.

\bibitem[{{Zheng} and {Borysow}(1995)}]{Zheng1995h2h2o1}
{Zheng}, C., {Borysow}, A., 1995. {Modeling of collision-induced infrared
  absorption spectra of H$_2$ pairs in the first overtone band at temperatures
  from 20 to 500 K}. Icarus 113, 84--90.

\end{thebibliography}

\end{document}